\documentclass{aa}  
\usepackage[english]{babel}
\usepackage{amsmath}
\usepackage{graphicx}
\usepackage{natbib}
\usepackage[colorlinks=true, allcolors=blue]{hyperref}
\usepackage{txfonts}
\begin{document}
 \title{HALO
 }
 \subtitle{II: Constraining Hubble constant $H_{0}$ through continuum delay fitting of Fairall~9}
 \titlerunning{Hubble constant constraint through AGN Light curves Observation}


\author{Amit Kumar Mandal \inst{1,\thanks{amitastro.am@gmail.com}}
\and
Francisco Pozo Nu\~nez \inst{2,\thanks{francisco.pozon@gmail.com}}
\and
Vikram Kumar Jaiswal \inst{1} 
\and 
Mohammad Hassan Naddaf\inst{3}
\and 
Bo\.zena Czerny \inst{1,\thanks{bcz@cft.edu.pl}}
\and
Swayamtrupta Panda\inst{4,\thanks{Gemini Science Fellow}}
\and
Paulina Karczmarek\inst{5}
\and
Grzegorz Pietrzy\' nski\inst{5,6}
\and
Shivangi Pandey\inst{1,7}
\and
{R. Edelson}\inst{8}
\and
B. M. Peterson\inst{9,\thanks{Deceased.}}
\and
{Michal Zaja\v{c}ek}\inst{10}
\and
{Alberto Floris}\inst{11, 12, 13}
\and
{Mary Loli Martínez-Aldama}\inst{6,14}
\and
{Michal Dov\v{c}iak}\inst{15}
\and
{Vladimir Karas}\inst{15}
\and
{Weronika Narloch}\inst{5}
\and
{Miros\l{}aw Kicia}\inst{5}
\and
{Marek G\'orski}\inst{5}
\and
{Miko\l{}aj Ka\l{}uszy\'nski}\inst{5}
\and
{Gergely Hajdu}\inst{5}
\and
{Piotr Wielg\'orski}\inst{5}
\and
{Bart\l{}omiej Zgirski}\inst{6}
\and
{Cezary Ga\l{}an}\inst{5}
\and
{Wojciech Pych}\inst{5}
\and
{Rados\l{}aw Smolec}\inst{5}
\and
{Ricardo Salinas}\inst{5}
\and
Henryka Netzel-I\l{}kiewicz\inst{5}
\and
{Karolina B\k{a}kowska}\inst{16}
\and
{Wolfgang Gieren}\inst{6,17}
\and
{Pierre Kervella}\inst{18,19}
}

   \institute{Center for Theoretical Physics, Polish Academy of Sciences, Al. Lotnik\'ow 32/46, 02-668 Warsaw, Poland
\and
Astroinformatics, Heidelberg Institute for Theoretical Studies, Schloss-Wolfsbrunnenweg 35, 69118 Heidelberg, Germany
    \and 
    Institut d'Astrophysique et de Géophysique, Université de Liège Allée du six août 19c, B-4000 Liège (Sart-Tilman), Belgium
    \and
    International Gemini Observatory/NSF NOIRLab, Casilla 603, La Serena, Chile
    \and
    Nicolaus Copernicus Astronomical Center, Polish Academy of Sciences, Bartycka 18, 00-716 Warszawa, Poland
    \and
    Universidad de Concepci\'on, Departamento de Astronom\'ia, Casilla 160 \textendash C, Concepci\'on, Chile
    \and
    Aryabhatta Research Institute of Observational Sciences, Nainital\textendash263001, Uttarakhand, India
    \and
    Eureka Scientific Inc., 2452 Delmer St., Suite 100, Oakland, CA 94602, USA
    \and
    c/o Tracy L. Turner, 205 South Prospect Street, Granville, OH 43023, USA
    \and
    Department of Theoretical Physics and Astrophysics, Faculty of Science, Masaryk University, Kotlá\v{r}ská 2, 611 37 Brno, Czech Republic
    \and
    Department of Physics University of Crete, Voutes University Campus, 70013 Heraklion, Greece 
    \and 
    Institute of Astrophysics, FORTH, N.Plastira 100, Vassilika Vouton, 70013 Heraklion, Greece 
    \and
    National Institute for Astrophysics (INAF), Astronomical Observatory of Padova, IT-35122 Padova, Italy
    \and
    Millennium Nucleus on Transversal Research and Technology to Explore Supermassive Black Holes (TITANs), Chile
    \and
    Astronomical Institute of the Czech Academy of Sciences, Bo\v{c}n\'i II 1401, CZ-14100 Prague, Czech Republic
    \and
    Institute of Astronomy, Faculty of Physics, Astronomy and Informatics, Nicolaus Copernicus University, ul. Grudzi\k{a}dzka 5, 87-100 Toru\'n, Poland
    \and
    Millenium Institute of Astrophysics, Avenue Libertador Bernardo O’Higgins 340, Casa Central, Santiago, Chile
    \and
    LIRA, Observatoire de Paris, Université PSL, Sorbonne Université, Université Paris Cité, CY Cergy Paris Université, CNRS, 92190 Meudon, France
    \and
    French-Chilean Laboratory for Astronomy, IRL 3386, CNRS and U. de Chile, Casilla 36-D, Santiago, Chile
    }
    \date{}

 \abstract
 {The Hubble tension remains one of the most significant unresolved problems in modern cosmology. A key question is whether it may arise from underestimated systematic uncertainties in the different measurement techniques. In this context, new independent methods are of exceptional importance.} {We therefore pursue a novel approach to determining the Hubble constant, $H_{0}$ based on continuum time delay and spectral energy distribution (SED) modeling in active galactic nuclei (AGNs). Unlike conventional techniques, this method is entirely independent of the cosmic distance ladder and does not require cross-calibration against other distance indicators. As a result, it enables a direct determination of $H_{0}$, free from the arbitrary normalizations that often affect indirect measurements.}{We conducted a dedicated monitoring campaign of the Seyfert galaxy Fairall~9 and further developed the {\tt H0RIZON-AGN} model to interpret the resulting observations. The model incorporates the effects of radiation reprocessing in the surrounding cold accretion disk, enabling a more realistic description of the observed continuum delays.}{Through the simultaneous modeling of the continuum lag-spectrum and the broadband SED of Fairall~9, we derived a Hubble constant of $H_{0}=72.4_{-3.7}^{+3.4} \, \rm km \, s^{-1} \, Mpc^{-1}$. Achieving a measurement precision of approximately 5\% from a single source demonstrates the considerable potential of this method for independent determinations of the Hubble constant.}{Our determination of $H_{0}$ is broadly consistent, within the current uncertainties, with both early- and late-Universe measurements. Future applications of the method to larger datasets, particularly those provided by the Vera Rubin Observatory, are expected to reduce the uncertainty to below 1\%, thereby establishing this approach as a powerful independent probe of the Hubble tension.}

\keywords{galaxies: active – galaxies: distances and redshifts – galaxies: nuclei – galaxies: photometry – quasars: emission lines – galaxies: Seyfert
               }
\maketitle

\section{Introduction}

Light-echo studies of the continuum emission from active galactic nuclei (AGNs) have long been proposed as a direct method for measuring the Hubble constant, $H_{0}$ \citep{collier1999}. The appeal of this approach lies in its apparent simplicity, e.g., within the framework of the standard Keplerian accretion disk model, the monochromatic luminosity is directly related to the disk size at the corresponding wavelength. In principle, this relationship allows a determination of cosmological distances independent of the black hole mass and accretion rate, which cancel out, and without reference to the cosmic distance ladder.

Despite this promise, the first attempt to derive $H_{0}$ from continuum time delays was unsuccessful. Applying the method to NGC 7469,  \citet{collier1999} obtained $H_{0}=42 \pm 9 \, \rm km \, s^{-1} \, Mpc^{-1}$ for an assumed inclination of $i = 45^{\circ}$, roughly a factor of two below the contemporary consensus value of $72 \pm 9 \, \rm km \, s^{-1} \, Mpc^{-1}$ from the cosmic distance ladder \citep{2001ApJ...553...47F}. Building on this work, \citet{cackett2007} fitted the wavelength-dependent time delays and optical spectral energy distributions (SEDs) of 14 AGNs. Their analysis likewise yielded a substantially underestimated value of $H_{0}=44 \pm 5 \, \rm km \, s^{-1} \, Mpc^{-1}$.

Subsequent continuum reverberation mapping (continuum--RM) campaigns shifted attention from the inferred value of $H_{0}$ to a more fundamental discrepancy in the accretion-disk size itself. While observations successfully recovered the theoretically predicted dependence of the time delay, $\tau$, on wavelength, $\lambda$, namely $\tau \propto \lambda^{4/3}$, the normalization of the relation was found to be systematically larger than expected by a factor of approximately 3--6 \citep{2014MNRAS.444.1469M, 2016ApJ...821...56F, pozo2019, 2023ApJ...947...62K, 2022ApJ...940...20G, 2025ApJ...985...30M}. This result implies that AGN accretion disks are apparently significantly larger than predicted by the standard thin-disk model.

Interestingly, evidence for this discrepancy had already emerged from quasar microlensing studies \citep{rauch1991, 2010ApJ...712.1129M, 2011ApJ...729...34B, 2016AN....337..356C, 2019MNRAS.483.2275L, 2024SSRv..220...14V}. Measurements of disk sizes inferred from microlensing consistently indicated emission regions larger than theoretical expectations. Moreover, attempts to reconcile the observations with the standard model through color-correction factors applied to the disk emission proved insufficient to eliminate the discrepancy. The consistency of the results obtained from both continuum--RM and microlensing therefore points to a persistent and unresolved challenge for the standard theory of AGN accretion disks.

A key clue to the origin of this discrepancy emerged from studies of emission from the broad line region (BLR). The BLR does not merely produce the prominent broad emission lines characteristic of AGN spectra; it also reprocesses a fraction of the ionizing radiation into a diffuse continuum through free-free, free-bound emission and Thomson scattering \citep{korista2001}. Although the contribution of the BLR to the continuum had been considered in earlier studies \citep{davidson1976}, those investigations primarily focused on the physical conditions and emission processes within the BLR itself rather than on its impact on continuum--RM measurements.

The importance of BLR continuum reprocessing for accretion disk lag studies was emphasized by \citet{korista2019}, who argued that this additional emission component must influence continuum time delay measurements. Observational evidence for this effect was also identified in the lag-spectra of NGC~4593 \citep{cackett2018} and Fairall~9 \citep{halo1}, which exhibit a pronounced excess in the $U/u$- band associated with the Balmer jump at 3646 {\AA} \citep{Cacket_Sci2021}. Further support came from the detection of systematically longer lags in photometric bands contaminated by broad emission lines \citep{lawther2018, 2025ApJ...985...30M}, indicating that a significant fraction of the measured delays may originate from BLR reprocessing rather than from the accretion disk alone. These findings motivated a series of theoretical and observational studies aimed at quantifying the impact of BLR contamination on continuum time delay measurements  \citep[e.g.,][]{2019NatAs...3..251C,netzer2022,pozo2023,jaiswal2023,netzer2024}.

With a plausible solution to the accretion disk size discrepancy now emerging, it is timely to revisit the original idea of measuring the Hubble constant from AGN continuum time delays. As a first step in this direction, we recently carried out a pilot study of the extensively monitored AGN NGC~5548 \citep{jaiswal2025}. By simultaneously fitting its average SED and wavelength-dependent inter-band time delays, we obtained a value of the Hubble constant of $H_{0}=66.9^{+10.6}_{-2.1} \, \rm km \, s^{-1} \, Mpc^{-1}$. In the present paper, we extend this approach to a second source, Fairall~9, a relatively bright Seyfert~1.2 galaxy.

Fairall~9 is particularly well suited for such a study because of its brightness, strong variability, and relatively unobscured view of the central engine. The source was identified by \citet{fairall1977} as one of 150 bright and compact galactic nuclei discovered in the ESO Fast Blue Survey. It was soon recognized as an extreme AGN owing to its high bolometric luminosity \citep{hawley1978}. Early ultraviolet observations revealed pronounced variability, with flux changes reaching a factor of 33 at 1335 {\AA}  \citep{clavel1989}, while showing comparatively little spectral variability. Over the following decades, Fairall~9 became the subject of numerous multiwavelength investigations \citep[e.g.,][]{chapman1985,koratkar1989,rodriguez1997,patrick2011,noda2013}, including spectropolarimetric observations with the VLT \citep{jiang2021}.

These studies have established Fairall~9 as one of the clearest examples of a 'bare AGN'. The source exhibits little evidence for intrinsic absorption, with its far-ultraviolet continuum remaining observable down to wavelengths as short as $\sim 880$ {\AA} in the rest-frame \citep{zheng1995}. In addition, it shows no signatures of a warm absorber \citep{Emmanoulopoulos2011}. The absence of significant obscuration makes Fairall~9 an excellent laboratory for investigating the intrinsic properties of the accretion flow and its surrounding reprocessing regions.

The source has also been the target of extensive RM campaigns. Most recently, \citet{edelson2024} reported the results of an exceptionally dense monitoring program with {\it Swift}, which obtained near-daily observations over a period of 1.8 years. Earlier, \citet{hernandez2020} analyzed the first year of the {\it Swift} campaign supplemented with ground-based observations, thereby extending the wavelength coverage to longer optical bands.

In this study, we present new optical photometric monitoring of Fairall~9 obtained with a combination of intermediate- and broad- band filters. These observations were combined with archival {\it Swift} data and analyzed in our previous work \citep{halo1}, where we compared the variability properties derived from the two datasets and established the observational foundation for the present study. Building on these results, we model both the lag-spectrum and the broadband SED of Fairall~9, accounting for emission from the accretion disk, contribution from the warm corona, reprocessing by the BLR, and the dusty torus. Finally, we use this self-consistent framework to derive an independent estimate of the Hubble constant from the continuum--RM properties of the source. The paper is organized as follows: Sect.~\ref{ss:obs} outlines the observations and data acquired for Fairall~9. Sect.~\ref{ss:model} details the model developed to jointly fit AGN lag-spectra and SEDs for estimating $H_{0}$. The results are presented in Sect.~\ref{ss:result}, followed by a discussion in Sect.~\ref{ss:dis}. Finally, Sect.~\ref{ss:summ} summarizes the main findings.

\section{Observations and data of Fairall~9}
\label{ss:obs}

Fairall~9 is located in the local Universe at a redshift of $z = 0.046145 \pm 5.70 \times 10^{-5}$ retrieved from NASA/IPAC Extragalactic Database\footnote{\url{https://ned.ipac.caltech.edu/}} (NED). Therefore, we applied a correction for the Milky Way’s peculiar velocity, which has a significant impact at low redshift \citep[e.g.,][]{2025PhRvD.111h3545C, 2025PhRvD.112d3516C}. The correction was computed using the NED Velocity Correction Calculator\footnote{\url{https://ned.ipac.caltech.edu/help/velc_help.html}}, which accounts for Galactic rotation, the motion of the Milky Way within the Local Group, the infall of the Local Group toward the Local Supercluster, and motion relative to the CMB rest-frame. This yielded a peculiar-velocity-corrected redshift for Fairall~9 of $z = 0.045740$.

\subsection{Optical/UV data}

In 2024, we initiated a dedicated photometric monitoring program, 'Hubble constant constraints through AGN Light curve Observations' (HALO), with the primary goal of constraining $H_0$ and helping to address the long-standing Hubble tension \citep{halo1}. The central aim of HALO is to jointly model inter-band continuum time delays and SEDs for a carefully selected sample of AGNs spanning a wide range of luminosities and redshifts. By combining these complementary observables within a unified framework, the project seeks to provide an independent and robust estimate of $H_0$.

To achieve this, HALO observations are being conducted with the 60 cm Cerro Murphy Observatory (OCM) in Chile using four Str\"omgren intermediate-band filters ($u$, $v$, $b$, and $y$), and one Johnson--Cousins filter in $I$ band. As part of this program, we have successfully completed the observations and data reduction for our first target, Fairall~9.

To complement our new observations, we incorporated additional multi-wavelength photometric data for Fairall~9 from the literature. In particular, we used the extensive \textit{Swift} dataset compiled by \citet{edelson2024}, which includes hard and soft X-ray light curves from the X-Ray Telescope (XRT), together with UV--optical light curves in the $W2$, $M2$, $W1$, $U$, $B$, and $V$ bands obtained with the Ultraviolet/Optical Telescope (UVOT). These data substantially extend the wavelength coverage of our monitoring campaign.

In addition to the photometric data, we assembled archival spectroscopic observations to construct a comprehensive broadband view of the source. This dataset includes a UV spectrum obtained with the \textit{Hubble Space Telescope} (\textit{HST}) Faint Object Spectrograph on 1993 January 21 \citep{edelson2024}, as well as an optical spectrum using the European Southern Observatory (ESO) 1.5 m telescope equipped with the Boller \& Chivens spectrograph and a CCD detector \citep{1997ApJS..112..271S}. These optical observations were part of the broader international monitoring effort conducted in conjunction with the \textit{International Ultraviolet Explorer} \citep[IUE;][]{1997ApJS..110....9R}.

The observational strategy, data reduction procedures, lag analysis used to construct the wavelength-dependent lag-spectrum, and the UV--optical spectral properties of Fairall~9 were presented in detail in \citet{halo1}. Building on that foundation, we further compiled additional UV-to-X-ray photometric measurements from NED to construct the broadband SED used in the present analysis.

\subsection{X-ray data}

X-ray observations are crucial both for probing the physical conditions in the innermost regions of the accretion flow and for constructing broadband SED. Fairall~9 was observed for 130 ks with \textit{XMM-Newton} \citep{Emmanoulopoulos2011}. The resulting spectrum showed no evidence for a warm absorber, but revealed a prominent relativistically blurred reflection component originating from the inner accretion disk. This reflection, produced by irradiation from the X-ray corona, enabled constraints to be placed on the black hole spin. The broadband spectral fit yielded a hard X-ray photon index of $\Gamma = 2.01^{+0.01}_{-0.02}$. In addition, relatively narrow Fe K$\alpha$ and K$\beta$ emission lines were detected, indicating reflection from more distant material \citep{Emmanoulopoulos2011,liu2020}.

Subsequent joint observations with \textit{XMM-Newton}, \textit{Suzaku}, and \textit{NuSTAR} provided further insight into the corona structure, revealing the presence of a warm corona with a temperature of approximately 0.5 keV \citep{liu2020}. The existence of this warm corona component was later supported by the analysis of \textit{NICER} observations obtained between 2018 and 2021 \citep{partington2024}. That study found a hard X-ray photon index of $\Gamma \sim 2.05$, confirmed the presence of only a narrow Fe K$\alpha$ emission line, and provided additional evidence for the warm corona. Notably, the photon index derived from the \textit{NICER} data is consistent with the \textit{XMM-Newton} measurement of $\Gamma = 2.01$ \citep{Emmanoulopoulos2011} obtained roughly a decade earlier, indicating that the hard X-ray spectral shape of Fairall~9 remains stable despite its pronounced flux variability. This stability supports our treatment of $\Gamma$ as a fixed parameter (see Table~\ref{tab:param}).

\begin{table}
\centering
 \caption{Parameters utilized in {\tt H0RIZON-AGN} code to model the lag-spectrum and SED of Fairall~9.}
\label{tab:param}
\resizebox{9cm}{!}{
\fontsize{11pt}{11pt}\selectfont
\begin{tabular}{lcr} \hline \hline

\\
Description & parameter & value  \\
 &      \\
(1) & (2) & (3)
\\ \hline
 &     \\
  & fixed parameter & \\

Black hole mass & $M_{\rm BH}$ &  $2.18 \times 10^8$ \\ 
Inclination angle of the disk/BLR system & $i$ & $35^{(a)}$ \\
Photon index of hard X-ray power-law & $\Gamma$ & $2.01^{(b)}$ \\
Warm corona inner radius & $r_{\rm ISCO}$ & 6 \\
Outer cold disk radius & $r_{\rm out}$ & $10000$ \\
Inner radius of torus & $R_{\rm dust}$ & 70 \\
Color correction & $f_{\rm col}$ & 1 \\

 \hline \\

 & model fitted parameter & \\
 Lamp luminosity & $L_x$ & $ 6.74 \times 10^{44}$ \\
 Corona height & $h$ & $11.26$ \\ 
 Warm corona temperature & $T_{\rm WC}$ & $4.58\times10^6$ \\
 Warm corona optical depth & $\tau_{\rm WC}$ & 22.08 \\
 Inner cold disk radius & $r_{\rm transition}$ & $252.38$ \\
 Eddington ratio & $\dot{m}$ & $0.0224$ \\ 
 BLR radius & $R_{\rm BLR}^{\star}$ &  $17.4^{(c)}$ \\
 BLR contribution & $f_{\rm BLR}$ & 9.6$\%$ \\ 
 Torus contribution & $f_{\rm dust}$ & 12.9$\%$ \\
 Starlight & & $1.05 \times 10^{-16}$  \\
 
\\

\hline

\end{tabular}
}
\vspace{0.01cm}

\tablefoot{Columns are: (1) Description of the parameter used in the model. (2) parameter, and (3) value of the parameter. Black hole mass in unit of $M_{\odot}$, measured from $R_{\rm BLR}$ fiducial value and H$\beta$ line dispersion ($\sigma_{\mathrm{line,H\beta}}$) with virial factor $f_{BLR}=4.47$ \citep{2015ApJ...801...38W}; inclination angle in degree, the value is the inner-disk inclination from X-ray reflection and is assumed common to the disk and BLR; $r_{\rm ISCO}$, $r_{\rm transition}$, $r_{\rm out}$, and $h$ are in unit of $r_g$; $R_{\rm BLR}$, and $R_{\rm dust}$ are in light-days; lamp luminosity in erg s$^{-1}$; and starlight normalization is given in erg s$^{-1}$ cm$^{-2}$ \AA$^{-1}$ at 5100 {\AA}. The fitted values represent the best-fit for a luminosity distance of 190.6 Mpc. $\star$ $R_{\rm BLR}$ is treated as a semi-free parameter, the fiducial value is mentioned: it is effectively varied through changes in $\dot{m}$, while remaining fixed at the fiducial value during the {\tt Cloudy} computations, since this has a negligible impact on the {\tt Cloudy} results (see Sect.~\ref{dis:asump_blr} for details). References are: (a) \citet{2012ApJ...758...67L}, (b) \citet{Emmanoulopoulos2011}, and (c) \citet{peterson2004}.}

\end{table}

\begin{figure*}
    \centering

    \resizebox{9.5cm}{5cm}{\includegraphics{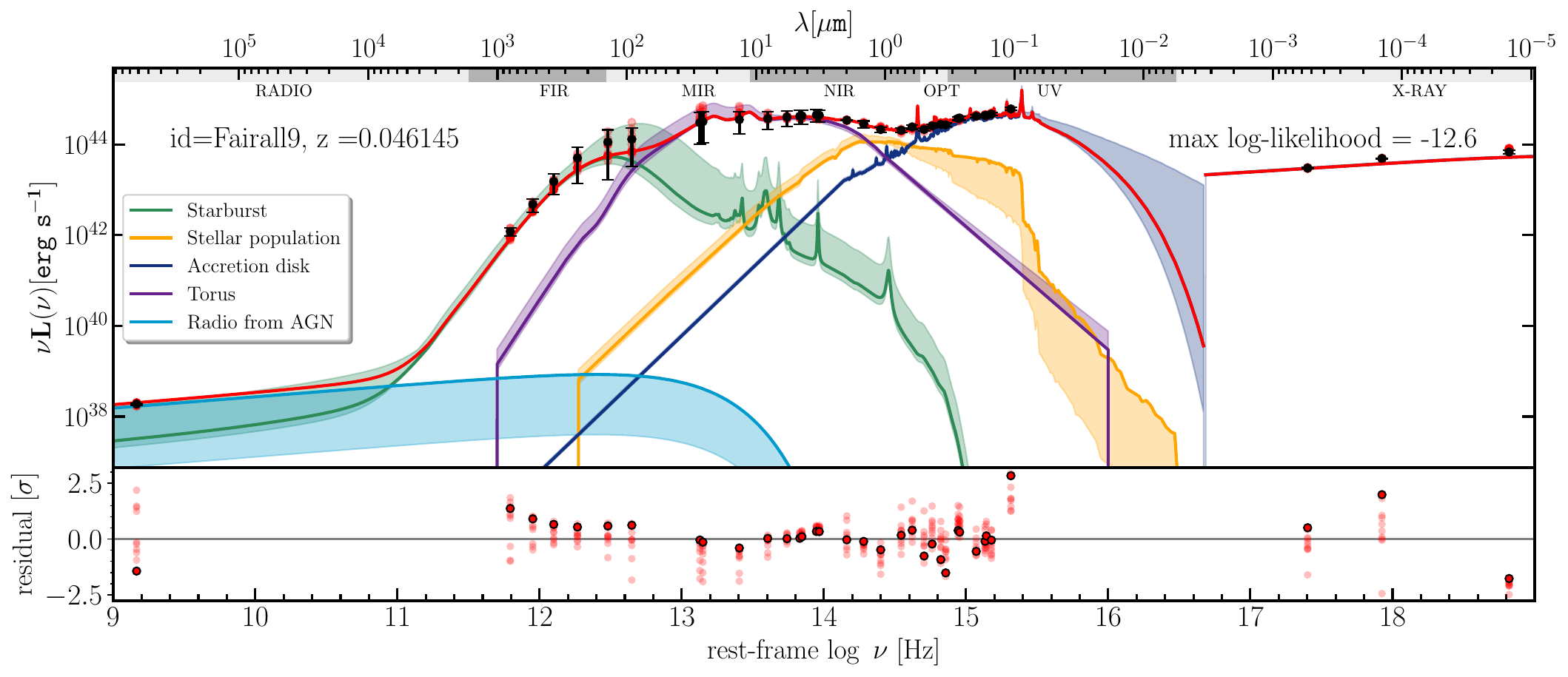}}
      \resizebox{8.5cm}{4.7cm}{\includegraphics{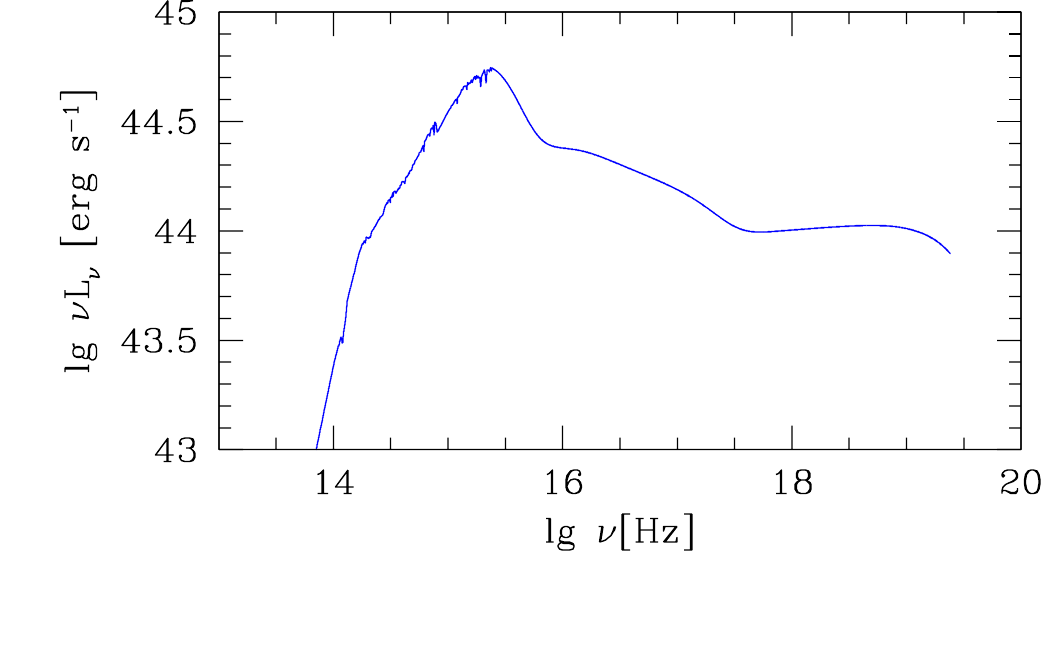}}
    \caption{Fairall~9 SED. Left: Best-fit SED obtained with {\sc AGNFitter-RX} \citep{AGNfitter_2024}. The contributions from the stellar population, cold dust emission, torus (\texttt{CAT3D}, \citealt{Honig_Kishimoto_2017}), accretion disk, and radio components are shown by the yellow, green, purple, dark-blue, and light-blue solid curves, respectively. The shaded regions represent the ranges spanned by 100 random realizations drawn from the posterior probability distributions. The total model SED is shown in red. Observed fluxes and their uncertainties are plotted as black points, while the corresponding model fluxes are indicated by red points. The residuals for the individual realizations are displayed in the lower panel. The maximum log-likelihood value of the fit is reported in the upper-right corner. Right: Full SED adopted for the {\tt Cloudy} simulations, extending from the Optical/UV band to hard X-rays. The SED was constructed by combining the intrinsic AGN emission components derived above, excluding the torus and starburst contributions$-$with the {\tt AGNSED} model of \citet{kubota2018} to provide a physically motivated description of the high-energy continuum. The cosmological parameters assumed in the construction of the SED are $H_{0} = 70 \, \rm km \, s^{-1} \, Mpc^{-1}$ and $\Omega_{\rm m}$ = 0.3.}
    \label{fig:sed}
\end{figure*}

\subsection{Global parameters and SED of Fairall~9}

Our methodology for determining $H_{0}$ requires a broadband SED for both spectral fitting and BLR emissivity modeling. In addition, we fix the black hole mass ($M_{\rm BH}$) and the inclination angle ($i$), as these parameters are strongly degenerate with the accretion rate when constrained by the observed source flux.

For Fairall~9, we adopted $M_{\rm BH} = 2.18 \times 10^8 \, M_{\odot}$ inferred from spectroscopic--RM \citep{peterson2004} and a fiducial Eddington ratio ($\dot{m}$) of 0.028 \citep{halo1}, consistent with the moderate Fe II-to-H$\beta$ intensity ratio of $R_{\rm FeII} \sim 0.5$ \citep{marziani2010, 2024A&A...689A.321F}. The relatively low global UV-X-ray spectral index measured between 2500 {\AA} and 2 keV, $\alpha_{\rm ox} = 1.26$ \citep{wilkes1994}, further indicates that the source is X-ray bright.

There is a range of estimates for the inclination angle of Fairall~9 in the literature. While a low inclination of $i \sim 11^{\circ}$ has been reported \citep{2017ApJ...835..226K}, this appears to be an outlier; in contrast, X-ray spectral modeling yields $i \sim 35^{\circ}$ \citep{2012ApJ...758...67L}, consistent with the independent constraint of $34^{+5}_{-3}{}^{\circ}$ from Fe K$\alpha$ modeling \citep{patrick2011}. We therefore adopted $i = 35^{\circ}$, corresponding to the inner-disk inclination inferred from X-ray reflection and Fe K$\alpha$ analyses. Assuming that the accretion disk and the BLR inferred from the Failed Radiatively Accelerated Dusty Outflow \citep[FRADO;][]{czhr2011} model are coplanar, we applied this inclination to both components. By contrast, the warm-corona response is assumed to be independent of inclination, as its geometry is not expected to be planar. We summarize these global model parameters in Table~\ref{tab:param}.

 The broadband SED was constructed using archival photometric measurements collected predominantly from NED. Because these data were obtained over many epochs during which the source exhibited substantial variability, the assembled photometry does not correspond to a single flux state. To produce a coherent SED representative of the conditions during our OCM monitoring campaign, we applied a wavelength-independent (grey) rescaling that anchors the archival photometry to the mean OCM flux level \citep[see][and Sect.~\ref{ss:sed}]{halo1}. The resulting SED is shown in the left panel of Figure~\ref{fig:sed}.

For the SED analysis, we employed {\sc AGNfitter-rx} \citep{AGNfitter_2016, AGNfitter_2024}, a Bayesian MCMC extension of {\sc AGNfitter} that models broadband emission from the radio to X-ray regime by decomposing the observed SED into physically motivated AGN and host-galaxy components. The AGN continuum was represented using the \texttt{THB21} template \citep{Temple2021}, which describes the optical--UV accretion-disk emission with a broken power law and includes contributions from hot dust ($1$--$3 \, \mu$m) and broad and narrow emission lines. The AGN infrared emission was modeled with the \texttt{CAT3D} templates \citep{Honig_Kishimoto_2017}, which describe a clumpy dusty torus accompanied by a polar wind, enabling constraints on the geometry and relative contributions of the equatorial and polar dust components.

The host-galaxy stellar emission was fitted using the \texttt{BC03\_metal} library \citep{Bruzual2003}, allowing variations in stellar age, metallicity, star-formation history, and dust attenuation following the \citet{Calzetti2000} reddening law. Far-infrared emission associated with star formation was modeled using the \texttt{S17} templates of \citet{Schreiber2018}, which incorporate redshift-dependent dust temperatures calibrated on main-sequence galaxies.

Radio synchrotron emission was described by a power law, $L_{\nu}\propto\nu^{-\alpha}$, adopting a fixed spectral index of $\alpha=0.75$ owing to the availability of only a single band radio measurement \citep{Baan2006}. The X-ray component was constrained through the empirical $\alpha_{\rm ox}$--$L_{2500\AA}$ relation of \citet{Lusso2017}, linking the UV luminosity at 2500 \AA\ to the 2 keV emission. We adopted the default {\sc AGNfitter-rx} priors \citep{AGNfitter_2024} and performed the inference with the \texttt{ultranest} nested sampling algorithm \citep{Buchner_2016, Buchner_2019}, using 100 walkers and two burn-in phases of 25,000 steps each to constrain the relative contributions of the AGN and host-galaxy components.

We employed {\sc AGNfitter-rx} for SED modeling to identify the primary radiative components of the source. In this decomposition, a starburst contribution was found to play a non-negligible role in the infrared regime; however, this component is not relevant for the incident continuum that irradiates the BLR and was therefore excluded from the BLR illumination model. The X-ray portion of the SED, on the other hand, was not well constrained within our fit due to the lack of high-quality far-UV and soft X-ray observational coverage. To mitigate this limitation, we supplemented the observed SED with the theoretical model of \citet{kubota2018}, which provides a physically motivated connection across the UV–X-ray gap and remained consistent with available X-ray spectral constraints. Based on the resulting decomposition, we constructed a final incident continuum for BLR calculations by retaining only those components that directly contribute to BLR irradiation. In addition, both the dust emission component and the single band radio data point were also excluded, since they do not significantly irradiate the BLR, although dust component was kept in the spectral fitting. The resulting SED used in our {\tt Cloudy} computations is shown in the right panel of Figure~\ref{fig:sed}. Overall, the adopted SED is broadly consistent with that proposed by \citet{hagen2023} for the same source.

The broadband SED shown in Figure~\ref{fig:sed} was also used to construct the mean spectrum of Fairall~9 for subsequent spectral fitting (see Sect.~\ref{ss:sed}). This was achieved by combining the UV/optical spectrum from \citet{edelson2024} with representative SED anchor points that extend beyond the wavelength range covered by the available spectroscopic data. Both the SED points and the spectroscopic measurements were subsequently grey-shifted to ensure a consistent overall normalization with the photometric points obtained from our HALO campaign. For the constructed combined broadband SED, the spectroscopic data were sampled densely while carefully avoiding strong emission lines. A detailed description of the SED modeling procedure is provided later in Sect.~\ref{ss:sed}.

\section{Model: The {\tt H0RIZON-AGN} code}
\label{ss:model}

RM is a powerful technique for probing the innermost regions of AGNs, which remain far beyond the reach of direct imaging with current observational facilities. Despite substantial progress over the past decades, the detailed structure of the BLR and the physical origin of continuum inter-band time delays are still not fully understood.

Traditionally, these continuum lags have been interpreted within the framework of standard accretion-disk reprocessing, where the delay scales with wavelength as a simple power law \citep{SS1973}. However, this approach has encountered several difficulties. Most notably, it often leads to the apparent disk-size problem and cannot reproduce the prominent atomic features in AGN lag-spectra, which are naturally explained by diffuse BLR continuum emission \citep{korista2001,lawther2018}. Additional complexities arise from scattering within the BLR \citep{jaiswal2023}, contamination by emission lines in the filters used to trace the continuum, combined with contributions from the warm corona at shorter UV–optical wavelengths and from hot dust at longer optical wavelengths \citep{halo1}.

To address these open questions, we developed a comprehensive modeling framework, implemented in the code, namely '$H_{0}$ from uv-optical Reprocessing  and Ionization ZONes via AGN lag-spectrum and SED fitting' ({\tt H0RIZON-AGN}) code. This model combines a lamppost irradiation geometry with a warm corona, a standard cold accretion disk, a radiation-pressure-driven dusty BLR, and torus dust emission at longer optical wavelengths. By simultaneously fitting the AGN lag-spectrum and broadband SED, {\tt H0RIZON-AGN} provides a self-consistent framework for estimating the Hubble constant, $H_{0}$.
The first successful application of {\tt H0RIZON-AGN} was to NGC~5548, for which we obtained an estimate of $H_0 = 66.9^{+10.6}_{-2.1}\ \mathrm{km\ s^{-1}\ Mpc^{-1}}$ \citep{jaiswal2025}. In that study, we assumed that the central radiation was reprocessed by both the accretion disk and the BLR. In the present work, we extend this framework by additionally incorporating reprocessing by the warm corona and the dusty torus.
Building on this foundation, we construct a unified model within the {\tt H0RIZON-AGN} framework that incorporates all the principal emitting components of an AGN. These include a hot inner corona that produces the primary hard X-ray emission, a warm Comptonizing corona responsible for the soft X-ray excess, and a standard optically thick, geometrically thin cold accretion disk dominating the outer regions. The model also incorporates emission from the BLR, thermal emission from hot dust in the torus, and the stellar contribution of the host galaxy to the observed spectrum.
In the following sections, we describe each of these components in detail and discuss their respective roles in shaping both the wavelength-dependent lag-spectrum and the broadband SED.

\begin{figure}
    \centering
    \includegraphics[width=\columnwidth]{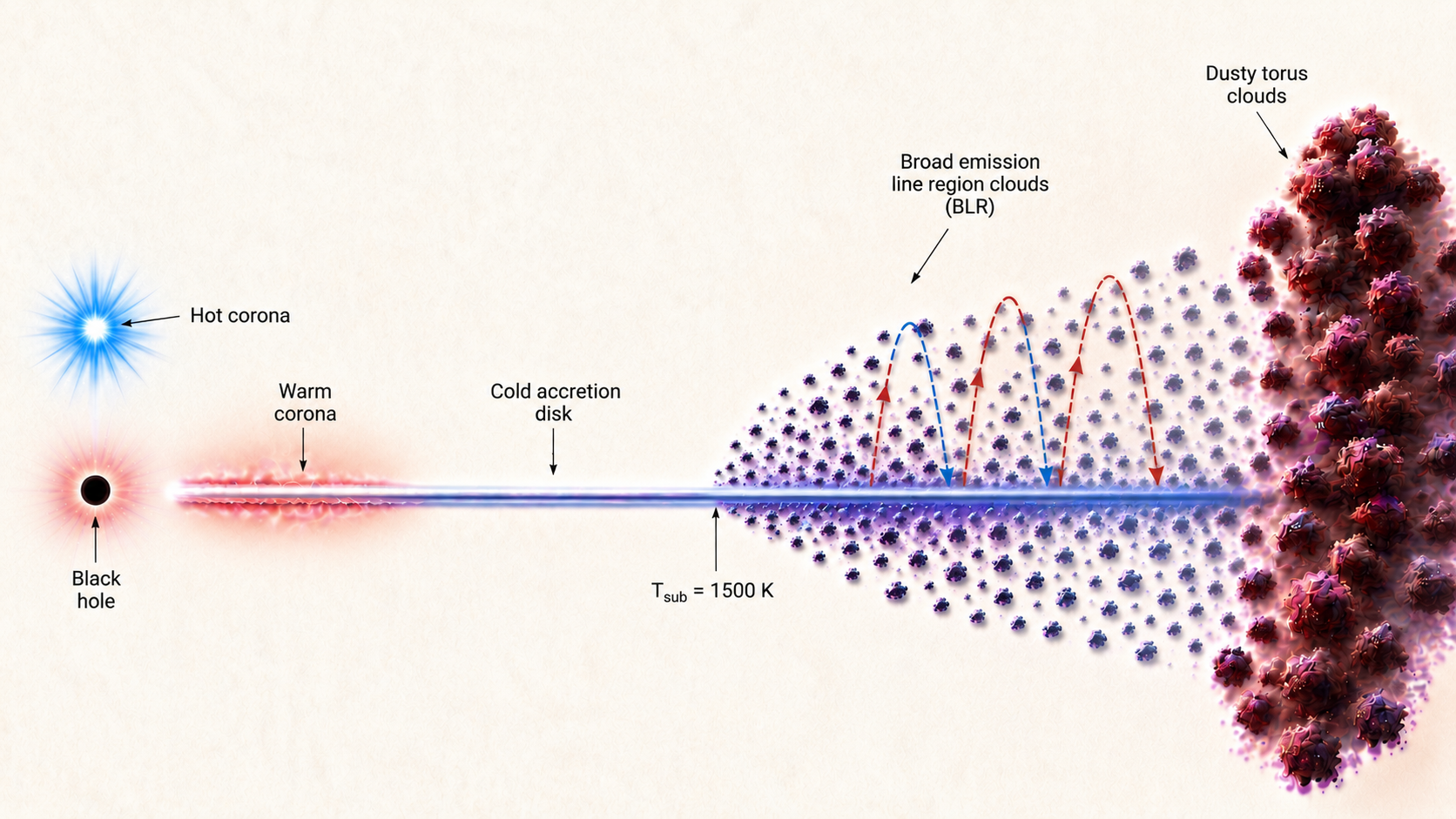}
    \caption{A schematic representation of the {\tt H0RIZON-AGN} framework (not to scale). This diagram illustrates the individual components of AGN included in our modeling approach, which collectively contribute to the observed continuum lag-spectrum and broadband SED.} 
    \label{fig:model}
\end{figure}

\subsection{Hot corona}

In our model, the hot corona is represented using the lamppost approximation, in which all hard X-ray emission is assumed to originate from a single point located along the accretion disk symmetry axis. This idealization significantly simplifies the computation of disk irradiation. In practice, it provides results that are broadly similar to those obtained for an extended spherical corona, provided that its center coincides with the lamppost location.

Previous studies have explored the impact of relaxing the point-source assumption by considering a finite corona size. However, the resulting changes in irradiation patterns depend sensitively on general relativistic (GR) effects, as demonstrated in \citet{dovciak2016}. Since GR effects are not included in our present framework, the lamppost approximation remains a sufficiently accurate and internally consistent description for our purposes.

The hot corona component is characterized by several parameters. Among these, the corona height, $h$ is treated as a global free parameter and is constrained directly from the data. The remaining spectral properties are fixed and include the observed hard X-ray flux, the photon index, $\Gamma$, which characterizes the slope of the hard X-ray power-law ($\epsilon_{\rm PL}(\lambda)$), as well as the low- and high-energy cut-offs.

We do not, however, compute the corona luminosity in a self-consistent manner coupled to the accretion disk structure. This represents a simplification relative to models such as \citet{kubota2018}, where the corona luminosity is explicitly coupled to the inner disk radius. In our approach, we fix the inner disk radius, $r_{\rm ISCO}$ at $6 \, r_g$, corresponding to the Schwarzschild solution for a non-rotating black hole (see details below). Within this framework, any excess energy dissipated in the hot corona can be effectively interpreted as being associated with accretion flow onto a spinning black hole. A more self-consistent treatment of disk–corona coupling is beyond the scope of this work, particularly given that the available data do not allow for a unique constraint of such additional model complexity.

Finally, the lamppost is assumed to emit radiation isotropically. Consequently, while a fraction of the hard X-ray emission propagates directly toward the observer, a significant portion is intercepted by the accretion disk and its surrounding medium, where it contributes to irradiation and reprocessing.

\subsection{Warm corona}

The soft X-ray excess is a common feature in the spectra of bright AGNs and is now widely interpreted as emission from a warm corona covering the inner region of the cold accretion disk \citep[e.g.,][]{petrucci2020,ballantyne2024}. Unlike the hot corona responsible for the primary hard X-ray emission, the warm corona is characterized by a much lower temperature, typically $T_{\rm WC} \sim$ a few times $10^6$ K, but a substantially higher optical depth, with $\tau_{\rm WC} \sim 10$ -- $20$. It is thought to form in the upper layers of the accretion disk, possibly as a consequence of radiation-pressure-driven instabilities, where enhanced dissipation leads to efficient Comptonization of the underlying disk photons \citep[e.g.,][]{rozanska2015, palit2024}. 

This component is included in the models of \citet{kubota2018}. However, to describe an extended warm corona region, we adopt the formulation of \citet{czerny2003}, in which Comptonization is computed independently at each disk radius based on the local disk flux. In the absence of a well-established theory for radial stratification of the warm corona, we assume that both the corona temperature, $T_{\rm WC}$, and the optical depth, $\tau_{\rm WC}$ are constant with radius and treat them as global model parameters. For energy conservation, the accretion power in the corona region is reduced by the corresponding Compton amplification factor, which is determined self-consistently by $T_{\rm WC}$, and $\tau_{\rm WC}$. We further assume that the inner radius of the disk covered by the warm corona is set at $6 \, r_g$, and extends to a transition radius $r_{\rm{transition}}$, beyond which the disk remains in its standard, un-Comptonized (bare) state. 

The emission from the warm corona is primarily powered by internal dissipation of energy. However, it can also be affected by irradiation from the hot corona. This additional coupling was not included in \citet{jaiswal2025}. In the present work, we incorporate it as a simple reflection (Thomson scattering), justified by the highly ionized state of the warm corona, which makes absorption negligible, although it may still contribute to the production of highly ionized Fe lines \citep{petrucci2020, palit2024}.

\subsection{Cold accretion disk}

The outer region of the accretion disk is described using the standard Shakura–Sunyaev thin-disk model \citep{SS1973}. This component is characterized by  $M_{BH}$, and dimensionless accretion rate, $\dot m$, which is defined in terms of the Eddington ratio. In our setup, $M_{BH}$ is fixed based on constraints from BLR time delays and emission-line widths, while $\dot m$ is treated as a free parameter. We also fix $i$ of the system. These three parameters, $M_{BH}$, $\dot m$, and $i$ are highly degenerate, since the optical continuum normalization in a standard thin disk depends on their combined effect.

The irradiation of the outer accretion disk by the central lamppost source is treated in a simplified manner by assuming complete thermalization of the incident flux. Under this assumption, the reprocessed radiation is locally added to the intrinsic disk emission, and the total output at each radius is computed as blackbody radiation. We do not apply any color correction, as the outer regions of AGN disks are expected to be weakly ionized, ensuring efficient thermalization of the radiation. Finally, the response function for a bare disk, $\psi_{\rm d}(\lambda,t)$, is obtained by integrating the Planck function over the disk surface. A more detailed description of the cold disk modeling and its implementation in the code can be found in \citet{jaiswal2025}.

\subsection{BLR reprocessing}

Previous continuum--RM campaigns have firmly established that the observed inter-band lags are not produced solely by thermal reprocessing in the accretion disk. In particular, they reveal significant contributions from the BLR, most notably through Balmer continuum emission and possible contamination from the Paschen jump \citep{GuoH_etal_2022, pandey2023}. These components can introduce excess delays around 3646 {\AA} and 8206 {\AA}, respectively, because they originate in the more extended BLR rather than in the accretion disk itself.

Clear observational evidence for such effects has been reported in several well-studied AGNs. For example, an excess lag in the $U/u$ band has been detected in both NGC~4593 \citep{cackett2018} and Fairall~9 \citep{halo1}, consistent with a substantial contribution from the Balmer continuum. In addition, the lag-spectrum of Fairall~9 shows evidence for possible Paschen-jump contamination in the $I$ band, beyond the contribution already attributed to hot dust emission.

More broadly, all photometric bands are expected to be affected, to some extent, by diffuse continuum emission from the BLR arising from free-free and free-bound recombination processes. Consequently, observed AGN lag-spectra often deviate from the simple predictions of standard thin-disk reprocessing models and instead exhibit distinct atomic signatures. Therefore, a physically robust interpretation of continuum delays requires accounting not only for disk reprocessing, but also for the diffuse BLR continuum and any contamination by emission lines in the continuum-tracing filters. Moreover, additional contribution from Fe II emission origination from BLR can contribute to the continuum delays.

Therefore, to incorporate the BLR contribution in a physically self-consistent manner, we adopt the FRADO model \citep{czhr2011, naddaf2021, naddaf2022}, a physically motivated dynamical framework for the BLR. Further theoretical support for the model is offered by \citet{2025MNRAS.544.4532O}.

\subsubsection{BLR structure as inferred from FRADO}

The FRADO model provides a physically motivated explanation for the origin of the low-ionization BLR in AGNs, which is primarily responsible for the emission of lines such as H$\beta$, H$\alpha$, Mg II, and Fe II. In this framework, the BLR forms at radii where the effective temperature of the accretion disk atmosphere falls below the dust sublimation temperature, $T_{\rm {sub}}$ $\sim$ 1500 K, allowing dust to condense. The high opacity of this dusty gas enables radiation pressure from the disk to efficiently lift material above the disk surface. Consequently, the characteristic BLR radius is set by the condition, $T_{\rm {eff}}(R_{\rm {BLR}}) \, \sim \, T_{\rm {sub}}$. This naturally connects the BLR radius, $R_{\rm {BLR}}$ to the underlying accretion-disk structure and leads directly to the observed reverberation-mapped radius--luminosity relation. We allow $R_{\rm BLR}$ to vary through its dependence on the accretion rate, $\dot{m}$ (see Sect.~\ref{dis:asump_blr} and Figure~\ref{fig:blr_lg} for further details). To model this structure in a self-consistent way, we employ the numerical framework developed by \citet{naddaf2021}, which includes a detailed treatment of the wavelength-dependent dust grain cross-sections.



\subsubsection{Emissivity profile from {\tt Cloudy}}
\label{ss:emsivity}

The emission from individual BLR clouds must be processed through the BLR emissivity profile ($\epsilon_{\rm BLR} (\lambda)$) before being observed. To model the resulting spectral shape, we perform photoionization calculations using {\tt Cloudy} (version C23.00; \citealt{2023RMxAA..59..327C}). In practice, the full 3-D BLR cloud distribution is simplified by representing it with a single effective cloud placed at the radius corresponding to the characteristic reverberation time delay (e.g., the H$\beta$ lag).

For Fairall~9, we therefore adopt an incident ionizing luminosity of  $\log L$ (erg/s) = 44.97, corresponding to a reference luminosity distance of 204.6 Mpc assuming $H_{0}=70 \, \rm km \, s^{-1} \, Mpc^{-1}$, which is later varied to account for different luminosity distances, and position the representative cloud at a BLR distance of $\log r$ (cm) = 16.64, consistent with the H$\beta$ time delay measured from spectroscopic--RM \citep{peterson2004}. The gas is assumed to have a constant hydrogen density of $\log n_{\rm H}$ (cm$^{-3}$) = 12, and a column density of $\log N_{\rm H}$ (cm$^{-2}$) = 23.5. The adopted luminosity is consistent with the source’s bolometric output inferred from broadband SED fitting. We further assume a metallicity of five times the solar value, following our previous modeling of NGC~5548 \citep{jaiswal2025}. We note that while NGC~5548 was better described by a slightly lower density of $\log n_{\rm H}$ (cm$^{-3}$) = 11, the higher density $\log n_{\rm H}$ (cm$^{-3}$) = 12 adopted here is more appropriate for Fairall~9, as will be further justified in Sect.~\ref{dis:cloud} (also see Figure~\ref{fig:emv_den}).

Modeling the Fe II emission with {\tt Cloudy} presents a challenge because the Fairall~9 exhibits unusually strong UV Fe II emission despite its low Eddington ratio. To address this, we adopt empirical Fe II templates, using the UV template of \citet{2001ApJS..134....1V} and \citet{2006ApJ...650...57T}, together with the optical template of \citet{1992ApJS...80..109B}, over the 1900--4000 {\AA} wavelength range. This Fe II component is then added to the {\tt Cloudy}-generated spectrum, with its relative normalization adjusted to better reproduce the observed data in this spectral region. The resulting combined spectrum is used for the broadband SED fitting. In addition, the same Fe II component is employed in the time-delay analysis, under the assumption that the Fe II emission originates from BLR clouds similar to those producing the H$\beta$ emission lines \citep{panda2018, panda2019, 2021ApJ...918...29M, 2023A&A...678A.189P, 2024A&A...683A.140Z}.

\subsection{Torus reprocessing}
\label{ss:torus}

This component was not included in \citet{jaiswal2025}. However, we detect a significant excess lag in the $I$-band ($\lambda_{\rm eff}=8105$ {\AA}) in the lag-spectrum of Fairall~9 \citep{halo1}. Simple tests indicate that additional emission begins to contribute to the continuum time delays at wavelengths longer than $\sim 7000$ {\AA}, suggesting a likely origin in the dusty/molecular torus. Motivated by this, we incorporate a dust reprocessing component associated with the torus into our model.

The 3-D structure of the torus remains highly uncertain. Previous studies have explored torus reverberation in moderate detail; for example, \citet{2020ApJ...891...26A,2017ApJ...843....3A} developed the {\tt TORMAC} code, assuming a geometry similar to the {\tt CLUMPY} torus model of \citet{2008ApJ...685..147N, 2008ApJ...685..160N}. These works demonstrate that the resulting dust transfer functions are highly asymmetric, typically characterized by a sharp rise followed by an extended tail. Similarly, \citet{2022MNRAS.516.4898G} adopted a parametric dust distribution and constructed the corresponding transfer function through Markov Chain Monte Carlo modeling, using optical ground-based and {\it Kepler} light curves together with {\it Spitzer} 3.6 and 4.5 $\mu$m observations of Z229$-$15.

In our case, however, the evidence for dust contamination is limited to the longest wavelength, namely the $I$-band, and thus affects only a single lag measurement in the Fairall~9 lag-spectrum. Introducing a fully physical torus model would significantly increase the number of free parameters and lead to strong degeneracies. Instead, we adopt a phenomenological approach and approximate the torus transfer function, $\psi_{\rm dust}(t)$, with a half-Gaussian profile. This choice is intended to reproduce the observed asymmetry. The adopted function is characterized by its mean delay, which approximately corresponds to the inner radius of the torus, and its width ($\sigma_{\rm dust}$), representing the radial extent of the dust distribution. We further set $\sigma_{\rm dust}$ to be approximately 0.3 times the mean delay, under the assumption that dust contamination within the optical bands arises predominantly from the innermost region of the torus. Subsequently, $\psi_{\rm dust}(t)$ is modulated by the torus emissivity profile ($\epsilon_{\rm dust}(\lambda)$) to construct wavelength dependent torus response. We treat the torus covering factor ($f_{\rm dust}$) as a free parameter inferred from the data fitting. 

\begin{figure*}
    \centering
    \includegraphics[width=0.33\linewidth]{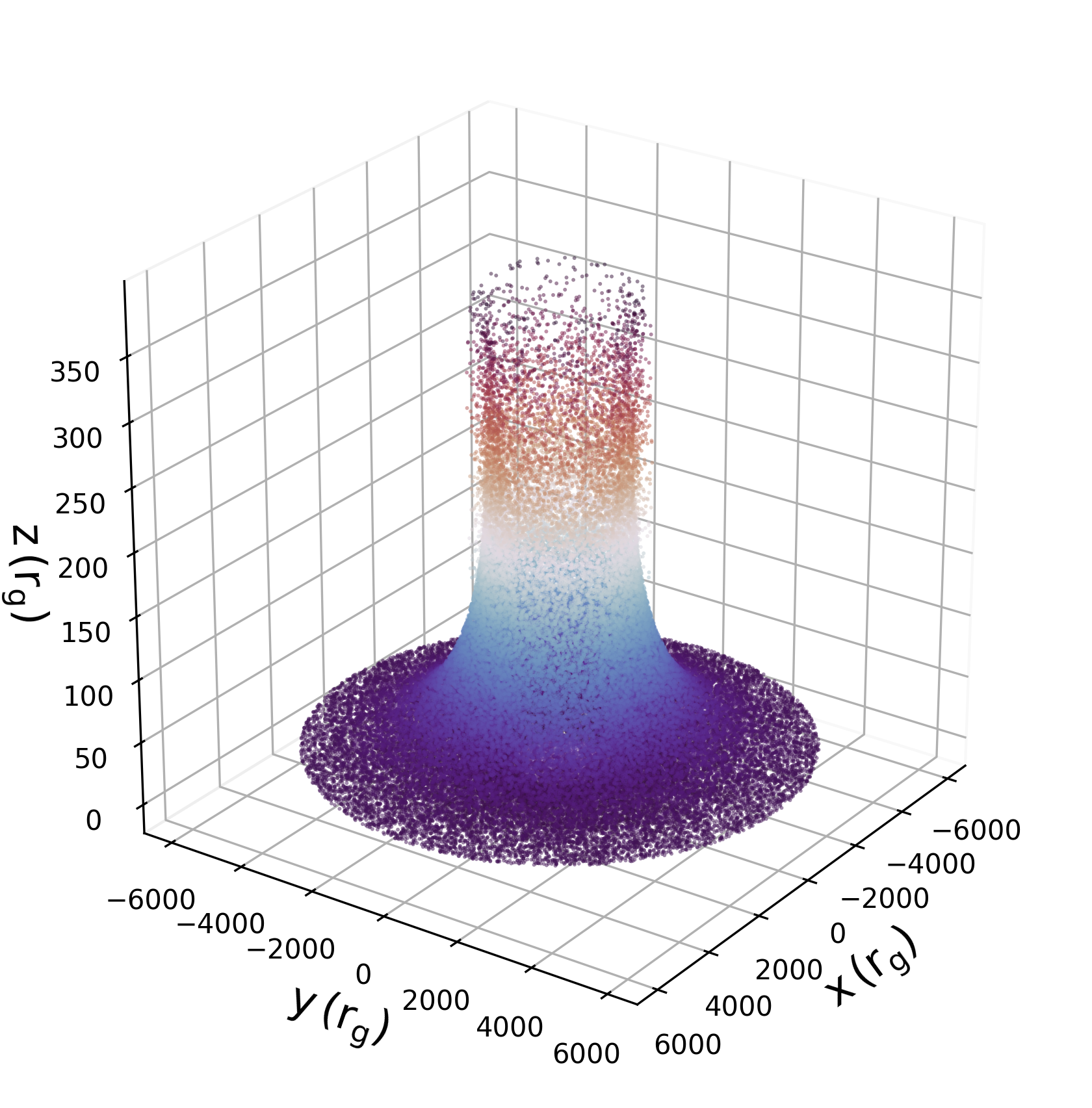}
    \includegraphics[width=0.33\linewidth]{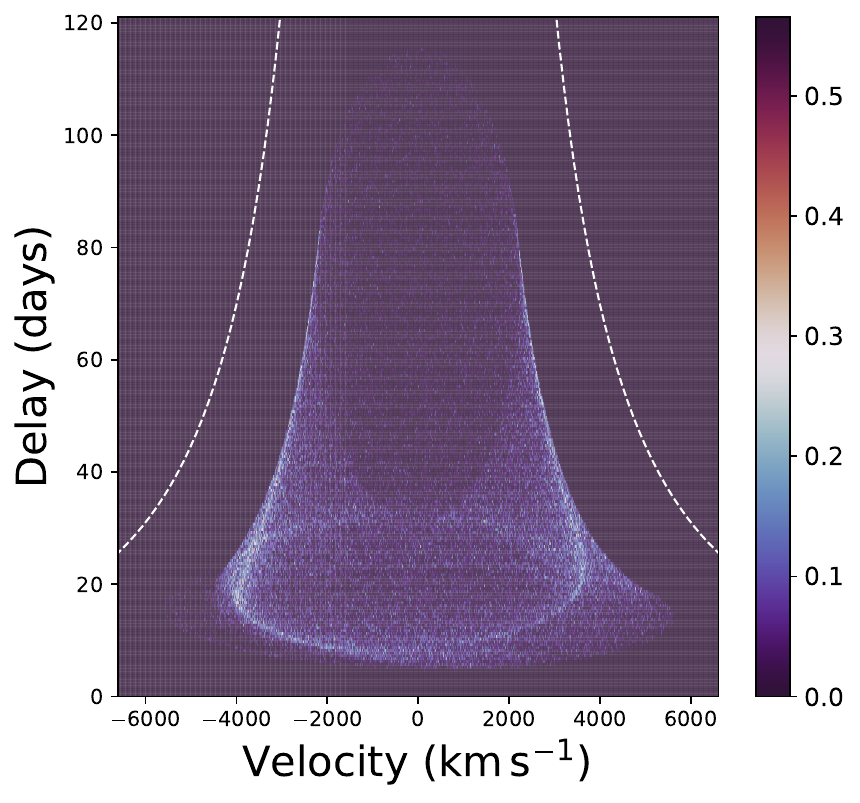}
    \includegraphics[width=0.33\linewidth]{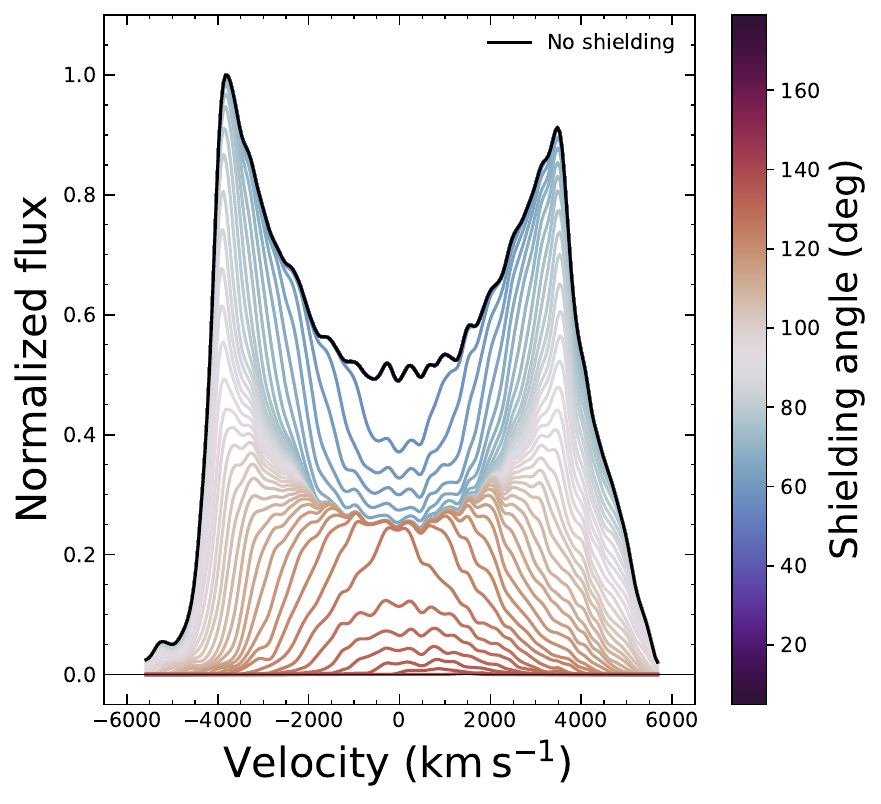}
    
\caption{Representative BLR properties of Fairall~9 from FRADO. Left: 3-D distribution of BLR cloud positions inferred from the FRADO model, assuming $M_{\mathrm{BH}} = 2.18 \times 10^8 \, M_{\odot}$, $\dot{m} = 0.0316$ (one of the three adopted accretion rates), and metallicity $Z = 5$ in solar unit. The axes are given in units of  $r_g$. Middle: 2-D velocity–delay map of Fairall~9 constructed from the FRADO model for an inclination angle of $i = 35^{\circ}$. The color scale represents the number density of BLR clouds in each velocity–delay bin. The white dashed curve denotes the virial envelope corresponding to Keplerian disk-like rotation, defined by $v^2 \times \tau$ = constant for $M_{\mathrm{BH}} = 2.18 \times 10^8 \, M_{\odot}$. Right: Emission-line profiles predicted by the FRADO model for different shielding angles. The black curve shows the line profile obtained without shielding.}
    \label{fig:cloud}
\end{figure*}

\subsection{Starlight contribution in shaping SED}
 
Starlight contributes to the observed flux in AGNs as a non-variable component. Although it does not influence the measured inter-band time delays, it plays an essential role in accurately modeling the broadband SED. In the case of Fairall~9, the presence of host-galaxy starlight is clearly evident. Host morphology in NED is given as Sc, but  \citet{jiang2021} mentioned more likely SB0a on a visual inspection. A comparison with a scaled Sb-galaxy template shows good agreement with the inferred host contribution across the optical bands \citep{halo1}. Motivated by this consistency, we represent the stellar emission using the Sb-galaxy template of \citet{kinney1996}. In our model, the corresponding normalization factor, $A_{\rm star}$, is treated as a free parameter and is determined through fitting.

\subsection{Combined time delay from model fitting}

Our {\tt H0RIZON-AGN} framework self-consistently incorporates a lamppost irradiation geometry, an intermediate warm Comptonizing region, a standard cold accretion disk, a radiation-pressure-driven dusty BLR, and torus dust emission that contributes at longer optical wavelengths. Accordingly, the total response function is constructed as the sum of the individual responses from each of these components, as expressed in the following equation.

\begin{equation}
\begin{split}
\psi_{\rm tot}(\lambda,t) &= \psi_{\rm d}(\lambda,t) + \epsilon_{\rm PL}(\lambda) \,\psi_{\rm WC}(t) \\
&+ f_{\rm BLR} \,  \epsilon_{\rm BLR}(\lambda)\int_{t_0}^{t_{\max}} \psi_{\rm d}(\lambda,t-t')\,\psi_{\rm BLR}(t')\,dt' \\
&+ f_{\rm dust} \, \epsilon_{\rm dust}(\lambda)\int_{t_0}^{t_{\max}} \psi_{\rm d}(\lambda,t-t')\,\psi_{\rm dust}(t')\,dt'
\end{split}
\label{eq:tf}
\end{equation}

Using this formalism, we compute the combined response function, $\psi_{\rm tot}(\lambda,t)$, over a grid of 100 wavelengths spanning the range from 1000 {\AA} to 10000 {\AA}. To allow direct comparison with observations, these model response functions are then convolved with the transmission curves, $R(\lambda)$, of the relevant photometric filters. This procedure yields the filter-averaged time delays at the corresponding effective wavelengths ($\lambda_{\rm eff}$) or band-centers, which are computed as:

\begin{equation}
\label{eq:delay}
\tau(\lambda) = \frac{\int \int t\psi_{\rm tot}(\lambda,t)R(\lambda)dtd\lambda}{\int \int 
\psi_{\rm tot}(\lambda,t) R(\lambda)dtd\lambda}
\end{equation}

Finally, to model the observed data of Fairall~9, we utilize the full set of parameters listed in Table~\ref{tab:param}, ensuring that all components defined in Equation~\ref{eq:tf} are consistently included. The parameter space is constrained simultaneously by the observed lag-spectrum and the SED, and the best-fitting solution is obtained through $\chi^2$ minimization following the methodology described in \citet{jaiswal2025}. In Figure~\ref{fig:model}, we present a schematic representation of our modeling approach implemented in the {\tt H0RIZON-AGN} code. This figure illustrates the various components of AGN, including the hard X-ray emitting hot corona, the warm corona, the cold accretion disk, and the outer BLR and torus reprocessing regions. Together, these components shape the AGN UV–optical continuum time delays as well as the broadband SED.

\section{Results}
\label{ss:result}

In this section, we present the results of the simultaneous lag-spectrum and SED modeling of Fairall~9 performed with the {\tt H0RIZON-AGN} code. A key feature of our approach is that all AGN components are modeled self-consistently and constrained through a joint fit to the lag-spectrum and SED.

The model incorporates the warm-corona and accretion-disk responses, a BLR transfer function constructed from physically motivated dynamical modeling within the FRADO framework combined with the BLR emissivity profile computed using {\tt Cloudy}, and a torus transfer function based on simplified first-order assumptions. All of these components are treated within a unified framework and optimized simultaneously. The joint fit treats the accretion rate and luminosity distance as key free parameters, allowing the luminosity distance, $D_{\rm L}$ $-$ and consequently the Hubble constant, $H_{0}$ $-$ to be constrained independently. The best-fitting model parameters are obtained through the simultaneous optimization of all AGN components against both the lag-spectrum and SED.

\subsection{Modeling warm corona}

To describe the temporal response due to Comptonization in a warm, optically thick ($\tau_{\text{op-depth}} > 10$), and relatively cool ($kT \sim 0.1$–1 keV) corona, we adopted a simple parameterization using a half-Gaussian response function, $\psi_{\rm WC}(t)$. This choice was motivated by its qualitative similarity to the asymmetric transfer functions expected from a flat accretion disk in a lamppost geometry \citep[e.g.,][]{kammoun_data2021,jaiswal2025}. Specifically, we set the peak of the half-Gaussian at a time delay corresponding to five times the corona height, and the width ($\sigma_{\rm WC}$) was also parameterized by the height. To guide this prescription, we performed a simple test assuming a lamppost corona illuminating the disk and examined how the irradiating flux spreads with radius, approximating the distribution with a Gaussian profile. From this, we found that the width scales approximately with the corona height as $\sigma_{\rm WC} \sim 0.76 \, h$. We treated $h$ as a free parameter in the model. Nevertheless, unlike the disk transfer function, we did not explicitly tie this response function to the viewing angle or transition radius, since the geometry of the warm corona is not expected to be planar. Finally $\psi_{\rm WC}(t)$ was modulated by the spectral shape of the hard-Xray power-law $\epsilon_{\rm PL}(\lambda)$. Such a treatment is physically motivated by growing observational and theoretical evidence for reprocessing within the warm corona. In particular, signatures of warm-corona reflection have recently been identified in high-quality observations \citep[see, e.g., the analysis of the Seyfert galaxy PG~1426+015 by][]{walton2025}, while their properties have also been investigated in theoretical studies \citep[e.g.,][]{ballantyne2024}.

\subsection{Modeling cold accretion disk}

In this setup, we assumed that the inner radius of the cold accretion disk, $r_{\rm transition}$, began at the outer boundary of the warm corona, while the outer radius of the cold disk was taken to be sufficiently large and fixed at $r_{\rm out} \, \sim \, 10^{4} \, r_g$. We treated $r_{\rm transition}$ as a free parameter in the model and varied it together with the warm-corona parameters ($\tau_{\rm WC}$, $T_{\rm WC}$) and the Eddington ratio to obtain the best fit to both the broadband SED, including the X-ray emission, and the lag-spectrum of Fairall~9. Using these best-fitting parameters, we then constructed the cold disk response function, $\psi_{\rm d}$.

\begin{figure}
    \centering
    \includegraphics[width=\columnwidth]{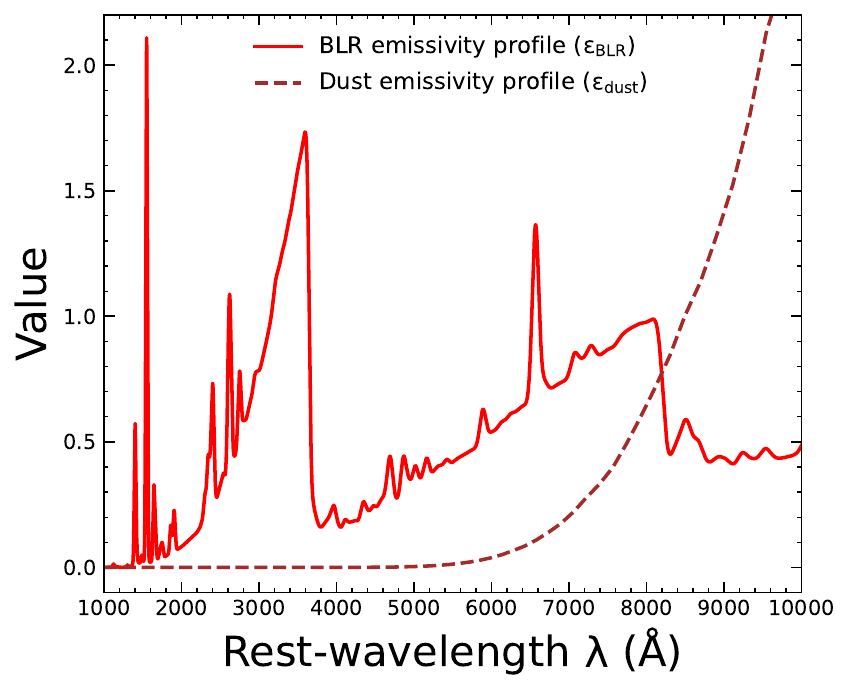}
    \caption{Emissivity profiles. BLR emissivity profile (ratio of the reprocessed to the incident continuum) of a BLR cloud centered at $\log r$ (cm) = 16.64, with an incident ionizing luminosity of $\log L$ (erg/s) = 44.97, a hydrogen density of $\log n_H$ (cm$^{-3}$) = 12, a column density of $\log N_H$ (cm$^{-2}$) = 23.5, and $f_{\rm BLR}=1$, shown by the solid red line computed using {\tt Cloudy}. The corresponding dust emissivity profile, assuming a dust sublimation temperature of $T_{\rm sub}=1500$ K, and $f_{\rm dust}=1$, is shown by the dashed brown line.} 
    \label{fig:emv_cld}
\end{figure}

\begin{figure*}
    \centering
    \includegraphics[scale=0.29]{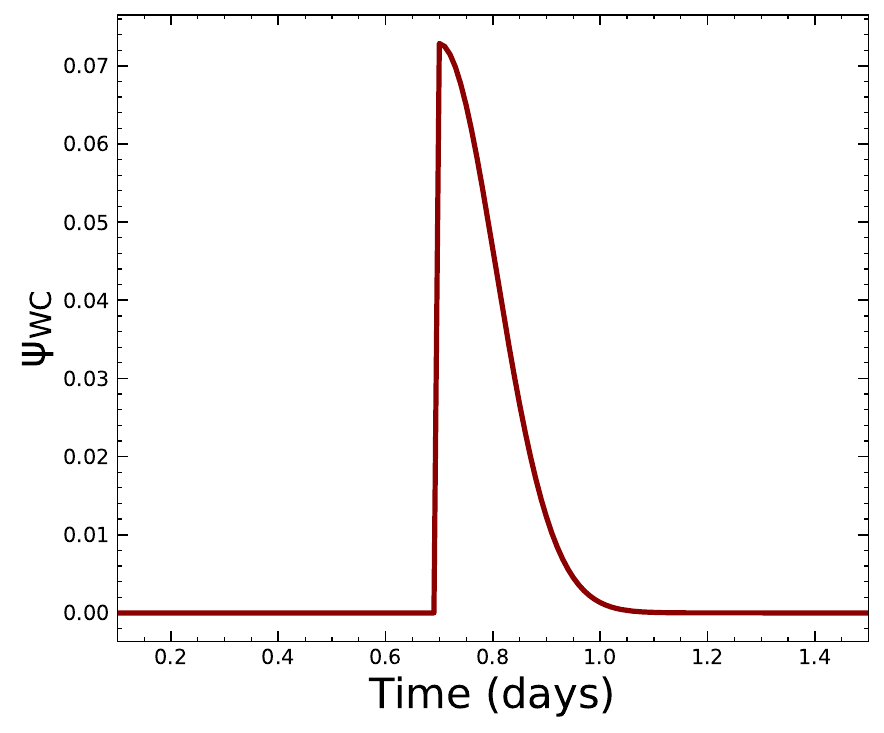}
    \includegraphics[scale=0.29]{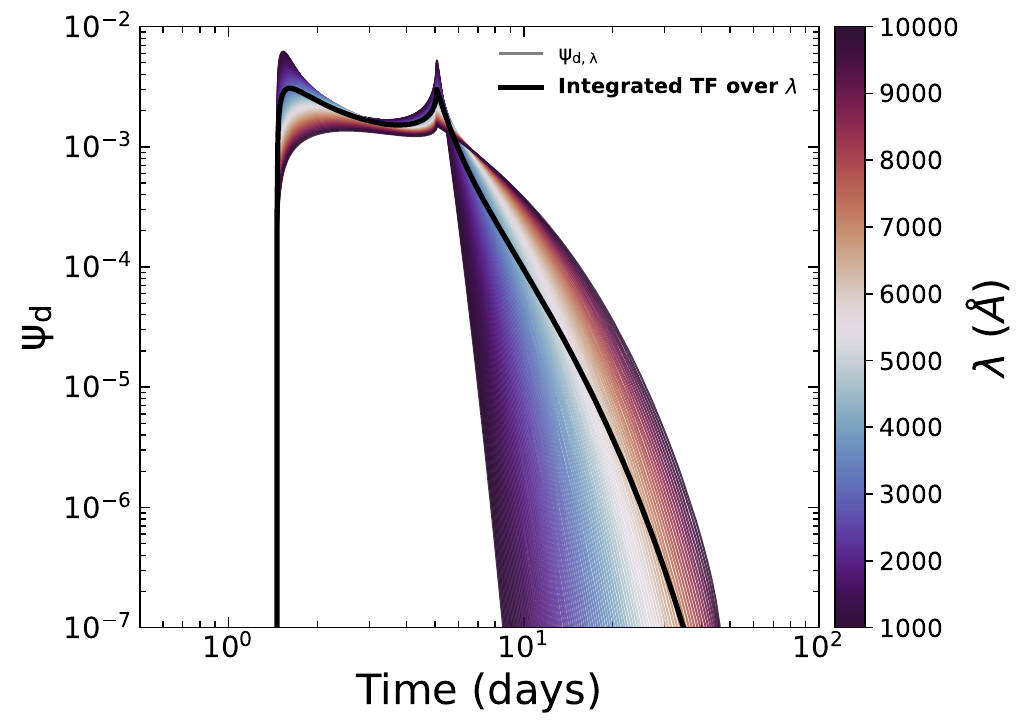}
    \includegraphics[scale=0.29]{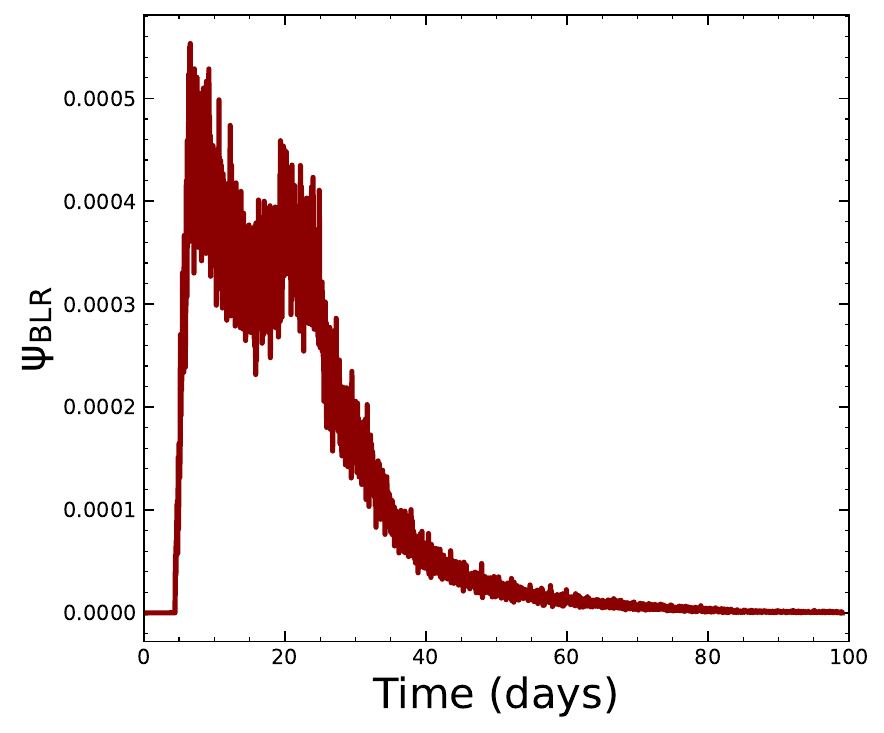}
    \includegraphics[scale=0.29]{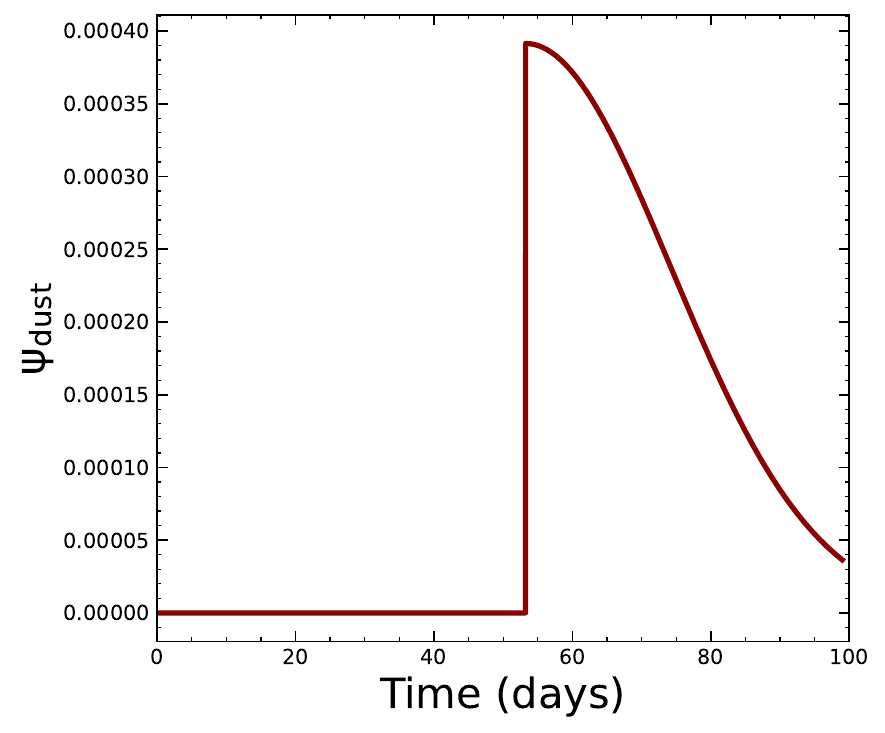}
    \caption{Transfer functions of the different emitting regions in the central parts of Fairall~9 for the best-fit luminosity distance, $D_{\rm L} = $ 190.6 Mpc. From left to right: (i) the transfer function of the warm corona; (ii) transfer functions of the cold accretion disk at different wavelengths, with the wavelength-integrated disk response function shown by the black line; (iii) the BLR response function obtained from FRADO assuming an inclination angle of $i=35^{\circ}$; and (iv) the torus response function.
    } 
    \label{fig:all_tf}
\end{figure*}

\subsection{BLR modeling using FRADO}
\label{ss:blrmod}

We employed the FRADO BLR dynamical model, which allows us to reproduce key observables of AGN BLRs, including low-ionization emission lines such as H$\beta$, the BLR transfer function ($\psi_{\rm BLR}(t)$), time delays, and the mean spectrum excluding detailed line intensities. Here, we adopted a dust sublimation temperature of $T_{\rm sub} \sim 1500 \, \mathrm{K}$ \citep{barvainis1987} to define the inner boundary of the BLR. 

The 3-D cloud distribution and their corresponding velocity fields within the BLR were fully determined by the FRADO model for $M_{\rm BH} = 2.18 \times 10^8 \, M_{\odot}$, and $i = 35^{\circ}$. We further assumed a metallicity of five times the solar value, consistent within uncertainties with the estimate of \citet{2024A&A...689A.321F}, and explored three representative Eddington ratios, $\dot{m} = 0.0199$, $0.0316$, and $0.0398$, which are characteristic of Fairall~9 \citep{marziani2010, jiang2021}. Note that we refer to $\dot{m} = 0.028$ as the fiducial value inferred from the monochromatic luminosity at 5100 {\AA} and $M_{\rm BH}$ \citep{halo1}, while $\dot{m} = 0.0316$ is adopted as a representative value for illustrating the 3-D BLR cloud distribution. In the subsequent analyses, the best-fit value of $\dot{m}$ is determined from the {\tt H0RIZON-AGN} model fitting. We set the BLR covering factor ($f_{\rm BLR}$) as free parameter in the model.

The representative distribution of BLR clouds for Fairall~9 is shown in the left panel of Figure~\ref{fig:cloud}. In this model, clouds are launched from the accretion-disk surface but remain strongly confined to its vicinity, with the vertical-to-radial extent limited to $\boldsymbol{z/R}\sim$3\%, where $\boldsymbol{z}$ is the height above the disk plane and $\boldsymbol{R}$ is the radial distance from the black hole. Such a limited vertical extension arises naturally from the low Eddington ratio of the source, resulting in a BLR geometry that resembles a geometrically thin, mildly puffed-up disk. The inner edge of the BLR is located at a radius of $\sim$1220~$r_g$. At this radius, the intrinsic Keplerian velocity is $\sim$8583 km s$^{-1}$, which corresponds to a projected line-of-sight velocity of $\sim$4923 km s$^{-1}$ for the adopted inclination angle. This projected velocity is broadly consistent with the observed H$\beta$ FWHM measurements for Fairall~9, such as $6901 \pm 707$ km s$^{-1}$ reported by \citet{peterson2004}, $5900 \pm 650$ km s$^{-1}$ reported by \citet{2020A&A...640A..39C}, and $ 5958 \pm 596$ reported by \citet{2024A&A...689A.321F}.

In the middle panel of Figure~\ref{fig:cloud}, we present the representative 2-D velocity-delay map computed for an inclination angle of $35^\circ$. The resulting structure indicates that the BLR gas is predominantly virialized and is well described by a simple flat Keplerian disk model. This follows from the fact that vertical velocities are negligible, no radial outflows are included, and no selective shielding effects are assumed in the model. As a result, the kinematics are governed primarily by rotational motion. The most responsive BLR gas is concentrated at time delays between approximately 6 and 40 days. However, a weaker but extended response is also present, reaching delays of up to $\sim$100 days. The corresponding velocity distribution spans up to $\pm 6000$ km s$^{-1}$, which is consistent with emission from low-ionization line–emitting clouds, such as those producing the H$\beta$ line.

The 3-D spatial and dynamical distribution of the FRADO clouds allows us to construct the corresponding broad emission-line profile, which we consider representative of the H$\beta$ emission. Since the observed profile depends strongly on inclination through the projection of cloud velocities along the line of sight, we introduced a shielding angle, $\theta_{\rm shield}$, measured with respect to the direction opposite to the observer. Clouds satisfying $\theta < \theta_{\rm shield}$ were removed, representing either obscuration of their emission or blockage of the ionizing radiation reaching them, while clouds outside the shielding cone remained visible. Consequently, small shielding angles affect only a limited far-side region of the BLR, whereas larger shielding angles progressively obscure a substantial fraction of the emitting clouds.

The right panel of Figure~\ref{fig:cloud} presents the resulting emission-line profiles for different shielding angles, with the black curve corresponding to the unshielded case where all BLR clouds contribute to the emission. As the shielding angle increases, progressively larger portions of the far-side BLR are removed, modifying the velocity distribution of the emitting gas and altering both the shape and symmetry of the line profile. In particular, stronger shielding suppresses one side of the velocity distribution, weakening the double-peaked structure and gradually transforming the profile into a smoother single-peaked shape. This demonstrates that anisotropic obscuration within the BLR can significantly modify the observed broad-line morphology and the apparent kinematic signature of the rotating BLR, even when the intrinsic cloud dynamics remain unchanged. In our modeling, however, we adopted the unshielded scenario, in which all BLR clouds contributed to the emission line, yielding an FWHM of $\sim 8000$ km s$^{-1}$, broadly consistent with the observed H$\beta$ FWHM reported by \citet{peterson2004}.

Next, we derived $\psi_{\rm BLR}(t)$ by projecting the delay-map onto time axis. Note that $\psi_{\rm BLR}(t)$ was constructed at certain $\dot{m}$ from the simulated transfer function corresponding to 3-D dynamical cloud distributions for adopted three different values of $\dot{m}$, as discussed earlier. Since variations in $\dot{m}$ naturally alter the characteristic BLR radius, $R_{\rm BLR}$, the latter was also effectively treated as a free parameter in constructing $\psi_{\rm BLR}(t)$. Finally, we convolved $\psi_{\rm BLR}(t)$ with the BLR emissivity profile, $\epsilon_{\rm BLR}(\lambda)$, computed using {\tt Cloudy} and shown by the solid red line in Figure~\ref{fig:emv_cld}, to obtain the wavelength-dependent BLR response.

\subsection{Modeling dusty torus}

This code modeled the dusty torus in Fairall~9 as a causal, spatially extended reprocessor of radiation, in which the dust response to a delta-function continuum flash is described by a normalized right-truncated Gaussian transfer function, $\psi_{\rm dust}(t)$. In this framework, the mean time delay is directly associated with the characteristic dust reverberation radius, $R_{\rm dust}$, which was taken to be approximately four times the BLR radius \citep{2006ApJ...639...46S, 2014ApJ...788..159K, 2020MNRAS.491.4615K, 2024ApJ...968...59M}, corresponding to $R_{\rm dust} \sim 70$ light-days. The width of the transfer function was parametrized as $\sigma_{\rm dust} = 0.3 \, \times \, R_{\rm dust}$, as discussed in Sect.~\ref{ss:torus}, thereby encoding the radial extent and geometric smearing of the torus response. Next, we constructed the spectral emissivity of the dust, $\epsilon_{\rm dust}(\lambda)$, assuming $T_{\rm sub} \sim 1500 \, \mathrm{K}$ to characterize the reprocessed dust emission,  following \citet{halo1}. The resulting profile is shown by the dashed brown line in Figure~\ref{fig:emv_cld}. Finally, the wavelength-dependent dust response was obtained by convolving $\psi_{\rm dust}(t)$ with $\epsilon_{\rm dust}(\lambda)$.

\subsection{Combined transfer function of Fairall~9}
\label{ss:tf_tot}

We present transfer functions for different emitting regions in the central engine of Fairall~9 in Figure~\ref{fig:all_tf}, evaluated at the best-fit model parameters. The left-most panel shows the warm corona transfer function, $\psi_{\rm WC}(t)$, which describes a geometrically smeared reflection response of the inner accretion flow due to irradiation from a compact hot corona located at a height $h = 11.26 \, r_g$ above the black hole.

The second panel from the left shows the accretion disk response functions, $\psi_{\rm d}(\lambda, t)$, at different wavelengths. As expected from light-travel time effects in an extended cold disk, the transfer functions become progressively broader toward longer wavelengths. This behavior results from the radial temperature structure of the disk, where the decreasing temperature with increasing radius causes longer-wavelength emission to arise from a broader range of disk radii. Along with this wavelength-dependent broadening, all response functions share a common minimum delay. This behavior arises from the temperature-dependent spectral distribution of the disk emission. Although the innermost disk region predominantly emits at shorter wavelengths, with its blackbody spectrum peaking in the UV, its high temperature allows it to contribute across the entire wavelength range considered here. Consequently, the earliest reprocessing signal is present in all wavelength-dependent responses and corresponds to the near side of the innermost disk boundary (azimuthal angle $\phi = 0$, $r = r_{\rm transition}$). In the adopted lamppost geometry, the minimum delay is determined by the combined effects of $r_{\rm transition}$, $h$, and $i$ \citep[see][for details of the geometry]{jaiswal2025}.

The second panel from the right presents the BLR transfer function, $\psi_{\rm BLR}(t)$, inferred from the FRADO model for Fairall~9. Its characteristic double-peaked structure arises naturally from geometry, e.g., the first peak at short delays ($\sim 6.7$ days) corresponds to BLR clouds located on the near side of the black hole relative to the observer, while the second peak at $\sim 20$ days arises from clouds on the far side. Since no shielding effects are included in this construction, the two-peak structure is expected. The resulting median delay of $\sim 18.74$ days is in excellent agreement with the observed H$\beta$ lag of $17.4^{+3.2}_{-4.3}$ days \citep{peterson2004}.

Finally, the right-most panel shows the dusty torus transfer function, $\psi_{\rm dust}(t)$, representing the response of the dust distribution that primarily contributes at larger optical ($\gtrsim 6000$ {\AA}) and NIR wavelengths. We note that this construction is based on simplified assumptions about the dust geometry; in reality, the dust distribution may extend to significantly longer delays. However, such extended components are unlikely to contribute appreciably to the UV–optical lag-spectrum of Fairall~9 considered in this work.

\begin{figure}
    \centering
    \includegraphics[width=\columnwidth]{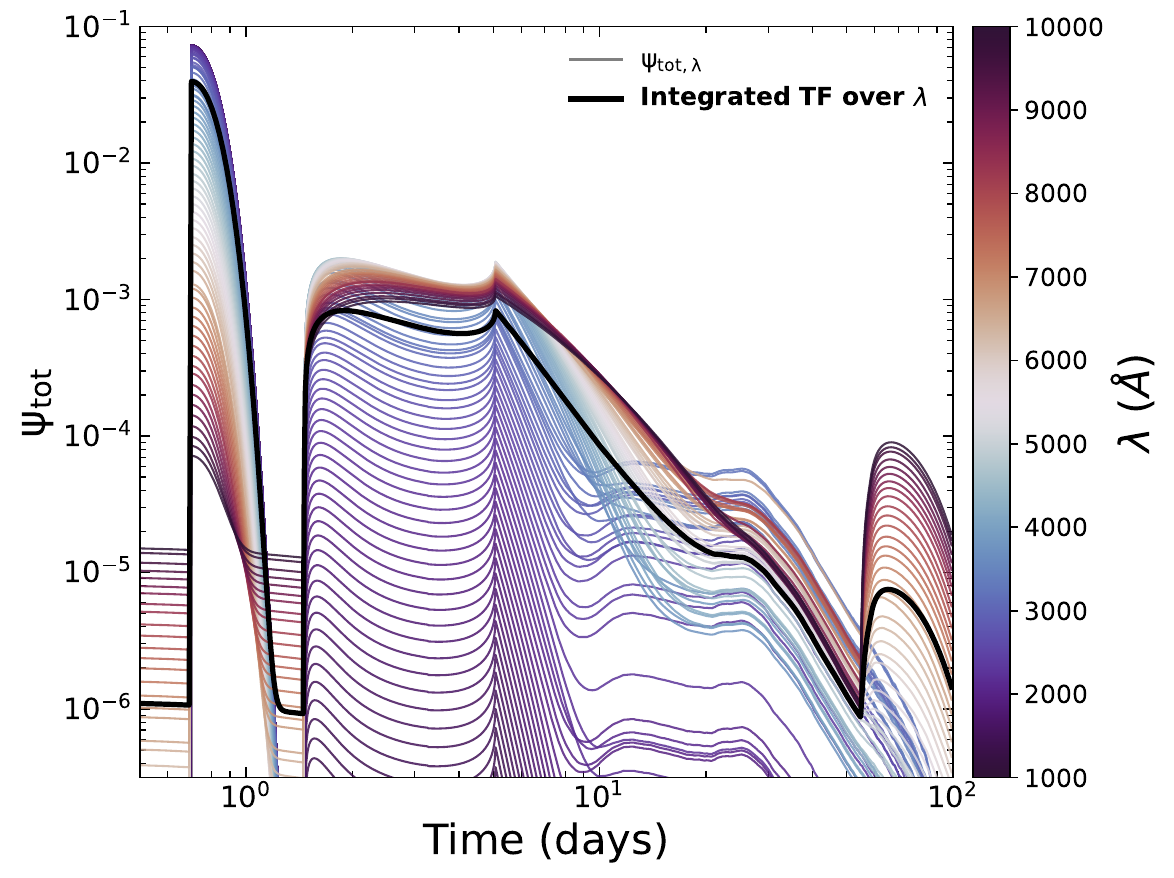}
    \caption{Total transfer functions, $\psi_{\rm tot}(\lambda,t)$, at different wavelengths. The color coding follows the wavelength labels. The wavelength-integrated total transfer function, $\psi_{\rm tot}(t)$, is plotted as the black curve for illustrative purposes. Parameters used: $M_{\text{BH}}$ = $2.18 \times10^{8}M_{\odot}$, Eddington ratio $\dot{m}$ = 0.0224, $L_X = 6.74 \times10^{44}$ erg s$^{-1}$, $h = 11.26 \, r_g$, $T_{\rm WC} = 4.58 \times 10^6$ K,  $\tau_{\rm WC} = 22.08$, $r_{\rm transition} = 252.38 \, r_g$, $r_{\rm out} = 10000 \, r_g$, $f_{\rm BLR} = 9.6 \%$,  $f_{\rm dust} = 12.9 \%$, and  $i = 35^{\circ}$.
    } 
    \label{fig:comb_tf}
\end{figure}

\begin{table}[h!]
\centering

 \caption{Fairall~9 lag-spectrum fitting results.}
 \label{tab:lag_spc}

\resizebox{9cm}{!}{
\fontsize{8pt}{8pt}\selectfont
\begin{tabular}{ccccr} \hline \hline

band/filter & $\lambda_{\rm eff}$/band center &$\tau$ & $\tau_{\rm mod}$ 
\\ 
 & ({\AA}) &  (days) & (days) \\
(1) & (2) & (3) & (4) 
\\ \hline \\

{\it Swift}-$W2$ & $2055$   & $-4.61_{-1.26}^{+1.29}$ & $-3.85$ \\
{\it Swift}-$M2$ & $2246$   & $-2.46_{-1.58}^{+1.82}$ & $-3.85$ \\
{\it Swift}-$W1$ & $2580$   & $-1.78_{-1.59}^{+1.63}$ & $-2.99$ \\
{\it Swift}-$U$ & $3463$   & $-0.04_{-1.20}^{+1.20}$ & $-0.65$ \\
Str\"omgren-$u$ & $3468$   & $-0.00_{-0.88}^{+0.92}$ & $0.00$ \\
Str\"omgren-$v$ & $4120$   & $-0.21_{-1.03}^{+0.93}$ & $0.00$ \\
{\it Swift}-$B$ & $4350$   & $-0.44_{-1.18}^{+0.86}$ & $-1.12$ \\
Str\"omgren-$b$ & $4668$   & $-0.83_{-0.89}^{+0.91}$ & $-0.69$ \\
{\it Swift}-$V$ & $5425$   & $0.19_{-0.16}^{+0.16}$ & $0.26$ \\
Str\"omgren-$y$ & $5460$  & $0.40_{-0.72}^{+1.21}$ & $0.28$ \\
Johnson--Cousins-$I$ & $8105$  & $7.04_{-1.43}^{+1.09}$ & $6.43$ \\

\\ 
\hline
\end{tabular}
}

\tablefoot{Columns are: (1) name of the OCM from HALO and \textit{Swift} bands/filters, (2) effective wavelength for OCM filter or band center for \textit{Swift}  (observed-frame for the source), (3) rest-frame inter-band time-delays with respect to $u/U$-band (OCM/\textit{Swift}) from ICCF method with their 1$\sigma$ lower and upper uncertainties, and (4) model recovered time delays. The observed data are retrieved from \citet{halo1}. Note that in Figure~\ref{fig:lag_spec}, the rest-frame inter-band time delays are shown relative to the shortest {\it Swift}-$W2$ band for presentation purposes.}

\end{table}

After constructing the transfer functions for the individual emitting components, we combined them using Equation~\ref{eq:tf} to obtain the total wavelength-dependent transfer function, $\psi_{\rm tot}(\lambda,t)$. In this formulation, the relative contributions of the BLR and the dusty torus are controlled by the free parameters $f_{\rm BLR}$ and $f_{\rm dust}$, respectively, which are constrained through a best-fit to the observed lag-spectrum and the SED (see Sect.~\ref{ss:sed})  of Fairall~9. The resulting model-recovered total transfer functions are shown in Figure~\ref{fig:comb_tf} at different wavelengths. The total transfer functions exhibit a distinctly multi-modal structure, reflecting the superposition of contributions from physically distinct emitting regions. In particular, the warm corona dominates the response at the shortest UV--optical wavelengths, while the cold accretion disk governs the maximum UV--optical range. The BLR introduces delayed, broadly distributed components, and the dusty torus contributes predominantly at the longest optical wavelengths. Together, these components give rise to the complex wavelength-dependent temporal behavior observed in the continuum reverberation signal of Fairall~9.

\subsection{Lag-spectrum fitting of Fairall~9}

We then derived the model-predicted time delays from the total transfer function, $\psi_{\rm tot}(\lambda,t)$, at the effective wavelengths / band-centers of the filters used to construct the observed Fairall~9 lag-spectrum. This was achieved by convolving $\psi_{\rm tot}(\lambda,t)$ with the corresponding filter response curves using Equation~\ref{eq:delay}, thereby obtaining the filter-averaged time delays at the reference wavelengths. The observed time delays used for fitting the Fairall~9 lag-spectrum, together with the corresponding model-recovered delays, are listed in Table~\ref{tab:lag_spc}.

\begin{figure}
    \centering
    \includegraphics[width=\columnwidth]{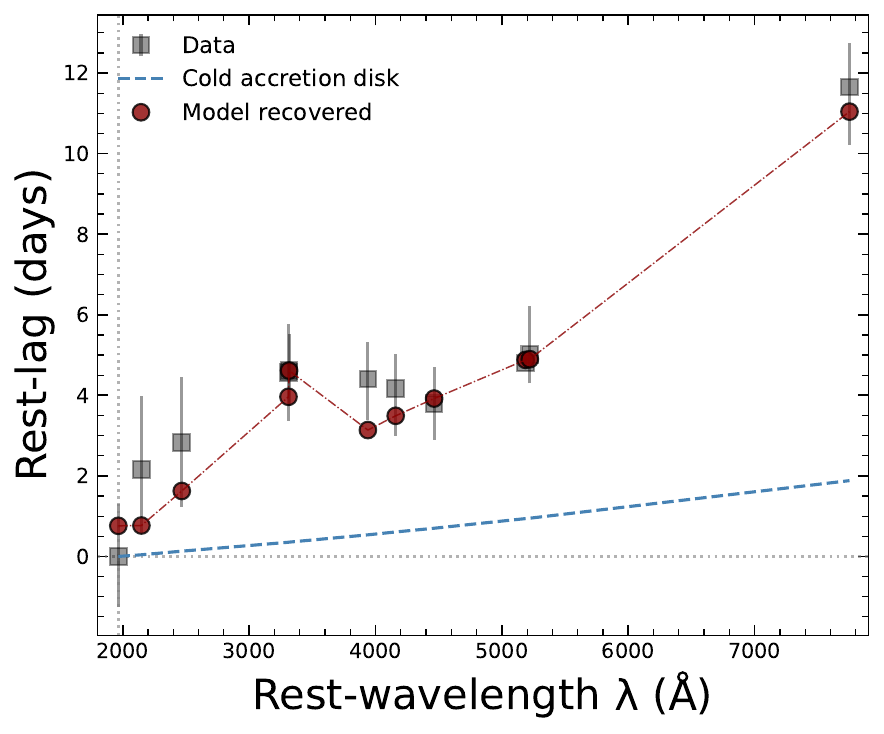}
    \caption{ Lag-spectrum modeling of Fairall~9 for the best-fit luminosity distance, $D_{\rm L} = $ 190.6 Mpc. The black square points with error bars represent the observed data, while the red circular points show the model-recovered time delays. The red dot-dashed line connects the model-recovered points to guide the eye. The blue dashed line represents the lag--wavelength dependence expected from a  cold accretion disk. The vertical and horizontal dotted lines indicate the rest-frame reference wavelength and zero rest-frame lag, respectively. Note that the inter-band delays are shown relative to the {\it Swift}-$W2$ band, which is adopted as the reference wavelength for illustration purposes.
    } 
    \label{fig:lag_spec}
\end{figure}

An example of the recovered model time delays, overplotted on the observed lag-spectrum of Fairall~9, is presented in Figure~\ref{fig:lag_spec} for the best-fit model corresponding to a luminosity distance of $D_{\rm L}$ = 190.6~Mpc. Overall, our model successfully reproduces the observed time delays within the associated error bars.

For comparison, \citet{halo1} modeled the Fairall~9 lag-spectrum using both a relativistic accretion disk model \citep{kammoun2021, kammoun2023} and a modified radiation-pressure-confined (RPC) model \citep{netzer2022} that additionally incorporates dust emission. The relativistic accretion disk model naturally accounts for GR effects, which are expected to be most important in the far-UV and extreme-UV regimes ($\lambda \lesssim 2000$--$3000~{\AA}$). However, this model neglects contributions to the observed continuum delays from both the warm corona and more extended regions, such as the BLR. In contrast, the RPC model assumes that the UV--optical continuum lags are governed primarily by BLR properties, with only a negligible contribution from the accretion disk; \citet{halo1} further extended this framework by including dust emission.

Despite these modifications, \citet{halo1} found that neither the relativistic accretion disk model nor the RPC-based model was able to accurately reproduce the observed lag-spectrum of Fairall~9. In contrast, our physically motivated {\tt H0RIZON-AGN} model successfully recovers the observed lag-spectrum, highlighting the importance of simultaneously accounting for both the inner accretion flow and the larger-scale reprocessing regions in AGN continuum reverberation modeling.

\subsection{SED fitting of Fairall~9}
\label{ss:sed}

\begin{figure}
    \centering
    \includegraphics[width=\columnwidth]{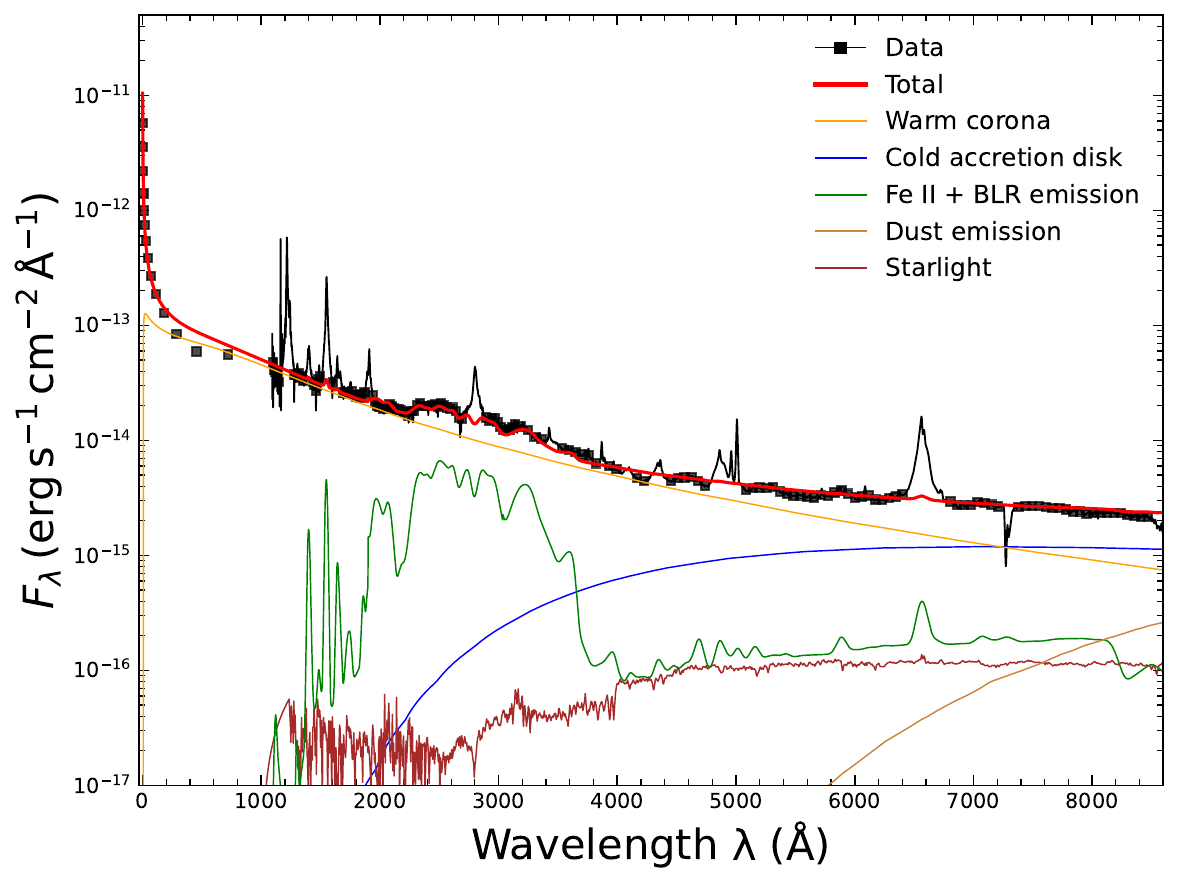}
    \caption{Best fit of the broadband SED of Fairall~9 for luminosity distance, $D_{\rm L} = $ 190.6 Mpc. The observational data are marked in black, the total model in red, the components of the model are the following: warm corona (orange), cold outer disk (blue), BLR and Fe II (green), dusty torus(peru), and  starlight (brown).
    } 
    \label{fig:spectrum}
\end{figure}

The best-fitting broadband spectrum is presented in Figure~\ref{fig:spectrum}. To clearly illustrate the Optical/UV emission, we display the full spectral range, extending from the X-ray to the near-infrared, on a linear scale. The hard X-ray emission (at wavelengths of a few \AA) is dominated by the power-law component, whereas the UV emission and the majority of the optical continuum originate primarily from the warm corona. The contribution from the outer cold accretion disk gradually increases toward longer wavelengths and becomes comparable to that of the warm corona at approximately 7000~\AA. Beyond this wavelength, emission from the dusty torus also becomes significant. In addition, the host-galaxy starlight contributes noticeably across the optical band and remains detectable down to $\sim 4000$~\AA, as is commonly observed in luminous AGNs.

A particularly noteworthy aspect of the spectral decomposition is the BLR component. In this source, the expected signatures of the Balmer edge are largely obscured by strong Fe~II emission. Such prominent Fe~II features are typically associated with high-Eddington-ratio AGNs \citep{2001marziani,2011dong,panda2019, 2025Univ...11...69M}, with \citet{bruhweiler2008} highlighting the extreme case of I~Zw~1. Although Fairall~9 is not characterized by a high Eddington ratio, it nevertheless exhibits a substantial Fe~II contribution, especially in the vicinity of the Mg~II emission line. Consequently, this component was explicitly included in our spectral modeling. The resulting best-fit AGN continuum model, shown as the solid red line in Figure~\ref{fig:spectrum}, accurately reproduces the observed AGN continuum represented by the black square data points. The individual components contributing to the total continuum are also displayed in different colors, illustrating their respective roles across the broadband spectrum.

\begin{figure}
    \centering
    \includegraphics[width=\columnwidth]{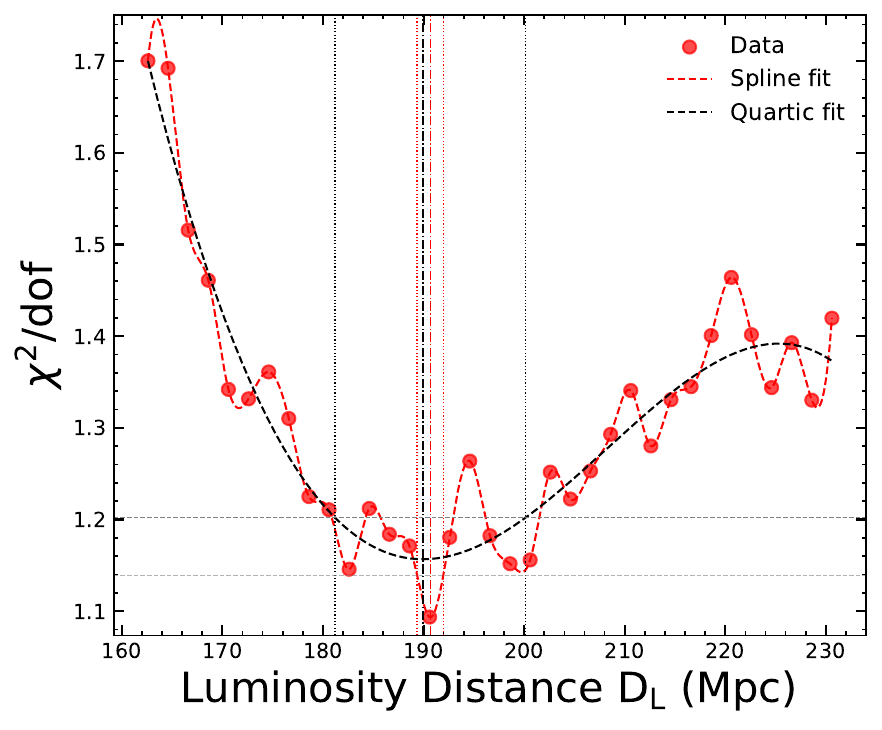}
    \includegraphics[width=\columnwidth]{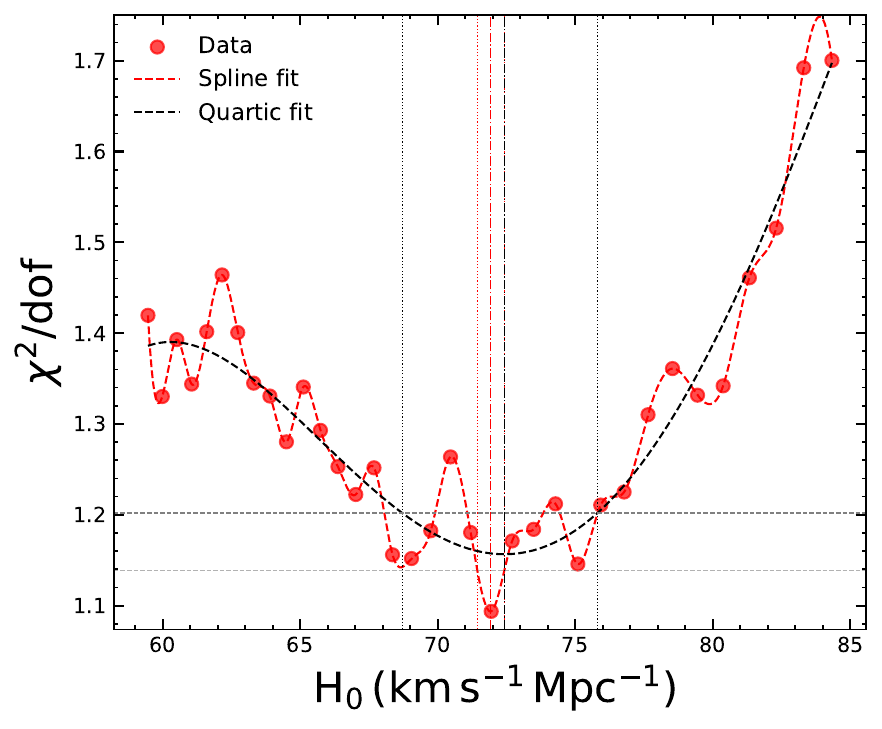}
    \caption{Simultaneous lag-spectrum and SED fitting results for Fairall~9.
Top: Best-fit $\chi^2/{\rm dof}$ as a function of the luminosity distance, $D_{\rm L}$, marginalized over all other model parameters.
Bottom: Same as the top panel, but shown as a function of the inferred Hubble constant, $H_{0}$. The red dashed and black dashed curves represent spline and quartic fits, respectively, to the data shown as red circular points. The vertical red and black dotted-dashed lines indicate the luminosity distance (and, in the bottom panel, the corresponding Hubble constant) at which $\chi^2/{\rm dof}$ is minimized for the spline and quartic fits, respectively. The associated $1\sigma$ uncertainties are indicated by the vertical dotted lines.
    } 
    \label{fig:chi}
\end{figure}

\subsection{Distance estimation via simultaneous lag-spectrum and SED fitting}

In our model, we fixed $M_{\rm BH}$, $i$, $\Gamma$, $r_{\rm ISCO}$, $r_{\rm out}$, $R_{\rm dust}$, and $f_{\rm col}$, since these parameters were either independent of the luminosity distance or were independently constrained from observations. In particular, $M_{\rm BH}$ was obtained from spectroscopic--RM, $\Gamma$ from the X-ray spectral slope, and $i$ from SED modeling. However, because our primary goal was to infer the source distance, we explicitly avoided relying on quantities that were themselves degenerate with or dependent on an assumed distance scale.

To ensure consistency, we therefore allowed the Eddington accretion rate $\dot{m}$ to vary over a broad range, effectively capturing the uncertainty associated with different possible luminosity distances. Building on this framework, we computed models on a grid of pre-selected luminosity distances. For each assumed distance, we then optimized the remaining free parameters listed in Table~\ref{tab:param} by minimizing the corresponding $\chi^2$ statistic.

For the SED fitting, we selected a set of representative continuum points spanning the soft X-ray to optical wavelength range while avoiding regions contaminated by prominent emission lines. To account for uncertainties in the spectral modeling, we adopted a uniform 10\% error on the selected continuum fluxes.

For the lag-spectrum fitting, we used the observed continuum lag-spectrum of Fairall~9, together with the reported delay uncertainties from \citet{halo1}. We then simultaneously fitted both the SED and the lag-spectrum over a grid of luminosity distances and computed the corresponding combined reduced $\chi^2$ values for each model realization.

The results of the joint lag-spectrum and SED fitting are shown in the top panel of Figure~\ref{fig:chi} in luminosity-distance space. In the $\chi^2/{\rm dof}$--$D_{\rm L}$ plane, each point corresponds to a model evaluated with the same $\dot{m}$ and all other parameters consistently applied to both the lag-spectrum and SED fits. The global minimum within the adopted luminosity-distance grid occurs at $D_{\rm L}=190.6$ Mpc. The corresponding best-fit lag-spectrum and SED are presented in Figures~\ref{fig:lag_spec} and \ref{fig:spectrum}, respectively.

Since the $\chi^2/{\rm dof}(D_{\rm L})$ profile exhibited several secondary minima in addition to the primary minimum near $D_{\rm L}=190.6$ Mpc, we further refined the luminosity-distance estimate by reconstructing the continuous $\chi^2/{\rm dof}(D_{\rm L})$ profile from the discrete sampled points and identifying its minimum numerically. To assess the robustness of the result, we employed two independent interpolation and fitting approaches.

First, we applied a cubic interpolating spline that passes exactly through all sampled points, thereby providing a smooth, continuous representation of the $\chi^2/{\rm dof}(D_{\rm L})$ profile without imposing any predefined analytic form. The minimum of the interpolated curve was then determined numerically, while the $1\sigma$ confidence interval was estimated using the criterion $\chi^2/{\rm dof} \leq \chi^2_{\rm min}/{\rm dof} + 1/22$ for 22 degrees of freedom.

Second, we modeled the $\chi^2/{\rm dof}(D_{\rm L})$ profile using a weighted quartic polynomial, with the coefficients determined through weighted least-squares regression. The fitted polynomial was evaluated on the same dense luminosity-distance grid, and both the minimum and its confidence bounds were obtained using the same statistical criterion.

The resulting spline interpolation, shown by the red dashed curve in Figure~\ref{fig:chi}, reproduces all sampled points exactly and is therefore highly sensitive to local fluctuations and numerical irregularities in the profile. Consequently, the inferred minimum closely follows the local structure of the sampled data, yielding $D_{\rm L}=190.7_{-1.3}^{+1.3}$ Mpc. In contrast, the weighted quartic fit, shown by the black dashed curve, imposes a smooth global functional form on the $\chi^2/{\rm dof}$--$D_{\rm L}$ relation, effectively suppressing small-scale variations and providing a more stable estimate of the underlying statistical trend. This approach yields $D_{\rm L}=189.9_{-8.7}^{+10.2}$ Mpc. Although the quartic-fit minimum differs slightly from the raw-data minimum, it likely provides a more robust characterization of the global trend and more conservative uncertainty estimates.

 The corresponding joint lag-spectrum and SED fitting results are also presented in terms of the Hubble constant, $H_{0}$, in the bottom panel of Figure~\ref{fig:chi}. Since the redshift of Fairall 9 is relatively small, the Hubble constant is approximated using the relation $H_{0}=zc/D_{\rm L}$. Applying this conversion to the fiducial quartic-fit distance yields $H_{0}=72.4_{-3.7}^{+3.4} \, \rm km \, s^{-1} \, Mpc^{-1}$. For completeness, the corresponding spline-based results are also shown in the same panel for illustrative purposes.


\section{Discussion}
\label{ss:dis}

The primary objective of our HALO program is to constrain the Hubble constant through the simultaneous modeling of AGN continuum lag-spectra and SEDs. To achieve this goal, we have developed a comprehensive physical framework, implemented in the {\tt H0RIZON-AGN} code, that self-consistently integrates a lamppost irradiated standard accretion disk, an inner hot flow embedded within a warm corona, a BLR structure governed by the FRADO model, and potential dust contamination originating from the torus.

The model is characterized by a broad set of physical parameters, including the black hole mass, inclination angle, hard X-ray power-law slope, accretion rate, lamppost height, cloud density, warm corona temperature and optical depth, BLR cloud covering factor, dust covering factor, and several additional quantities. In total, the model contains 17 parameters, of which 10 are treated as free parameters and constrained through fitting to the observational data, while the remaining 7 are held fixed throughout the fitting procedure (see Table~\ref{tab:param} for reference). A key feature of our methodology is the simultaneous fitting of the mean spectrum and the wavelength-dependent continuum lags, allowing both spectral and timing information to jointly constrain the model parameters.

The first successful application of this methodology was carried out using archival observations of NGC~5548 \citep{jaiswal2025}. Encouraged by the results of that pilot study, we now applied the same framework to Fairall~9, the first source monitored within the dedicated HALO observational campaign.

For both objects, the joint SED and continuum lag-spectrum analysis yields independent constraints on the Hubble constant. The inferred values of $H_{0}=72.4_{-3.7}^{+3.4} \, \rm km \, s^{-1} \, Mpc^{-1}$ for Fairall~9 and $66.9^{+10.6}_{-2.1} \, \rm km \, s^{-1} \, Mpc^{-1}$ for NGC~5548 \citep{jaiswal2025} are mutually consistent within their respective uncertainties. Importantly, both measurements are consistent within 2$\sigma$ with the \textit{Planck} measurement based on the $\Lambda$CDM cosmological model, $H_0 = 67.4 \pm 0.5$ km s$^{-1}$ Mpc$^{-1}$ \citep{Planck2020}. Although these results demonstrate the potential of the method, the current uncertainties remain relatively large, amounting to approximately $\sim$ 5\% for Fairall~9 and $\sim$ 10\% for NGC~5548. Consequently, the present measurements do not yet provide sufficient precision to address the Hubble-tension problem, namely the reported discrepancy between the values of the Hubble constant inferred from observations of the early and late Universe \citep[see, e.g.,][]{eleonora_2022}. Nevertheless, the results obtained so far provide an important proof of concept and motivate the extension of the HALO program to a larger sample of AGNs. 

The higher precision achieved for Fairall~9 compared to NGC~5548 may, at least in part, results from the use of sub-day cadence intermediate-band photometry from HALO observations, which minimizes contamination from BLR emission lines. Combined with the near daily-cadence, high-quality {\it Swift} data, this yields cleaner measurements of the intrinsic continuum variability, as discussed in \citet{halo1}. More generally, our current results represent a substantial advance over the early attempts to determine the Hubble constant from AGN continuum time delays \citep[e.g.,][]{collier1999,cackett2007}. This progress stems from several developments, most notably the inclusion of continuum reprocessing within the BLR and the adoption of a comprehensive numerical framework in place of the simplified analytical treatments used previously.

Despite these improvements, the precision of our method remains well below that achieved by the most mature cosmological probes. A significant part of this difference arises from the limited number of sources studied so far. For example, strong-lensing analyses that combine measurements from multiple systems have reached uncertainties below 2\% \citep[e.g.,][]{wong2020}. Other direct methods have also demonstrated impressive precision. In particular, the determination of $H_0$ from the strongly lensed supernova Refsdal achieved a comparable level of accuracy using a single object \citep{Kelly2023Refsdal}. Even higher precision is obtained by large-scale distance-ladder programs. For example, the Local Distance Network analysis of \citet{HoDN2025} reports a highly precise value of the Hubble constant and argues for a tension with the early-Universe determination from the cosmic microwave background at a significance exceeding 7$\sigma$.

At the same time, these highly precise measurements rely on multi-step calibration procedures and complex scaling relations, which may be susceptible to systematic effects. For example, \citet{son2025} argued that a correlation between standardized Type Ia supernova magnitudes and progenitor age could introduce a significant redshift-dependent bias into supernova cosmology. Statistical assumptions may also play an important role. For instance, \citet{desmond2025} showed that accounting for the expected large-scale homogeneity of the galaxy distribution shifts the inferred value of $H_0$ toward lower values. In this context, direct methods such as the one developed here remain particularly valuable as independent and largely complementary probes of cosmic expansion. Considerable effort is currently being devoted to their development. One recent example is the work of \citet{wagner2025}, who mapped the local expansion field around M~81 and M~82 and derived a local value of $H_0 = 63 \pm 6$ km s$^{-1}$ Mpc$^{-1}$.

Nevertheless, the application of our method to precision cosmology will require substantially larger datasets and a more rigorous assessment of both statistical and systematic uncertainties. Expanding the AGN sample and refining the error budget will be essential steps toward achieving competitive constraints on the Hubble constant and fully exploiting the potential of AGN reverberation-based cosmology.

 \subsection{Error determination}

Given the complexity of the {\tt H0RIZON-AGN} framework, the model necessarily contains a large number of parameters. Uncertainties on the free parameters are estimated directly during the fitting procedure. However, a number of quantities listed in Table~\ref{tab:param} are held fixed, and additional assumptions are embedded within the underlying FRADO model \citep{naddaf2021}. Among these, one of the most important is the dust sublimation temperature, which is assumed to be 1500 K \citep[see discussion in][]{jaiswal2025}. Furthermore, the radiative transfer calculations involve additional cloud properties, such as the local gas density and column density.

The impact of these fixed parameters on the inferred value of the Hubble constant is not expected to be uniform. Some quantities are already tightly constrained by the available observations and therefore introduce only a limited source of uncertainty. For example, the hard X-ray spectral slope can be measured directly from the X-ray data. Other parameters are expected to have only a minor influence on the final results, such as the outer radius of the accretion disk. In the case of the local cloud density, we investigated its role in Sect.~\ref{dis:cloud} and showed that the predicted BLR structure is relatively insensitive to this parameter over the range of values generally considered plausible.

In principle, the effect of all fixed parameters and model assumptions on the determination of the Hubble constant could be quantified by allowing them to vary and marginalizing over their uncertainties. In practice, however, such an analysis is currently beyond the capabilities of the present implementation of the code and would require substantially greater computational resources. A comprehensive assessment of these additional sources of uncertainty therefore remains an important goal for future studies and will be essential for further improving the robustness of AGN-based measurements of the Hubble constant.

\subsection{Future data prospects}

We are monitoring a selected sample of AGNs with the OCM within the HALO program, spanning a broad range of luminosities and redshifts up to $z \sim 3$. In parallel, we are modeling several nearby AGNs with existing OCM monitoring data, using the same filter set and adjusting the observing cadence to each source's intrinsic luminosity. This strategy is expected to yield increasingly precise continuum time delay measurements.

To further expand the sample, we plan to incorporate high-quality continuum lag measurements available in the literature, covering approximately $\sim 15-18$ AGNs \citep{2017ApJ...840...41E, cackett2018, 2018ApJ...854..107F, 2019ApJ...870..123E, pozo2019, lobban2020,  cackett2020, 2021ApJ...922..151K, two_timescales2021, 2022ApJ...934L..37M, 2023ApJ...947...62K, 2023MNRAS.525.4524G, 2023MNRAS.523..545D, 2023ApJ...953..137M, 2024ApJ...974..271L, 2025MNRAS.541..642P, 2025A&A...700L...8P, 2026MNRAS.546ag025K, 2026ApJ...997..326F, 2026MNRAS.546ag067D, 2026ApJ..1003..196P, 2026MNRAS.548ag642M, 2026ApJ..1003..147M}, following the approach successfully applied to NGC~5548 \citep{jaiswal2025}. Combined with our OCM results, this will provide a homogeneous sample of more than 20 high-quality distance measurements, substantially reducing the uncertainty in the Hubble constant.

In addition, decomposing the measured delays into contributions from the accretion disk, BLR, and dusty torus will offer valuable insight into the structure of AGN and enable investigations of how these components depend on global source properties, such as black hole mass, inclination angle, and Eddington ratio.

The next major challenge will be to apply our methodology to the large AGN sample expected from the Vera Rubin Observatory. The potential of measuring continuum time delays with the Legacy Survey of Space and Time (LSST) has been extensively discussed in the literature \citep[e.g.,][]{homayouni2019, pozo2023, pozo2024}. Current forecasts predict continuum lag measurements for approximately 4,000--10,000 AGNs in the Deep Drilling Fields \citep[DDF;][]{2022ApJS..262...49K, 2026ApJ..1000..165L}, where the observing cadence will be sufficiently dense for reverberation mapping studies. These fields overlap with regions that already possess extensive multi-wavelength coverage, including X-ray observations. Since our modeling requires a well constrained broadband SED, a high-quality mean optical spectrum, and a spectroscopic redshift, many DDF sources are expected to meet these requirements \citep[e.g.,][]{Zhou2022}.

The availability of hundreds, and potentially thousands, of suitable AGNs offers the prospect of dramatically reducing statistical uncertainties, possibly enabling precision below 1\% on the Hubble constant. Realizing this potential, however, will require a careful assessment of systematic uncertainties associated with the model, as well as significant improvements in computational efficiency. An additional challenge will be achieving the photometric precision needed for continuum lag measurements, which may exceed the performance of the standard LSST data-processing pipeline.

Despite these challenges, the scientific prospects are highly promising. Although our model contains several free parameters and LSST will provide only six photometric bands, preliminary tests indicate that combining six inter-band delays (including the reference band delay) with a high-quality optical spectrum and broadband SED yields constraints on the Hubble constant that are only modestly weaker than those obtained with more extensive datasets. Nevertheless, comprehensive simulations and further validation of this approach will be essential before LSST data become available.

\section{Summary}
\label{ss:summ}

In this work, we present the AGN continuum lag-spectrum and SED fitting results for Fairall~9, obtained using our newly developed {\tt H0RIZON-AGN} code to estimate the Hubble constant, $H_{0}$. As the first target of the HALO AGN monitoring program, Fairall~9 was monitored photometrically with the OCM telescope, complemented by archival \textit{Swift} observations to construct the lag-spectrum. The broadband SED was assembled from photometric measurements retrieved from NED and UV--optical spectrum available in the literature. This analysis has proven highly successful and has generated several important findings, summarized below.

\begin{enumerate}
   
    \item We introduce {\tt H0RIZON-AGN}, a physically motivated multi-component model that combines a lamppost-irradiated accretion disk with emission from the warm corona and reprocessing by the BLR and dusty torus. By simultaneously fitting AGN continuum time delays and broadband SEDs, the model provides a self-consistent description of the observed emission across multiple wavelengths. This framework enables direct constraints on the luminosity distance and, in turn, an independent determination of the Hubble constant, $H_{0}$.

    \item  The BLR was modeled using the radiation-pressure-regulated dusty outflow model (FRADO), which successfully reproduced both the observed H$\beta$ reverberation lag and the emission line profile for Fairall~9. The agreement between the model predictions and observations provides independent validation of the adopted BLR structure and dynamics. Moreover, the adopted BLR structure combined with the emissivity profile derived from {\tt Cloudy} supports the inclusion of diffuse continuum emission from the BLR as a significant contributor to the observed inter-band continuum time delays, particularly through free-free, free-bound emission and Thomson scattering processes within the line-emitting gas.

    \item Our model successfully reproduced the continuum lag-spectrum of Fairall~9 by incorporating the coupled contributions of the cold accretion disk, warm corona, BLR, and dusty torus within a self-consistent framework. Emission from the warm corona predominantly accounts for the short-wavelength UV-optical lags, diffuse BLR continuum emission reproduces the excess delays around the Balmer jump at 3646 {\AA}, and dust reprocessing in the torus, along with the Paschen jump, explains the enhanced lag observed in the $I$ band. These components are linked through physically motivated transfer functions that describe their temporal response to variations in the central source. Together, they provide a comprehensive and physically grounded explanation of the observed wavelength-dependent continuum time delays.

     \item In addition to reproducing the observed lag-spectrum, our model successfully fitted the broadband SED of Fairall~9. By incorporating all relevant AGN emission and reprocessing components, including the cold accretion disk, warm corona, BLR, and dusty torus together with the contribution from host-galaxy starlight, we were able to recover the observed AGN continuum in a self-consistent manner. The successful simultaneous modeling of both the SED and the continuum time delays provides strong support for the physical completeness of the adopted framework.

    \item To avoid imposing a luminosity fixed by the redshift-based distance, we simultaneously fitted the lag-spectrum and SED of Fairall~9 across a range of Eddington ratios and luminosity distances. For each luminosity distance, the Eddington ratio and all remaining free parameters were optimized, allowing us to determine the overall best-fitting model. The resulting analysis yields a Hubble constant of $H_{0}=72.4_{-3.7}^{+3.4} \, \rm km \, s^{-1} \, Mpc^{-1}$. This estimate is in agreement, with the value previously derived from our analysis of NGC~5548 \citep{jaiswal2025} within the uncertainties, providing independent support for the robustness of the method.

    \item While the current uncertainties in the derived Hubble constant are still relatively large, about 5\% for Fairall~9 and 10\% for NGC~5548, these results highlight the viability of the method. Continued development of the underlying physical models, combined with applications to a larger sample of well-monitored AGNs, could yield significantly more precise and competitive determinations of $H_{0}$.

\end{enumerate}

In summary, this work establishes AGN continuum-lag and SED modeling as a physically grounded and independent route to measuring cosmological distances and the Hubble constant. The successful application of {\tt H0RIZON-AGN} model (code) to Fairall~9, together with our earlier results for NGC~5548, demonstrates both the robustness and scalability of the method. With ongoing HALO monitoring program, the incorporation of additional high-quality AGN datasets, and the unprecedented opportunities that will be provided by the LSST, this approach has the potential to evolve into a precision cosmological probe capable of delivering sub-percent constraints on $H_{0}$ while simultaneously advancing our understanding of AGN structure and physics.

\begin{acknowledgements}
This project has received funding from the European Research Council (ERC) under the European Union’s Horizon 2020 research and innovation program (grant agreement No. [951549]). The Czech-Polish Mobility program of the two Academies of Sciences, titled
“Appearance and dynamics of accretion onto black holes”, is greatly appreciated. VKJ acknowledges the OPUS-LAP/GA\v{C}R-LA bilateral project (2021/43/I/ST9/01352/OPUS
22 and GF23-04053L) funded by National Science Centre, Poland
under the OPUS call in the Weave programme. MHN acknowledges the financial support by the University of Liege under Special Funds for Research, IPD-STEMA Program. SP is supported by the international Gemini Observatory, a program of NSF NOIRLab, which is managed by the Association of Universities for Research in Astronomy (AURA) under a cooperative agreement with the U.S. National Science Foundation, on behalf of the Gemini partnership of Argentina, Brazil, Canada, Chile, the Republic of Korea, and the United States of America. BC and SP acknowledge the support
from COST Action CA21136 - Addressing observational tensions in cosmology with systematics and fundamental physics (CosmoVerse), supported by COST (European Cooperation in Science and Technology). FPN gratefully acknowledges the generous and invaluable support of the Klaus Tschira Foundation. We also acknowledge support from the Polish Ministry of Science and Higher Education grant 2024/WK/02. MZ received the support from the Czech Science Foundation Junior Star grant no. GM24-10599M. MLMA acknowledges financial support from Millennium Nucleus NCN2023\_002 (TITANs) and the China-Chile Joint Research Fund (CCJRF2310). Generative AI tools were used solely to assist in the graphical rendering of the schematic illustration presented in Figure~\ref{fig:model}. The underlying scientific concept, model design, interpretation, and final figure content were conceived, developed, and validated by the authors and constitute entirely original work.
\end{acknowledgements}

\bibliographystyle{aa}
\bibliography{main}

@ARTICLE{2024SSRv..220...14V,
       author = {{Vernardos}, G. and {Sluse}, D. and {Pooley}, D. and {Schmidt}, R.~W. and {Millon}, M. and {Weisenbach}, L. and {Motta}, V. and {Anguita}, T. and {Saha}, P. and {O'Dowd}, M. and {Peel}, A. and {Schechter}, P.~L.},
        title = "{Microlensing of Strongly Lensed Quasars}",
      journal = {\ssr},
         year = 2024,
        month = feb,
       volume = {220},
       number = {1},
          eid = {14},
        pages = {14},
          doi = {10.1007/s11214-024-01043-8},
archivePrefix = {arXiv},
       eprint = {2312.00931},
 primaryClass = {astro-ph.GA},
       adsurl = {https://ui.adsabs.harvard.edu/abs/2024SSRv..220...14V}
}

@ARTICLE{2019MNRAS.483.2275L,
       author = {{Li}, Ya-Ping and {Yuan}, Feng and {Dai}, Xinyu},
        title = "{Reconciling the quasar microlensing disc size problem with a wind model of active galactic nucleus}",
      journal = {\mnras},
         year = 2019,
        month = feb,
       volume = {483},
       number = {2},
        pages = {2275-2281},
          doi = {10.1093/mnras/sty3245},
archivePrefix = {arXiv},
       eprint = {1806.08496},
 primaryClass = {astro-ph.HE},
       adsurl = {https://ui.adsabs.harvard.edu/abs/2019MNRAS.483.2275L}
}

@ARTICLE{2016AN....337..356C,
       author = {{Chartas}, G. and {Rhea}, C. and {Kochanek}, C. and {Dai}, X. and {Morgan}, C. and {Blackburne}, J. and {Chen}, B. and {Mosquera}, A. and {MacLeod}, C.},
        title = "{Gravitational lensing size scales for quasars}",
      journal = {Astronomische Nachrichten},
         year = 2016,
        month = may,
       volume = {337},
       number = {4-5},
        pages = {356},
          doi = {10.1002/asna.201612313},
archivePrefix = {arXiv},
       eprint = {1509.05375},
 primaryClass = {astro-ph.HE},
       adsurl = {https://ui.adsabs.harvard.edu/abs/2016AN....337..356C}
}

@ARTICLE{2011ApJ...729...34B,
       author = {{Blackburne}, Jeffrey A. and {Pooley}, David and {Rappaport}, Saul and {Schechter}, Paul L.},
        title = "{Sizes and Temperature Profiles of Quasar Accretion Disks from Chromatic Microlensing}",
      journal = {\apj},
         year = 2011,
        month = mar,
       volume = {729},
       number = {1},
          eid = {34},
        pages = {34},
          doi = {10.1088/0004-637X/729/1/34},
archivePrefix = {arXiv},
       eprint = {1007.1665},
 primaryClass = {astro-ph.CO},
       adsurl = {https://ui.adsabs.harvard.edu/abs/2011ApJ...729...34B}
}

@ARTICLE{Zhou2022,
       author = {{Zou}, Fan and {Brandt}, W.~N. and {Chen}, Chien-Ting and {Leja}, Joel and {Ni}, Qingling and {Yan}, Wei and {Yang}, Guang and {Zhu}, Shifu and {Luo}, Bin and {Nyland}, Kristina and {Vito}, Fabio and {Xue}, Yongquan},
        title = "{Spectral Energy Distributions in Three Deep-drilling Fields of the Vera C. Rubin Observatory Legacy Survey of Space and Time: Source Classification and Galaxy Properties}",
      journal = {\apjs},
         year = 2022,
        month = sep,
       volume = {262},
       number = {1},
          eid = {15},
        pages = {15},
          doi = {10.3847/1538-4365/ac7bdf},
archivePrefix = {arXiv},
       eprint = {2206.06432},
 primaryClass = {astro-ph.GA},
       adsurl = {https://ui.adsabs.harvard.edu/abs/2022ApJS..262...15Z}
}

@ARTICLE{homayouni2019,
       author = {{Homayouni}, Y. and {Trump}, Jonathan R. and {Grier}, C.~J. and {Shen}, Yue and {Starkey}, D.~A. and {Brandt}, W.~N. and {Fonseca Alvarez}, G. and {Hall}, P.~B. and {Horne}, Keith and {Kinemuchi}, Karen and {I-Hsiu Li}, Jennifer and {McGreer}, Ian D. and {Sun}, Mouyuan and {Ho}, L.~C. and {Schneider}, D.~P.},
        title = "{The Sloan Digital Sky Survey Reverberation Mapping Project: Accretion Disk Sizes from Continuum Lags}",
      journal = {\apj},
         year = 2019,
        month = aug,
       volume = {880},
       number = {2},
          eid = {126},
        pages = {126},
          doi = {10.3847/1538-4357/ab2638},
archivePrefix = {arXiv},
       eprint = {1806.08360},
 primaryClass = {astro-ph.GA},
       adsurl = {https://ui.adsabs.harvard.edu/abs/2019ApJ...880..126H}
}

@ARTICLE{2022ApJS..262...49K,
       author = {{Kova{\v{c}}evi{\'c}}, Andjelka B. and {Radovi{\'c}}, Viktor and {Ili{\'c}}, Dragana and {Popovi{\'c}}, Luka {\v{C}}. and {Assef}, Roberto J. and {S{\'a}nchez-S{\'a}ez}, Paula and {Nikutta}, Robert and {Raiteri}, Claudia M. and {Yoon}, Ilsang and {Homayouni}, Yasaman and {Li}, Yan-Rong and {Caplar}, Neven and {Czerny}, Bozena and {Panda}, Swayamtrupta and {Ricci}, Claudio and {Jankov}, Isidora and {Landt}, Hermine and {Wolf}, Christian and {Kova{\v{c}}evi{\'c}-Doj{\v{c}}inovi{\'c}}, Jelena and {Laki{\'c}evi{\'c}}, Ma{\v{s}}a and {Savi{\'c}}, {\DJ}or{\dj}e V. and {Vince}, Oliver and {Simi{\'c}}, Sa{\v{s}}a and {{\v{C}}vorovi{\'c}-Hajdinjak}, Iva and {Mar{\v{c}}eta-Mandi{\'c}}, Sladjana},
        title = "{The LSST Era of Supermassive Black Hole Accretion Disk Reverberation Mapping}",
      journal = {\apjs},
         year = 2022,
        month = oct,
       volume = {262},
       number = {2},
          eid = {49},
        pages = {49},
          doi = {10.3847/1538-4365/ac88ce},
archivePrefix = {arXiv},
       eprint = {2208.06203},
 primaryClass = {astro-ph.IM},
       adsurl = {https://ui.adsabs.harvard.edu/abs/2022ApJS..262...49K}
}

@ARTICLE{2020ApJ...891...26A,
       author = {{Almeyda}, Triana and {Robinson}, Andrew and {Richmond}, Michael and {Nikutta}, Robert and {McDonough}, Bryanne},
        title = "{Modeling the Infrared Reverberation Response of the Circumnuclear Dusty Torus in AGNs: An Investigation of Torus Response Functions}",
      journal = {\apj},
         year = 2020,
        month = mar,
       volume = {891},
       number = {1},
          eid = {26},
        pages = {26},
          doi = {10.3847/1538-4357/ab6aa1},
archivePrefix = {arXiv},
       eprint = {2002.12823},
 primaryClass = {astro-ph.GA},
       adsurl = {https://ui.adsabs.harvard.edu/abs/2020ApJ...891...26A}
}

@ARTICLE{2017ApJ...843....3A,
       author = {{Almeyda}, Triana and {Robinson}, Andrew and {Richmond}, Michael and {Vazquez}, Billy and {Nikutta}, Robert},
        title = "{Modeling the Infrared Reverberation Response of the Circumnuclear Dusty Torus in AGNs: The Effects of Cloud Orientation and Anisotropic Illumination}",
      journal = {\apj},
         year = 2017,
        month = jul,
       volume = {843},
       number = {1},
          eid = {3},
        pages = {3},
          doi = {10.3847/1538-4357/aa7687},
archivePrefix = {arXiv},
       eprint = {1709.07011},
 primaryClass = {astro-ph.GA},
       adsurl = {https://ui.adsabs.harvard.edu/abs/2017ApJ...843....3A}
}

@ARTICLE{2006ApJ...639...46S,
       author = {{Suganuma}, Masahiro and {Yoshii}, Yuzuru and {Kobayashi}, Yukiyasu and {Minezaki}, Takeo and {Enya}, Keigo and {Tomita}, Hiroyuki and {Aoki}, Tsutomu and {Koshida}, Shintaro and {Peterson}, Bruce A.},
        title = "{Reverberation Measurements of the Inner Radius of the Dust Torus in Nearby Seyfert 1 Galaxies}",
      journal = {\apj},
         year = 2006,
        month = mar,
       volume = {639},
       number = {1},
        pages = {46-63},
          doi = {10.1086/499326},
archivePrefix = {arXiv},
       eprint = {astro-ph/0511697},
 primaryClass = {astro-ph},
       adsurl = {https://ui.adsabs.harvard.edu/abs/2006ApJ...639...46S}
}

@article{Kelly2023Refsdal,
  author  = {Kelly, Patrick L. and Rodney, Steven and Treu, Tommaso and
             Oguri, Masamune and Chen, Wenlei and Zitrin, Adi and
             Birrer, Simon and Bonvin, Vivien and Dessart, Luc and
             Diego, Jose M. and Filippenko, Alexei V. and Foley, Ryan J.
             and others},
  title   = {Constraints on the Hubble Constant from Supernova Refsdal's Reappearance},
  journal = {Science},
  year    = {2023},
  volume  = {380},
  number  = {6649},
  pages   = {eabh1322},
  doi     = {10.1126/science.abh1322}
}

@ARTICLE{wong2020,
       author = {{Wong}, Kenneth C. and {Suyu}, Sherry H. and {Chen}, Geoff C.-F. and {Rusu}, Cristian E. and {Millon}, Martin and {Sluse}, Dominique and {Bonvin}, Vivien and {Fassnacht}, Christopher D. and {Taubenberger}, Stefan and {Auger}, Matthew W. and et al.},
        title = "{H0LiCOW - XIII. A 2.4 per cent measurement of H$_{0}$ from lensed quasars: 5.3{\ensuremath{\sigma}} tension between early- and late-Universe probes}",
      journal = {\mnras},
         year = 2020,
        month = oct,
       volume = {498},
       number = {1},
        pages = {1420-1439},
          doi = {10.1093/mnras/stz3094},
archivePrefix = {arXiv},
       eprint = {1907.04869},
 primaryClass = {astro-ph.CO},
       adsurl = {https://ui.adsabs.harvard.edu/abs/2020MNRAS.498.1420W}
}

@ARTICLE{2011dong,
       author = {{Dong}, Xiao-Bo and {Wang}, Jian-Guo and {Ho}, Luis C. and {Wang}, Ting-Gui and {Fan}, Xiaohui and {Wang}, Huiyuan and {Zhou}, Hongyan and {Yuan}, Weimin},
        title = "{What Controls the Fe II Strength in Active Galactic Nuclei?}",
      journal = {\apj},
         year = 2011,
        month = aug,
       volume = {736},
       number = {2},
          eid = {86},
        pages = {86},
          doi = {10.1088/0004-637X/736/2/86},
archivePrefix = {arXiv},
       eprint = {0903.5020},
 primaryClass = {astro-ph.GA},
       adsurl = {https://ui.adsabs.harvard.edu/abs/2011ApJ...736...86D}
}

@ARTICLE{2001marziani,
       author = {{Marziani}, P. and {Sulentic}, J.~W. and {Zwitter}, T. and {Dultzin-Hacyan}, D. and {Calvani}, M.},
        title = "{Searching for the Physical Drivers of the Eigenvector 1 Correlation Space}",
      journal = {\apj},
         year = 2001,
        month = sep,
       volume = {558},
       number = {2},
        pages = {553-560},
          doi = {10.1086/322286},
archivePrefix = {arXiv},
       eprint = {astro-ph/0105343},
 primaryClass = {astro-ph},
       adsurl = {https://ui.adsabs.harvard.edu/abs/2001ApJ...558..553M}
}

@ARTICLE{bruhweiler2008,
       author = {{Bruhweiler}, F. and {Verner}, E.},
        title = "{Modeling Fe II Emission and Revised Fe II (UV) Empirical Templates for the Seyfert 1 Galaxy I Zw 1}",
      journal = {\apj},
         year = 2008,
        month = mar,
       volume = {675},
       number = {1},
        pages = {83-95},
          doi = {10.1086/525557},
       adsurl = {https://ui.adsabs.harvard.edu/abs/2008ApJ...675...83B}
}

@ARTICLE{2014ApJ...788..159K,
       author = {{Koshida}, Shintaro and {Minezaki}, Takeo and {Yoshii}, Yuzuru and {Kobayashi}, Yukiyasu and {Sakata}, Yu and {Sugawara}, Shota and {Enya}, Keigo and {Suganuma}, Masahiro and {Tomita}, Hiroyuki and {Aoki}, Tsutomu and {Peterson}, Bruce A.},
        title = "{Reverberation Measurements of the Inner Radius of the Dust Torus in 17 Seyfert Galaxies}",
      journal = {\apj},
         year = 2014,
        month = jun,
       volume = {788},
       number = {2},
          eid = {159},
        pages = {159},
          doi = {10.1088/0004-637X/788/2/159},
archivePrefix = {arXiv},
       eprint = {1406.2078},
 primaryClass = {astro-ph.GA},
       adsurl = {https://ui.adsabs.harvard.edu/abs/2014ApJ...788..159K}
}

@ARTICLE{2024ApJ...968...59M,
       author = {{Mandal}, Amit Kumar and {Woo}, Jong-Hak and {Wang}, Shu and {Rakshit}, Suvendu and {Cho}, Hojin and {Son}, Donghoon and {Stalin}, C.~S.},
        title = "{Revisiting the Dust Torus Size{\textendash}Luminosity Relation Based on a Uniform Reverberation-mapping Analysis}",
      journal = {\apj},
         year = 2024,
        month = jun,
       volume = {968},
       number = {2},
          eid = {59},
        pages = {59},
          doi = {10.3847/1538-4357/ad414d},
archivePrefix = {arXiv},
       eprint = {2403.01885},
 primaryClass = {astro-ph.GA},
       adsurl = {https://ui.adsabs.harvard.edu/abs/2024ApJ...968...59M}
}

@ARTICLE{2020MNRAS.491.4615K,
       author = {{Kokubo}, Mitsuru and {Minezaki}, Takeo},
        title = "{Rapid luminosity decline and subsequent reformation of the innermost dust distribution in the changing-look AGN Mrk 590}",
      journal = {\mnras},
         year = 2020,
        month = feb,
       volume = {491},
       number = {4},
        pages = {4615-4633},
          doi = {10.1093/mnras/stz3397},
archivePrefix = {arXiv},
       eprint = {1904.08946},
 primaryClass = {astro-ph.GA},
       adsurl = {https://ui.adsabs.harvard.edu/abs/2020MNRAS.491.4615K}
}

@ARTICLE{2008ApJ...685..160N,
       author = {{Nenkova}, Maia and {Sirocky}, Matthew M. and {Nikutta}, Robert and {Ivezi{\'c}}, {\v{Z}}eljko and {Elitzur}, Moshe},
        title = "{AGN Dusty Tori. II. Observational Implications of Clumpiness}",
      journal = {\apj},
         year = 2008,
        month = sep,
       volume = {685},
       number = {1},
        pages = {160-180},
          doi = {10.1086/590483},
archivePrefix = {arXiv},
       eprint = {0806.0512},
 primaryClass = {astro-ph},
       adsurl = {https://ui.adsabs.harvard.edu/abs/2008ApJ...685..160N}
}

@ARTICLE{2008ApJ...685..147N,
       author = {{Nenkova}, Maia and {Sirocky}, Matthew M. and {Ivezi{\'c}}, {\v{Z}}eljko and {Elitzur}, Moshe},
        title = "{AGN Dusty Tori. I. Handling of Clumpy Media}",
      journal = {\apj},
         year = 2008,
        month = sep,
       volume = {685},
       number = {1},
        pages = {147-159},
          doi = {10.1086/590482},
archivePrefix = {arXiv},
       eprint = {0806.0511},
 primaryClass = {astro-ph},
       adsurl = {https://ui.adsabs.harvard.edu/abs/2008ApJ...685..147N}
}

@ARTICLE{2022MNRAS.516.4898G,
       author = {{Guise}, E. and {H{\"o}nig}, S.~F. and {Gorjian}, V. and {Barth}, A.~J. and {Almeyda}, T. and {Pei}, L. and {Cenko}, S.~B. and {Edelson}, R. and {Filippenko}, A.~V. and {Joner}, M.~D. and {Laney}, C.~D. and {Li}, W. and {Malkan}, M.~A. and {Nguyen}, M.~L. and {Zheng}, W.},
        title = "{Dust reverberation mapping and light-curve modelling of Zw229-015}",
      journal = {\mnras},
         year = 2022,
        month = nov,
       volume = {516},
       number = {4},
        pages = {4898-4915},
          doi = {10.1093/mnras/stac2529},
archivePrefix = {arXiv},
       eprint = {2209.01409},
 primaryClass = {astro-ph.GA},
       adsurl = {https://ui.adsabs.harvard.edu/abs/2022MNRAS.516.4898G}
}

@ARTICLE{halo1,
       author = {{Mandal}, Amit Kumar and {Pozo Nu{\~n}ez}, Francisco and {Jaiswal}, Vikram Kumar and {Naddaf}, Mohammad Hassan and {Czerny}, Bo{\.z}ena and {Panda}, Swayamtrupta and {Karczmarek}, Paulina and {Pietrzy{\'n}ski}, Grzegorz and {Pandey}, Shivangi and {Peterson}, B.~M. and {Zaja{\v{c}}ek}, Michal and {Dov{\v{c}}iak}, Michal and {Karas}, Vladimir and {Narloch}, Weronika and {Kicia}, Miros{\l}aw and {G{\'o}rski}, Marek and {Ka{\l}uszy{\'n}ski}, Miko{\l}aj and {Hajdu}, Gergely and {Wielg{\'o}rski}, Piotr and {Zgirski}, Bart{\l}omiej and {Ga{\l}an}, Cezary and {Pych}, Wojciech and {Smolec}, Rados{\l}aw and {B{\k{a}}kowska}, Karolina and {Gieren}, Wolfgang and {Kervella}, Pierre},
        title = "{HALO: I. Photometric continuum reverberation mapping of Fairall 9}",
      journal = {\aap},
         year = 2026,
        month = feb,
       volume = {706},
          eid = {A176},
        pages = {A176},
          doi = {10.1051/0004-6361/202557795},
archivePrefix = {arXiv},
       eprint = {2512.13296},
 primaryClass = {astro-ph.GA},
       adsurl = {https://ui.adsabs.harvard.edu/abs/2026A&A...706A.176M}
}

@ARTICLE{partington2024,
       author = {{Partington}, Ethan R. and {Cackett}, Edward M. and {Edelson}, Rick and {Horne}, Keith and {Gelbord}, Jonathan and {Kara}, Erin and {Malacaria}, Christian and {Miller}, Jake A. and {Steiner}, James F. and {Sanna}, Andrea},
        title = "{Connecting the X-Ray/UV Variability of Fairall 9 with NICER: A Possible Warm Corona}",
      journal = {\apj},
         year = 2024,
        month = dec,
       volume = {977},
       number = {1},
          eid = {77},
        pages = {77},
          doi = {10.3847/1538-4357/ad8dc2},
archivePrefix = {arXiv},
       eprint = {2410.21432},
 primaryClass = {astro-ph.HE},
       adsurl = {https://ui.adsabs.harvard.edu/abs/2024ApJ...977...77P}
}

@ARTICLE{dovciak2016,
       author = {{Dov{\v{c}}iak}, M. and {Done}, C.},
        title = "{Minimum X-ray source size of the on-axis corona in AGN}",
      journal = {Astronomische Nachrichten},
         year = 2016,
        month = may,
       volume = {337},
       number = {4-5},
        pages = {441},
          doi = {10.1002/asna.201612327},
       adsurl = {https://ui.adsabs.harvard.edu/abs/2016AN....337..441D}
}

@ARTICLE{czerny2003,
       author = {{Czerny}, B. and {Niko{\l}ajuk}, M. and {R{\'o}{\.z}a{\'n}ska}, Agata and {Dumont}, A.-M. and {Loska}, Z. and {Zycki}, P.~T.},
        title = "{Universal spectral shape of high accretion rate AGN}",
      journal = {\aap},
         year = 2003,
        month = dec,
       volume = {412},
        pages = {317-329},
          doi = {10.1051/0004-6361:20031441},
archivePrefix = {arXiv},
       eprint = {astro-ph/0309242},
 primaryClass = {astro-ph},
       adsurl = {https://ui.adsabs.harvard.edu/abs/2003A&A...412..317C}
}

@ARTICLE{palit2024,
       author = {{Palit}, B. and {R{\'o}{\.z}a{\'n}ska}, A. and {Petrucci}, P.~O. and {Gronkiewicz}, D. and {Barnier}, S. and {Bianchi}, S. and {Ballantyne}, D.~R. and {Gianolli}, V.~E. and {Middei}, R. and {Belmont}, R. and {Ursini}, F.},
        title = "{X-ray view of dissipative warm corona in active galactic nuclei}",
      journal = {\aap},
         year = 2024,
        month = oct,
       volume = {690},
          eid = {A308},
        pages = {A308},
          doi = {10.1051/0004-6361/202450111},
archivePrefix = {arXiv},
       eprint = {2406.14378},
 primaryClass = {astro-ph.HE},
       adsurl = {https://ui.adsabs.harvard.edu/abs/2024A&A...690A.308P}
}

@ARTICLE{walton2025,
       author = {{Walton}, D.~J. and {Madathil-Pottayil}, A. and {Kosec}, P. and {Jiang}, J. and {Garcia}, J. and {Fabian}, A.~C. and {Pinto}, C. and {Buisson}, D.~J.~K. and {Parker}, M.~L. and {Alston}, W.~N. and {Reynolds}, C.~S.},
        title = "{The broad-band view of the bare Seyfert PG 1426+015: relativistic reflection, the soft excess, and the importance of oxygen}",
      journal = {\mnras},
         year = 2025,
        month = nov,
       volume = {543},
       number = {3},
        pages = {2633-2648},
          doi = {10.1093/mnras/staf1545},
archivePrefix = {arXiv},
       eprint = {2509.13411},
 primaryClass = {astro-ph.HE},
       adsurl = {https://ui.adsabs.harvard.edu/abs/2025MNRAS.543.2633W}
}

@ARTICLE{desmond2025,
       author = {{Desmond}, Harry and {Stiskalek}, Richard and {Najera}, Jose Antonio and {Banik}, Indranil},
        title = "{The subtle statistics of the distance ladder: On the distance prior and selection effects}",
      journal = {arXiv e-prints},
         year = 2025,
        month = nov,
          eid = {arXiv:2511.03394},
        pages = {arXiv:2511.03394},
          doi = {10.48550/arXiv.2511.03394},
archivePrefix = {arXiv},
       eprint = {2511.03394},
 primaryClass = {astro-ph.CO},
       adsurl = {https://ui.adsabs.harvard.edu/abs/2025arXiv251103394D}
}

@ARTICLE{ballantyne2024,
       author = {{Ballantyne}, D.~R. and {Sudhakar}, V. and {Fairfax}, D. and {Bianchi}, S. and {Czerny}, B. and {De Rosa}, A. and {De Marco}, B. and {Middei}, R. and {Palit}, B. and {Petrucci}, P.-O. and {R{\'o}{\.z}a{\'n}ska}, A. and {Ursini}, F.},
        title = "{Unveiling energy pathways in AGN accretion flows with the warm corona model for the soft excess}",
      journal = {\mnras},
         year = 2024,
        month = may,
       volume = {530},
       number = {2},
        pages = {1603-1623},
          doi = {10.1093/mnras/stae944},
archivePrefix = {arXiv},
       eprint = {2404.03040},
 primaryClass = {astro-ph.HE},
       adsurl = {https://ui.adsabs.harvard.edu/abs/2024MNRAS.530.1603B}
}

@ARTICLE{wagner2025,
       author = {{Wagner}, Jenny and {Benisty}, David and {Karachentsev}, Igor D.},
        title = "{The Binary Ballet: Mapping Local Expansion Around M81 \& M82}",
      journal = {arXiv e-prints},
         year = 2025,
        month = oct,
          eid = {arXiv:2510.24840},
        pages = {arXiv:2510.24840},
          doi = {10.48550/arXiv.2510.24840},
archivePrefix = {arXiv},
       eprint = {2510.24840},
 primaryClass = {astro-ph.CO},
       adsurl = {https://ui.adsabs.harvard.edu/abs/2025arXiv251024840W}
}

@ARTICLE{son2025,
       author = {{Son}, Junhyuk and {Lee}, Young-Wook and {Chung}, Chul and {Park}, Seunghyun and {Cho}, Hyejeon},
        title = "{Strong progenitor age bias in supernova cosmology ─ II. Alignment with DESI BAO and signs of a non-accelerating universe}",
      journal = {\mnras},
         year = 2025,
        month = nov,
       volume = {544},
       number = {1},
        pages = {975-987},
          doi = {10.1093/mnras/staf1685},
archivePrefix = {arXiv},
       eprint = {2510.13121},
 primaryClass = {astro-ph.CO},
       adsurl = {https://ui.adsabs.harvard.edu/abs/2025MNRAS.544..975S}
}

@ARTICLE{HoDN2025,
       author = {{H0DN Collaboration} and {Casertano}, Stefano and {Anand}, Gagandeep and {Anderson}, Richard I. and {Beaton}, Rachael and {Bhardwaj}, Anupam and {Blakeslee}, John P. and {Boubel}, Paula and {Breuval}, Louise and {Brout}, Dillon and {Cantiello}, Michele and {Cruz Reyes}, Mauricio and {Cs{\"o}rnyei}, Geza and {de Jaeger}, Thomas and {Dhawan}, Suhail and {Di Valentino}, Eleonora and {Galbany}, Llu{\'\i}s and {Gil-Mar{\'\i}n}, H{\'e}ctor and {Graczyk}, Dariusz and {Huang}, Caroline and {Jensen}, Joseph B. and {Kervella}, Pierre and {Leibundgut}, Bruno and {Lengen}, Bastian and {Li}, Siyang and {Macri}, Lucas and {{\"O}z{\"u}lker}, Emre and {Pesce}, Dominic W. and {Riess}, Adam and {Romaniello}, Martino and {Said}, Khaled and {Sch{\"o}neberg}, Nils and {Scolnic}, Dan and {Sicignano}, Teresa and {Skowron}, Dorota M. and {Uddin}, Syed A. and {Verde}, Licia and {Nota}, Antonella},
        title = "{The Local Distance Network: a community consensus report on the measurement of the Hubble constant at 1\% precision}",
      journal = {arXiv e-prints},
         year = 2025,
        month = oct,
          eid = {arXiv:2510.23823},
        pages = {arXiv:2510.23823},
          doi = {10.48550/arXiv.2510.23823},
archivePrefix = {arXiv},
       eprint = {2510.23823},
 primaryClass = {astro-ph.CO},
       adsurl = {https://ui.adsabs.harvard.edu/abs/2025arXiv251023823H}
}

@ARTICLE{liu2020,
       author = {{Liu}, Honghui and {Wang}, Haiyang and {Abdikamalov}, Askar B. and {Ayzenberg}, Dimitry and {Bambi}, Cosimo},
        title = "{Reflection Features in the X-Ray Spectrum of Fairall 9 and Implications for Tests of General Relativity}",
      journal = {\apj},
         year = 2020,
        month = jun,
       volume = {896},
       number = {2},
          eid = {160},
        pages = {160},
          doi = {10.3847/1538-4357/ab917a},
archivePrefix = {arXiv},
       eprint = {2004.11542},
 primaryClass = {gr-qc},
       adsurl = {https://ui.adsabs.harvard.edu/abs/2020ApJ...896..160L}
}

@ARTICLE{Emmanoulopoulos2011,
       author = {{Emmanoulopoulos}, D. and {Papadakis}, I.~E. and {McHardy}, I.~M. and {Nicastro}, F. and {Bianchi}, S. and {Ar{\'e}valo}, P.},
        title = "{An XMM-Newton view of the `bare' nucleus of Fairall 9}",
      journal = {\mnras},
         year = 2011,
        month = aug,
       volume = {415},
       number = {2},
        pages = {1895-1906},
          doi = {10.1111/j.1365-2966.2011.18834.x},
archivePrefix = {arXiv},
       eprint = {1104.0044},
 primaryClass = {astro-ph.CO},
       adsurl = {https://ui.adsabs.harvard.edu/abs/2011MNRAS.415.1895E}
}

@ARTICLE{2023ApJ...947...62K,
       author = {{Kara}, Erin and {Barth}, Aaron J. and {Cackett}, Edward M. and {Gelbord}, Jonathan and {Montano}, John and {Li}, Yan-Rong and {Santana}, Lisabeth and {Horne}, Keith and {Alston}, William N. and {Buisson}, Douglas and {Chelouche}, Doron and {Du}, Pu and {Fabian}, Andrew C. and {Fian}, Carina and {Gallo}, Luigi and {Goad}, Michael R. and {Grupe}, Dirk and {Gonz{\'a}lez Buitrago}, Diego H. and {Hern{\'a}ndez Santisteban}, Juan V. and {Kaspi}, Shai and {Hu}, Chen and {Komossa}, S. and {Kriss}, Gerard A. and {Lewin}, Collin and {Lewis}, Tiffany and {Loewenstein}, Michael and {Lohfink}, Anne and {Masterson}, Megan and {McHardy}, Ian M. and {Mehdipour}, Missagh and {Miller}, Jake and {Panagiotou}, Christos and {Parker}, Michael L. and {Pinto}, Ciro and {Remillard}, Ron and {Reynolds}, Christopher and {Rogantini}, Daniele and {Wang}, Jian-Min and {Wang}, Jingyi and {Wilkins}, Dan},
        title = "{UV-Optical Disk Reverberation Lags despite a Faint X-Ray Corona in the Active Galactic Nucleus Mrk 335}",
      journal = {\apj},
         year = 2023,
        month = apr,
       volume = {947},
       number = {2},
          eid = {62},
        pages = {62},
          doi = {10.3847/1538-4357/acbcd3},
archivePrefix = {arXiv},
       eprint = {2302.07342},
 primaryClass = {astro-ph.HE},
       adsurl = {https://ui.adsabs.harvard.edu/abs/2023ApJ...947...62K}
}

@ARTICLE{2026ApJ..1003..147M,
       author = {{Montano}, John W. and {Barth}, Aaron J. and {Horne}, Keith and {Cackett}, Edward M. and {De Rosa}, Gisella and {Homayouni}, Y. and {Kara}, Erin A. and {Kriss}, Gerard A. and {Landt}, Hermine and {Apolonio}, Gilvan G. and {Arav}, Nahum and {Boizelle}, Benjamin D. and {Dalla Bont{\`a}}, Elena and {Chelouche}, Doron and {Dehghanian}, Maryam and {Edelson}, Rick and {Ferland}, Gary J. and {Fian}, Carina and {Sobrino Figaredo}, Catalina and {Goad}, Michael R. and {Gonzalez-Buitrago}, Diego H. and {Guo}, Wei-Jian and {Hu}, Chen and {Ili{\'c}}, Dragana and {Joner}, Michael D. and {Kaspi}, Shai and {Kochanek}, Christopher S. and {Kova{\v{c}}evi{\'c}}, Andjelka B. and {Lewin}, Collin and {Li}, Sha-Sha and {Li}, Yan-Rong and {Liu}, Jun-Rong and {Miller}, Jake A. and {Neustadt}, Jack M.~M. and {Netzer}, Hagai and {Ochner}, Paolo and {Partington}, Ethan R. and {Pizzella}, Alessandro and {Plesha}, Rachel and {Popovi{\'c}}, Luka {\v{C}}. and {Sanmartim}, David and {Hern{\'a}ndez Santisteban}, Juan V. and {Vestergaard}, Marianne and {Wooley}, Jack H.~F. and {Yang}, Sen and {Yao}, Zhu-Heng and {Zaidouni}, Fatima},
        title = "{AGN STORM 2. XII. Ground-based Optical Photometry and Lag Measurements of Mrk 817}",
      journal = {\apj},
         year = 2026,
        month = jun,
       volume = {1003},
       number = {2},
          eid = {147},
        pages = {147},
          doi = {10.3847/1538-4357/ae6103},
archivePrefix = {arXiv},
       eprint = {2605.02875},
 primaryClass = {astro-ph.GA},
       adsurl = {https://ui.adsabs.harvard.edu/abs/2026ApJ..1003..147M}
}

@ARTICLE{2025A&A...700L...8P,
       author = {{Pozo Nu{\~n}ez}, F. and {Ba{\~n}ados}, E. and {Panda}, S. and {Heidt}, J.},
        title = "{Accretion disc reverberation mapping in a high-redshift quasar}",
      journal = {\aap},
         year = 2025,
        month = aug,
       volume = {700},
          eid = {L8},
        pages = {L8},
          doi = {10.1051/0004-6361/202555421},
archivePrefix = {arXiv},
       eprint = {2508.00209},
 primaryClass = {astro-ph.GA},
       adsurl = {https://ui.adsabs.harvard.edu/abs/2025A&A...700L...8P}
}

@ARTICLE{2026MNRAS.548ag642M,
       author = {{Marculewicz}, Marcin and {Hern{\'a}ndez Santisteban}, Juan V. and {Horne}, Keith and {Cackett}, Edward M. and {Landt}, Hermine and {Gelbord}, Jonathan and {Winkler}, Hartmut and {Vestergaard}, Marianne and {Barth}, Aaron J. and {Goad}, Michael and {Kaspi}, Shai and {Lira}, Paulina and {Onken}, Christopher A. and {Gonz{\'a}lez-Buitrago}, Diego H. and {Valenti}, Stefano},
        title = "{The unusually red delay spectrum of the low-mass black hole AGN NGC 4051 as revealed by intensive continuum reverberation mapping with the Las Cumbres Observatory}",
      journal = {\mnras},
         year = 2026,
        month = may,
       volume = {548},
       number = {3},
          eid = {stag642},
        pages = {stag642},
          doi = {10.1093/mnras/stag642},
archivePrefix = {arXiv},
       eprint = {2604.01405},
 primaryClass = {astro-ph.HE},
       adsurl = {https://ui.adsabs.harvard.edu/abs/2026MNRAS.548ag642M}
}

@ARTICLE{2023ApJ...953..137M,
       author = {{Miller}, Jake A. and {Cackett}, Edward M. and {Goad}, Michael R. and {Horne}, Keith and {Barth}, Aaron J. and {Romero-Colmenero}, Encarni and {Fausnaugh}, Michael and {Gelbord}, Jonathan and {Korista}, Kirk T. and {Landt}, Hermine and {Treu}, Tommaso and {Winkler}, Hartmut},
        title = "{Continuum Reverberation Mapping of Mrk 876 over Three Years with Remote Robotic Observatories}",
      journal = {\apj},
         year = 2023,
        month = aug,
       volume = {953},
       number = {2},
          eid = {137},
        pages = {137},
          doi = {10.3847/1538-4357/ace342},
archivePrefix = {arXiv},
       eprint = {2307.02630},
 primaryClass = {astro-ph.GA},
       adsurl = {https://ui.adsabs.harvard.edu/abs/2023ApJ...953..137M}
}

@ARTICLE{2023MNRAS.523..545D,
       author = {{Donnan}, Fergus R. and {Hern{\'a}ndez Santisteban}, Juan V. and {Horne}, Keith and {Hu}, Chen and {Du}, Pu and {Li}, Yan-Rong and {Xiao}, Ming and {Ho}, Luis C. and {Aceituno}, Jes{\'u}s and {Wang}, Jian-Min and {Guo}, Wei-Jian and {Yang}, Sen and {Jiang}, Bo-Wei and {Yao}, Zhu-Heng},
        title = "{Testing super-eddington accretion on to a supermassive black hole: reverberation mapping of PG 1119+120}",
      journal = {\mnras},
         year = 2023,
        month = jul,
       volume = {523},
       number = {1},
        pages = {545-567},
          doi = {10.1093/mnras/stad1409},
archivePrefix = {arXiv},
       eprint = {2302.09370},
 primaryClass = {astro-ph.GA},
       adsurl = {https://ui.adsabs.harvard.edu/abs/2023MNRAS.523..545D}
}

@ARTICLE{2023MNRAS.525.4524G,
       author = {{Gonz{\'a}lez-Buitrago}, D.~H. and {Garc{\'\i}a-D{\'\i}az}, Ma T. and {Pozo Nu{\~n}ez}, F. and {Guo}, Hengxiao},
        title = "{On the nature of the continuum reverberation of X-ray/UV and optical emission of IRAS 09149-6206}",
      journal = {\mnras},
         year = 2023,
        month = nov,
       volume = {525},
       number = {3},
        pages = {4524-4539},
          doi = {10.1093/mnras/stad2483},
archivePrefix = {arXiv},
       eprint = {2308.05433},
 primaryClass = {astro-ph.HE},
       adsurl = {https://ui.adsabs.harvard.edu/abs/2023MNRAS.525.4524G}
}

@ARTICLE{1997ApJS..112..271S,
       author = {{Santos-Lle{\'o}}, M. and {Chatzichristou}, E. and {de Oliveira}, C. Mendes and {Winge}, C. and {Alloin}, D. and {Peterson}, B.~M. and {Rodr{\'\i}guez-Pascual}, P.~M. and {Stirpe}, G.~M. and {Beers}, T. and {Bragaglia}, A. and {Claeskens}, J. -F. and {Federspiel}, M. and {Giannuzzo}, E. and {Gregorio-Hetem}, J. and {Mathys}, G. and {Salamanca}, I. and {Stein}, P. and {Stenholm}, B. and {Wilhelm}, R. and {Zanin}, C. and {Albrecht}, P. and {Calder{\'o}n}, J. and {Caretta}, C.~A. and {Carranza}, G. and {da Costa}, R.~D.~D. and {Diaz}, R. and {Dietrich}, M. and {Dottori}, H. and {Elizalde}, F. and {Goldes}, G. and {Ghosh}, K.~K. and {Maia}, M.~A.~G. and {Paolantonio}, S. and {de Oliveira Filho}, I. Rodrigues and {Rodriguez-Ardila}, A. and {Schmitt}, H.~R. and {Soundararajaperumal}, S. and {de Souza}, R.~E. and {Willmer}, C.~N.~A. and {Zheng}, W.},
        title = "{Steps toward Determination of the Size and Structure of the Broad-Line Region in Active Galactic Nuclei. X. Variability of Fairall 9 from Optical Data}",
      journal = {\apjs},
         year = 1997,
        month = oct,
       volume = {112},
       number = {2},
        pages = {271-283},
          doi = {10.1086/313046},
       adsurl = {https://ui.adsabs.harvard.edu/abs/1997ApJS..112..271S}
}

@ARTICLE{1997ApJS..110....9R,
       author = {{Rodr{\'\i}guez-Pascual}, P.~M. and {Alloin}, D. and {Clavel}, J. and {Crenshaw}, D.~M. and {Horne}, K. and {Kriss}, G.~A. and {Krolik}, J.~H. and {Malkan}, M.~A. and {Netzer}, H. and {O'Brien}, P.~T. and {Peterson}, B.~M. and {Reichert}, G.~A. and {Wamsteker}, W. and {Alexander}, T. and {Barr}, P. and {Blandford}, R.~D. and {Bregman}, J.~N. and {Carone}, T.~E. and {Clements}, S. and {Courvoisier}, T. -J. and {De Robertis}, M.~M. and {Dietrich}, M. and {Dottori}, H. and {Edelson}, R.~A. and {Filippenko}, A.~V. and {Gaskell}, C.~M. and {Huchra}, J.~P. and {Hutchings}, J.~B. and {Kollatschny}, W. and {Koratkar}, A.~P. and {Korista}, K.~T. and {Laor}, A. and {MacAlpine}, G.~M. and {Martin}, P.~G. and {Maoz}, D. and {McCollum}, B. and {Morris}, S.~L. and {Perola}, G.~C. and {Pogge}, R.~W. and {Ptak}, R.~L. and {Recondo-Gonz{\'a}lez}, M.~C. and {Rodr{\'\i}guez-Espinoza}, J.~M. and {Rokaki}, E.~L. and {Santos-Lle{\'o}}, M. and {Sekiguchi}, K. and {Shull}, J.~M. and {Snijders}, M.~A.~J. and {Sparke}, L.~S. and {Stirpe}, G.~M. and {Stoner}, R.~E. and {Sun}, W. -H. and {Wagner}, S.~J. and {Wanders}, I. and {Wilkes}, J. and {Winge}, C. and {Zheng}, W.},
        title = "{Steps toward Determination of the Size and Structure of the Broad-Line Region in Active Galactic Nuclei. IX. Ultraviolet Observations of Fairall 9}",
      journal = {\apjs},
         year = 1997,
        month = may,
       volume = {110},
       number = {1},
        pages = {9-20},
          doi = {10.1086/312996},
       adsurl = {https://ui.adsabs.harvard.edu/abs/1997ApJS..110....9R}
}

@ARTICLE{2025ApJ...985...30M,
       author = {{Mandal}, Amit Kumar and {Woo}, Jong-Hak and {Wang}, Shu},
        title = "{The Size of the Continuum Emission Region and Its Scaling Relations with Active Galactic Nucleus Luminosity and the Broad-line Region Size}",
      journal = {\apj},
         year = 2025,
        month = may,
       volume = {985},
       number = {1},
          eid = {30},
        pages = {30},
          doi = {10.3847/1538-4357/adc56e},
archivePrefix = {arXiv},
       eprint = {2502.19184},
 primaryClass = {astro-ph.GA},
       adsurl = {https://ui.adsabs.harvard.edu/abs/2025ApJ...985...30M}
}

@ARTICLE{2019NatAs...3..251C,
       author = {{Chelouche}, Doron and {Pozo Nu{\~n}ez}, Francisco and {Kaspi}, Shai},
        title = "{Direct evidence of non-disk optical continuum emission around an active black hole}",
      journal = {Nature Astronomy},
         year = 2019,
        month = jan,
       volume = {3},
        pages = {251-257},
          doi = {10.1038/s41550-018-0659-x},
       adsurl = {https://ui.adsabs.harvard.edu/abs/2019NatAs...3..251C}
}

@ARTICLE{2016ApJ...821...56F,
       author = {{Fausnaugh}, M.~M. and {Denney}, K.~D. and {Barth}, A.~J. and {Bentz}, M.~C. and {Bottorff}, M.~C. and {Carini}, M.~T. and {Croxall}, K.~V. and {De Rosa}, G. and {Goad}, M.~R. and {Horne}, Keith and {Joner}, M.~D. and {Kaspi}, S. and {Kim}, M. and {Klimanov}, S.~A. and {Kochanek}, C.~S. and {Leonard}, D.~C. and {Netzer}, H. and {Peterson}, B.~M. and {Schn{\"u}lle}, K. and {Sergeev}, S.~G. and {Vestergaard}, M. and {Zheng}, W. -K. and {Zu}, Y. and {Anderson}, M.~D. and {Ar{\'e}valo}, P. and {Bazhaw}, C. and {Borman}, G.~A. and {Boroson}, T.~A. and {Brandt}, W.~N. and {Breeveld}, A.~A. and {Brewer}, B.~J. and {Cackett}, E.~M. and {Crenshaw}, D.~M. and {Dalla Bont{\`a}}, E. and {De Lorenzo-C{\'a}ceres}, A. and {Dietrich}, M. and {Edelson}, R. and {Efimova}, N.~V. and {Ely}, J. and {Evans}, P.~A. and {Filippenko}, A.~V. and {Flatland}, K. and {Gehrels}, N. and {Geier}, S. and {Gelbord}, J.~M. and {Gonzalez}, L. and {Gorjian}, V. and {Grier}, C.~J. and {Grupe}, D. and {Hall}, P.~B. and {Hicks}, S. and {Horenstein}, D. and {Hutchison}, T. and {Im}, M. and {Jensen}, J.~J. and {Jones}, J. and {Kaastra}, J. and {Kelly}, B.~C. and {Kennea}, J.~A. and {Kim}, S.~C. and {Korista}, K.~T. and {Kriss}, G.~A. and {Lee}, J.~C. and {Lira}, P. and {MacInnis}, F. and {Manne-Nicholas}, E.~R. and {Mathur}, S. and {McHardy}, I.~M. and {Montouri}, C. and {Musso}, R. and {Nazarov}, S.~V. and {Norris}, R.~P. and {Nousek}, J.~A. and {Okhmat}, D.~N. and {Pancoast}, A. and {Papadakis}, I. and {Parks}, J.~R. and {Pei}, L. and {Pogge}, R.~W. and {Pott}, J. -U. and {Rafter}, S.~E. and {Rix}, H. -W. and {Saylor}, D.~A. and {Schimoia}, J.~S. and {Siegel}, M. and {Spencer}, M. and {Starkey}, D. and {Sung}, H. -I. and {Teems}, K.~G. and {Treu}, T. and {Turner}, C.~S. and {Uttley}, P. and {Villforth}, C. and {Weiss}, Y. and {Woo}, J. -H. and {Yan}, H. and {Young}, S.},
        title = "{Space Telescope and Optical Reverberation Mapping Project. III. Optical Continuum Emission and Broadband Time Delays in NGC 5548}",
      journal = {\apj},
         year = 2016,
        month = apr,
       volume = {821},
       number = {1},
          eid = {56},
        pages = {56},
          doi = {10.3847/0004-637X/821/1/56},
archivePrefix = {arXiv},
       eprint = {1510.05648},
 primaryClass = {astro-ph.GA},
       adsurl = {https://ui.adsabs.harvard.edu/abs/2016ApJ...821...56F}
}

@ARTICLE{2017ApJ...835..226K,
       author = {{King}, Ashley L. and {Lohfink}, Anne and {Kara}, Erin},
        title = "{AGN Coronae through a Jet Perspective}",
      journal = {\apj},
         year = 2017,
        month = feb,
       volume = {835},
       number = {2},
          eid = {226},
        pages = {226},
          doi = {10.3847/1538-4357/835/2/226},
archivePrefix = {arXiv},
       eprint = {1612.07784},
 primaryClass = {astro-ph.HE},
       adsurl = {https://ui.adsabs.harvard.edu/abs/2017ApJ...835..226K}
}

@ARTICLE{marziani2010,
       author = {{Marziani}, P. and {Sulentic}, J.~W. and {Negrete}, C.~A. and {Dultzin}, D. and {Zamfir}, S. and {Bachev}, R.},
        title = "{Broad-line region physical conditions along the quasar eigenvector 1 sequence}",
      journal = {\mnras},
         year = 2010,
        month = dec,
       volume = {409},
       number = {3},
        pages = {1033-1048},
          doi = {10.1111/j.1365-2966.2010.17357.x},
archivePrefix = {arXiv},
       eprint = {1007.3187},
 primaryClass = {astro-ph.CO},
       adsurl = {https://ui.adsabs.harvard.edu/abs/2010MNRAS.409.1033M}
}

@ARTICLE{2020A&A...640A..39C,
       author = {{Campitiello}, Samuele and {Celotti}, Annalisa and {Ghisellini}, Gabriele and {Sbarrato}, Tullia},
        title = "{Estimating black hole masses: Accretion disk fitting versus reverberation mapping and single epoch}",
      journal = {\aap},
         year = 2020,
        month = aug,
       volume = {640},
          eid = {A39},
        pages = {A39},
          doi = {10.1051/0004-6361/201936218},
archivePrefix = {arXiv},
       eprint = {1907.00986},
 primaryClass = {astro-ph.HE},
       adsurl = {https://ui.adsabs.harvard.edu/abs/2020A&A...640A..39C}
}

@ARTICLE{edelson2024,
       author = {{Edelson}, R. and {Peterson}, B.~M. and {Gelbord}, J. and {Horne}, K. and {Goad}, M. and {McHardy}, I. and {Vaughan}, S. and {Vestergaard}, M.},
        title = "{Intensive Broadband Reverberation Mapping of Fairall 9 with 1.8 yr of Daily Swift Monitoring}",
      journal = {\apj},
         year = 2024,
        month = oct,
       volume = {973},
       number = {2},
          eid = {152},
        pages = {152},
          doi = {10.3847/1538-4357/ad64d4},
archivePrefix = {arXiv},
       eprint = {2407.09445},
 primaryClass = {astro-ph.HE},
       adsurl = {https://ui.adsabs.harvard.edu/abs/2024ApJ...973..152E}
}

@ARTICLE{kammoun_data2021,
       author = {{Kammoun}, E.~S. and {Papadakis}, I.~E. and {Dov{\v{c}}iak}, M.},
        title = "{Modelling the UV/optical continuum time-lags in AGN}",
      journal = {\mnras},
         year = 2021,
        month = may,
       volume = {503},
       number = {3},
        pages = {4163-4171},
          doi = {10.1093/mnras/stab725},
archivePrefix = {arXiv},
       eprint = {2103.04892},
 primaryClass = {astro-ph.HE},
       adsurl = {https://ui.adsabs.harvard.edu/abs/2021MNRAS.503.4163K}
}

@ARTICLE{rozanska2015,
       author = {{R{\'o}{\.z}a{\'n}ska}, A. and {Malzac}, J. and {Belmont}, R. and {Czerny}, B. and {Petrucci}, P. -O.},
        title = "{Warm and optically thick dissipative coronae above accretion disks}",
      journal = {\aap},
         year = 2015,
        month = aug,
       volume = {580},
          eid = {A77},
        pages = {A77},
          doi = {10.1051/0004-6361/201526288},
archivePrefix = {arXiv},
       eprint = {1504.03160},
 primaryClass = {astro-ph.GA},
       adsurl = {https://ui.adsabs.harvard.edu/abs/2015A&A...580A..77R}
}

@ARTICLE{2023A&A...678A.189P,
       author = {{Prince}, Raj and {Zaja{\v{c}}ek}, Michal and {Panda}, Swayamtrupta and {Hryniewicz}, Krzysztof and {Kumar Jaiswal}, Vikram and {Czerny}, Bo{\.z}ena and {Trzcionkowski}, Piotr and {Bronikowski}, Mateusz and {Ra{\l}owski}, Mateusz and {Sobrino Figaredo}, Catalina and {Martinez-Aldama}, Mary Loli and {{\'S}niegowska}, Marzena and {{\'S}redzi{\'n}ska}, Justyna and {Bilicki}, Maciej and {Naddaf}, Mohammad-Hassan and {Pandey}, Ashwani and {Haas}, Martin and {Sarna}, Marek Jacek and {Pietrzy{\'n}ski}, Grzegorz and {Karas}, Vladimir and {Olejak}, Aleksandra and {Przy{\l}uski}, Robert and {Sefako}, Ramotholo R. and {Genade}, Anja and {Worters}, Hannah L. and {Koz{\l}owski}, Szymon and {Udalski}, Andrzej},
        title = "{Wavelength-resolved reverberation mapping of intermediate-redshift quasars HE 0413-4031 and HE 0435-4312: Dissecting Mg II, optical Fe II, and UV Fe II emission regions}",
      journal = {\aap},
         year = 2023,
        month = oct,
       volume = {678},
          eid = {A189},
        pages = {A189},
          doi = {10.1051/0004-6361/202346738},
archivePrefix = {arXiv},
       eprint = {2304.13763},
 primaryClass = {astro-ph.GA},
       adsurl = {https://ui.adsabs.harvard.edu/abs/2023A&A...678A.189P}
}

@ARTICLE{2024A&A...683A.140Z,
       author = {{Zaja{\v{c}}ek}, Michal and {Panda}, Swayamtrupta and {Pandey}, Ashwani and {Prince}, Raj and {Rodr{\'\i}guez-Ardila}, Alberto and {Jaiswal}, Vikram and {Czerny}, Bo{\.z}ena and {Hryniewicz}, Krzysztof and {Urbanowicz}, Maciej and {Trzcionkowski}, Piotr and {{\'S}niegowska}, Marzena and {Fa{\l}kowska}, Zuzanna and {Mart{\'\i}nez-Aldama}, Mary Loli and {Werner}, Norbert},
        title = "{UV FeII emission model of HE 0413‒4031 and its relation to broad-line time delays}",
      journal = {\aap},
         year = 2024,
        month = mar,
       volume = {683},
          eid = {A140},
        pages = {A140},
          doi = {10.1051/0004-6361/202348172},
archivePrefix = {arXiv},
       eprint = {2310.03544},
 primaryClass = {astro-ph.GA},
       adsurl = {https://ui.adsabs.harvard.edu/abs/2024A&A...683A.140Z}
}

@ARTICLE{Cacket_Sci2021,
       author = {{Cackett}, Edward M. and {Bentz}, Misty C. and {Kara}, Erin},
        title = "{Reverberation mapping of active galactic nuclei: from X-ray corona to dusty torus}",
      journal = {iScience},
         year = 2021,
        month = jun,
       volume = {24},
       number = {6},
        pages = {102557},
          doi = {10.1016/j.isci.2021.102557},
archivePrefix = {arXiv},
       eprint = {2105.06926},
 primaryClass = {astro-ph.GA},
       adsurl = {https://ui.adsabs.harvard.edu/abs/2021iSci...24j2557C}
}

@ARTICLE{lawther2018,
       author = {{Lawther}, D. and {Goad}, M.~R. and {Korista}, K.~T. and {Ulrich}, O. and {Vestergaard}, M.},
        title = "{Quantifying the diffuse continuum contribution of BLR Clouds to AGN Continuum Inter-band Delays}",
      journal = {\mnras},
         year = 2018,
        month = nov,
       volume = {481},
       number = {1},
        pages = {533-554},
          doi = {10.1093/mnras/sty2242},
archivePrefix = {arXiv},
       eprint = {1808.04798},
 primaryClass = {astro-ph.GA},
       adsurl = {https://ui.adsabs.harvard.edu/abs/2018MNRAS.481..533L}
}

@ARTICLE{korista2019,
       author = {{Korista}, K.~T. and {Goad}, M.~R.},
        title = "{Quantifying the impact of variable BLR diffuse continuum contributions on measured continuum interband delays}",
      journal = {\mnras},
         year = 2019,
        month = nov,
       volume = {489},
       number = {4},
        pages = {5284-5300},
          doi = {10.1093/mnras/stz2330},
archivePrefix = {arXiv},
       eprint = {1908.07757},
 primaryClass = {astro-ph.GA},
       adsurl = {https://ui.adsabs.harvard.edu/abs/2019MNRAS.489.5284K}
}

@ARTICLE{SS1973,
       author = {{Shakura}, N.~I. and {Sunyaev}, R.~A.},
        title = "{Reprint of 1973A\&A....24..337S. Black holes in binary systems. Observational appearance.}",
      journal = {\aap},
         year = 1973,
        month = jun,
       volume = {500},
        pages = {33-51},
       adsurl = {https://ui.adsabs.harvard.edu/abs/1973A&A....24..337S}
}

@ARTICLE{lobban2020,
       author = {{Lobban}, A.~P. and {Zola}, S. and {Pajdosz-{\'S}mierciak}, U. and {Braito}, V. and {Nardini}, E. and {Bhatta}, G. and {Markowitz}, A. and {Bachev}, R. and {Carosati}, D. and {Caton}, D.~B. and {Damljanovic}, G. and {D{\k{e}}bski}, B. and {Haislip}, J.~B. and {Hu}, S.~M. and {Kouprianov}, V. and {Krzesi{\'n}ski}, J. and {Porquet}, D. and {Pozo Nu{\~n}ez}, F. and {Reeves}, J. and {Reichart}, D.~E.},
        title = "{X-ray, UV, and optical time delays in the bright Seyfert galaxy Ark 120 with co-ordinated Swift and ground-based observations}",
      journal = {\mnras},
         year = 2020,
        month = may,
       volume = {494},
       number = {1},
        pages = {1165-1179},
          doi = {10.1093/mnras/staa630},
archivePrefix = {arXiv},
       eprint = {2002.12348},
 primaryClass = {astro-ph.HE},
       adsurl = {https://ui.adsabs.harvard.edu/abs/2020MNRAS.494.1165L}
}

@ARTICLE{2021ApJ...922..151K,
       author = {{Kara}, Erin and {Mehdipour}, Missagh and {Kriss}, Gerard A. and {Cackett}, Edward M. and {Arav}, Nahum and {Barth}, Aaron J. and {Byun}, Doyee and {Brotherton}, Michael S. and {De Rosa}, Gisella and {Gelbord}, Jonathan and {Hern{\'a}ndez Santisteban}, Juan V. and {Hu}, Chen and {Kaastra}, Jelle and {Landt}, Hermine and {Li}, Yan-Rong and {Miller}, Jake A. and {Montano}, John and {Partington}, Ethan and {Aceituno}, Jes{\'u}s and {Bai}, Jin-Ming and {Bao}, Dongwei and {Bentz}, Misty C. and {Brink}, Thomas G. and {Chelouche}, Doron and {Chen}, Yong-Jie and {Colmenero}, Encarni Romero and {Dalla Bont{\`a}}, Elena and {Dehghanian}, Maryam and {Du}, Pu and {Edelson}, Rick and {Ferland}, Gary J. and {Ferrarese}, Laura and {Fian}, Carina and {Filippenko}, Alexei V. and {Fischer}, Travis and {Goad}, Michael R. and {Gonz{\'a}lez Buitrago}, Diego H. and {Gorjian}, Varoujan and {Grier}, Catherine J. and {Guo}, Wei-Jian and {Hall}, Patrick B. and {Ho}, Luis C. and {Homayouni}, Y. and {Horne}, Keith and {Ili{\'c}}, Dragana and {Jiang}, Bo-Wei and {Joner}, Michael D. and {Kaspi}, Shai and {Kochanek}, Christopher S. and {Korista}, Kirk T. and {Kynoch}, Daniel and {Li}, Sha-Sha and {Liu}, Jun-Rong and {McHardy}, Ian M. and {McLane}, Jacob N. and {Mitchell}, Jake A.~J. and {Netzer}, Hagai and {Olson}, Kianna A. and {Pogge}, Richard W. and {Popovi{\'c}}, Luka {\v{C}}. and {Proga}, Daniel and {Storchi-Bergmann}, Thaisa and {Strasburger}, Erika and {Treu}, Tommaso and {Vestergaard}, Marianne and {Wang}, Jian-Min and {Ward}, Martin J. and {Waters}, Tim and {Williams}, Peter R. and {Yang}, Sen and {Yao}, Zhu-Heng and {Zastrocky}, Theodora E. and {Zhai}, Shuo and {Zu}, Ying},
        title = "{AGN STORM 2. I. First results: A Change in the Weather of Mrk 817}",
      journal = {\apj},
         year = 2021,
        month = dec,
       volume = {922},
       number = {2},
          eid = {151},
        pages = {151},
          doi = {10.3847/1538-4357/ac2159},
archivePrefix = {arXiv},
       eprint = {2105.05840},
 primaryClass = {astro-ph.HE},
       adsurl = {https://ui.adsabs.harvard.edu/abs/2021ApJ...922..151K}
}

@ARTICLE{cackett2020,
       author = {{Cackett}, Edward M. and {Gelbord}, Jonathan and {Li}, Yan-Rong and {Horne}, Keith and {Wang}, Jian-Min and {Barth}, Aaron J. and {Bai}, Jin-Ming and {Bian}, Wei-Hao and {Carroll}, Russell W. and {Du}, Pu and {Edelson}, Rick and {Goad}, Michael R. and {Ho}, Luis C. and {Hu}, Chen and {Khatu}, Viraja C. and {Luo}, Bin and {Miller}, Jake and {Yuan}, Ye-Fei},
        title = "{Supermassive Black Holes with High Accretion Rates in Active Galactic Nuclei. XI. Accretion Disk Reverberation Mapping of Mrk 142}",
      journal = {\apj},
         year = 2020,
        month = jun,
       volume = {896},
       number = {1},
          eid = {1},
        pages = {1},
          doi = {10.3847/1538-4357/ab91b5},
archivePrefix = {arXiv},
       eprint = {2005.03685},
 primaryClass = {astro-ph.HE},
       adsurl = {https://ui.adsabs.harvard.edu/abs/2020ApJ...896....1C}
}

@ARTICLE{hernandez2020,
       author = {{Hern{\'a}ndez Santisteban}, J.~V. and {Edelson}, R. and {Horne}, K. and {Gelbord}, J.~M. and {Barth}, A.~J. and {Cackett}, E.~M. and {Goad}, M.~R. and {Netzer}, H. and {Starkey}, D. and {Uttley}, P. and {Brandt}, W.~N. and {Korista}, K. and {Lohfink}, A.~M. and {Onken}, C.~A. and {Page}, K.~L. and {Siegel}, M. and {Vestergaard}, M. and {Bisogni}, S. and {Breeveld}, A.~A. and {Cenko}, S.~B. and {Dalla Bont{\`a}}, E. and {Evans}, P.~A. and {Ferland}, G. and {Gonzalez-Buitrago}, D.~H. and {Grupe}, D. and {Joner}, M.~D. and {Kriss}, G. and {LaPorte}, S.~J. and {Mathur}, S. and {Marshall}, F. and {Mehdipour}, M. and {Mudd}, D. and {Peterson}, B.~M. and {Schmidt}, T. and {Vaughan}, S. and {Valenti}, S.},
        title = "{Intensive disc-reverberation mapping of Fairall 9: first year of Swift and LCO monitoring}",
      journal = {\mnras},
         year = 2020,
        month = nov,
       volume = {498},
       number = {4},
        pages = {5399-5416},
          doi = {10.1093/mnras/staa2365},
archivePrefix = {arXiv},
       eprint = {2008.02134},
 primaryClass = {astro-ph.GA},
       adsurl = {https://ui.adsabs.harvard.edu/abs/2020MNRAS.498.5399H}
}

@ARTICLE{pozo2019,
       author = {{Pozo Nu{\~n}ez}, F. and {Gianniotis}, N. and {Blex}, J. and {Lisow}, T. and {Chini}, R. and {Polsterer}, K.~L. and {Pott}, J. -U. and {Esser}, J. and {Pietrzy{\'n}ski}, G.},
        title = "{Optical continuum photometric reverberation mapping of the Seyfert-1 galaxy Mrk509}",
      journal = {\mnras},
         year = 2019,
        month = dec,
       volume = {490},
       number = {3},
        pages = {3936-3951},
          doi = {10.1093/mnras/stz2830},
archivePrefix = {arXiv},
       eprint = {1912.10319},
 primaryClass = {astro-ph.GA},
       adsurl = {https://ui.adsabs.harvard.edu/abs/2019MNRAS.490.3936P}
}

@ARTICLE{2025PhRvD.111h3545C,
       author = {{Cao}, Shulei and {Mandal}, Amit Kumar and {Zaja{\v{c}}ek}, Michal and {Czerny}, Bo{\.z}ena and {Ratra}, Bharat},
        title = "{Standardizing reverberation-mapped H{\ensuremath{\alpha}} and H{\ensuremath{\beta}} active galactic nuclei using radius-luminosity relations involving monochromatic and broad H{\ensuremath{\alpha}} luminosities}",
      journal = {\prd},
         year = 2025,
        month = apr,
       volume = {111},
       number = {8},
          eid = {083545},
        pages = {083545},
          doi = {10.1103/PhysRevD.111.083545},
archivePrefix = {arXiv},
       eprint = {2412.19665},
 primaryClass = {astro-ph.GA},
       adsurl = {https://ui.adsabs.harvard.edu/abs/2025PhRvD.111h3545C}
}

@ARTICLE{2025PhRvD.112d3516C,
       author = {{Cao}, Shulei and {Mandal}, Amit Kumar and {Zaja{\v{c}}ek}, Michal and {Ratra}, Bharat},
        title = "{Standardizing a larger, higher-quality, homogeneous sample of reverberation-mapped H{\ensuremath{\beta}} active galactic nuclei using the broad-line region radius-luminosity relation}",
      journal = {\prd},
         year = 2025,
        month = aug,
       volume = {112},
       number = {4},
          eid = {043516},
        pages = {043516},
          doi = {10.1103/w1qx-4v84},
archivePrefix = {arXiv},
       eprint = {2506.17422},
 primaryClass = {astro-ph.GA},
       adsurl = {https://ui.adsabs.harvard.edu/abs/2025PhRvD.112d3516C}
}

@ARTICLE{1992ApJS...80..109B,
       author = {{Boroson}, Todd A. and {Green}, Richard F.},
        title = "{The Emission-Line Properties of Low-Redshift Quasi-stellar Objects}",
      journal = {\apjs},
         year = 1992,
        month = may,
       volume = {80},
        pages = {109},
          doi = {10.1086/191661},
       adsurl = {https://ui.adsabs.harvard.edu/abs/1992ApJS...80..109B}
}

@ARTICLE{2006ApJ...650...57T,
       author = {{Tsuzuki}, Yumihiko and {Kawara}, Kimiaki and {Yoshii}, Yuzuru and {Oyabu}, Shinki and {Tanab{\'e}}, Toshihiko and {Matsuoka}, Yoshiki},
        title = "{Fe II Emission in 14 Low-Redshift Quasars. I. Observations}",
      journal = {\apj},
         year = 2006,
        month = oct,
       volume = {650},
       number = {1},
        pages = {57-79},
          doi = {10.1086/506376},
archivePrefix = {arXiv},
       eprint = {astro-ph/0606040},
 primaryClass = {astro-ph},
       adsurl = {https://ui.adsabs.harvard.edu/abs/2006ApJ...650...57T}
}

@ARTICLE{2001ApJS..134....1V,
       author = {{Vestergaard}, M. and {Wilkes}, B.~J.},
        title = "{An Empirical Ultraviolet Template for Iron Emission in Quasars as Derived from I Zwicky 1}",
      journal = {\apjs},
         year = 2001,
        month = may,
       volume = {134},
       number = {1},
        pages = {1-33},
          doi = {10.1086/320357},
archivePrefix = {arXiv},
       eprint = {astro-ph/0104320},
 primaryClass = {astro-ph},
       adsurl = {https://ui.adsabs.harvard.edu/abs/2001ApJS..134....1V}
}

@ARTICLE{rauch1991,
       author = {{Rauch}, Kevin P. and {Blandford}, Roger D.},
        title = "{Microlensing and the Structure of Active Galactic Nucleus Accretion Disks}",
      journal = {\apjl},
         year = 1991,
        month = nov,
       volume = {381},
        pages = {L39},
          doi = {10.1086/186191},
       adsurl = {https://ui.adsabs.harvard.edu/abs/1991ApJ...381L..39R}
}

@ARTICLE{2018ApJ...854..107F,
       author = {{Fausnaugh}, M.~M. and {Starkey}, D.~A. and {Horne}, Keith and {Kochanek}, C.~S. and {Peterson}, B.~M. and {Bentz}, M.~C. and {Denney}, K.~D. and {Grier}, C.~J. and {Grupe}, D. and {Pogge}, R.~W. and {De Rosa}, G. and {Adams}, S.~M. and {Barth}, A.~J. and {Beatty}, Thomas G. and {Bhattacharjee}, A. and {Borman}, G.~A. and {Boroson}, T.~A. and {Bottorff}, M.~C. and {Brown}, Jacob E. and {Brown}, Jonathan S. and {Brotherton}, M.~S. and {Coker}, C.~T. and {Crawford}, S.~M. and {Croxall}, K.~V. and {Eftekharzadeh}, Sarah and {Eracleous}, Michael and {Joner}, M.~D. and {Henderson}, C.~B. and {Holoien}, T.~W. -S. and {Hutchison}, T. and {Kaspi}, Shai and {Kim}, S. and {King}, Anthea L. and {Li}, Miao and {Lochhaas}, Cassandra and {Ma}, Zhiyuan and {MacInnis}, F. and {Manne-Nicholas}, E.~R. and {Mason}, M. and {Montuori}, Carmen and {Mosquera}, Ana and {Mudd}, Dale and {Musso}, R. and {Nazarov}, S.~V. and {Nguyen}, M.~L. and {Okhmat}, D.~N. and {Onken}, Christopher A. and {Ou-Yang}, B. and {Pancoast}, A. and {Pei}, L. and {Penny}, Matthew T. and {Poleski}, Rados{\l}aw and {Rafter}, Stephen and {Romero-Colmenero}, E. and {Runnoe}, Jessie and {Sand}, David J. and {Schimoia}, Jaderson S. and {Sergeev}, S.~G. and {Shappee}, B.~J. and {Simonian}, Gregory V. and {Somers}, Garrett and {Spencer}, M. and {Stevens}, Daniel J. and {Tayar}, Jamie and {Treu}, T. and {Valenti}, Stefano and {Van Saders}, J. and {Villanueva}, S., Jr. and {Villforth}, C. and {Weiss}, Yaniv and {Winkler}, H. and {Zhu}, W.},
        title = "{Continuum Reverberation Mapping of the Accretion Disks in Two Seyfert 1 Galaxies}",
      journal = {\apj},
         year = 2018,
        month = feb,
       volume = {854},
       number = {2},
          eid = {107},
        pages = {107},
          doi = {10.3847/1538-4357/aaaa2b},
archivePrefix = {arXiv},
       eprint = {1801.09692},
 primaryClass = {astro-ph.GA},
       adsurl = {https://ui.adsabs.harvard.edu/abs/2018ApJ...854..107F}
}

@ARTICLE{2019ApJ...870..123E,
       author = {{Edelson}, R. and {Gelbord}, J. and {Cackett}, E. and {Peterson}, B.~M. and {Horne}, K. and {Barth}, A.~J. and {Starkey}, D.~A. and {Bentz}, M. and {Brandt}, W.~N. and {Goad}, M. and {Joner}, M. and {Korista}, K. and {Netzer}, H. and {Page}, K. and {Uttley}, P. and {Vaughan}, S. and {Breeveld}, A. and {Cenko}, S.~B. and {Done}, C. and {Evans}, P. and {Fausnaugh}, M. and {Ferland}, G. and {Gonzalez-Buitrago}, D. and {Gropp}, J. and {Grupe}, D. and {Kaastra}, J. and {Kennea}, J. and {Kriss}, G. and {Mathur}, S. and {Mehdipour}, M. and {Mudd}, D. and {Nousek}, J. and {Schmidt}, T. and {Vestergaard}, M. and {Villforth}, C.},
        title = "{The First Swift Intensive AGN Accretion Disk Reverberation Mapping Survey}",
      journal = {\apj},
         year = 2019,
        month = jan,
       volume = {870},
       number = {2},
          eid = {123},
        pages = {123},
          doi = {10.3847/1538-4357/aaf3b4},
archivePrefix = {arXiv},
       eprint = {1811.07956},
 primaryClass = {astro-ph.HE},
       adsurl = {https://ui.adsabs.harvard.edu/abs/2019ApJ...870..123E}
}

@ARTICLE{cackett2018,
       author = {{Cackett}, Edward M. and {Chiang}, Chia-Ying and {McHardy}, Ian and {Edelson}, Rick and {Goad}, Michael R. and {Horne}, Keith and {Korista}, Kirk T.},
        title = "{Accretion Disk Reverberation with Hubble Space Telescope Observations of NGC 4593: Evidence for Diffuse Continuum Lags}",
      journal = {\apj},
         year = 2018,
        month = apr,
       volume = {857},
       number = {1},
          eid = {53},
        pages = {53},
          doi = {10.3847/1538-4357/aab4f7},
archivePrefix = {arXiv},
       eprint = {1712.04025},
 primaryClass = {astro-ph.HE},
       adsurl = {https://ui.adsabs.harvard.edu/abs/2018ApJ...857...53C}
}

@ARTICLE{2025MNRAS.541..642P,
       author = {{Prince}, Raj and {Hern{\'a}ndez Santisteban}, Juan V. and {Horne}, Keith and {Gelbord}, J. and {McHardy}, Ian and {Edelson}, R. and {Onken}, C.~A. and {Donnan}, F.~R. and {Vestergaard}, M. and {Kaspi}, S. and {Winkler}, H. and {Cackett}, E.~M. and {Landt}, H. and {Barth}, A.~J. and {Treu}, T. and {Valenti}, S. and {Lira}, P. and {Chelouche}, D. and {Romero Colmenero}, E. and {Goad}, M.~R. and {Gonzalez-Buitrago}, D.~H. and {Kara}, E. and {Villforth}, C.},
        title = "{Echo mapping of the black hole accretion flow in NGC 7469}",
      journal = {\mnras},
         year = 2025,
        month = jul,
       volume = {541},
       number = {1},
        pages = {642-661},
          doi = {10.1093/mnras/staf983},
archivePrefix = {arXiv},
       eprint = {2506.06731},
 primaryClass = {astro-ph.GA},
       adsurl = {https://ui.adsabs.harvard.edu/abs/2025MNRAS.541..642P}
}

@ARTICLE{2026ApJ..1003..196P,
       author = {{Pan}, Yu and {Guo}, Hengxiao and {Liu}, Chenxu and {Chen}, Xinlei and {Fang}, Yuan and {Zhang}, Jinghua and {Zuo}, Wenwen and {Edwards}, Philip G. and {Stevens}, Jamie and {Fu}, Manqi and {Sun}, Mouyuan and {Cai}, Zhen-yi and {Du}, Guowang and {Zou}, Xingzhu and {Wang}, Tao and {Zhu}, Xufeng and {Liu}, Xiangkun and {Liu}, Xiaowei},
        title = "{The Intermediate-mass Black Hole Reverberation Mapping Project: Stable Optical Continuum Lags of an Intermediate-mass Black Hole in the Dwarf Galaxy NGC 4395 over Years}",
      journal = {\apj},
         year = 2026,
        month = jun,
       volume = {1003},
       number = {2},
          eid = {196},
        pages = {196},
          doi = {10.3847/1538-4357/ae6857},
archivePrefix = {arXiv},
       eprint = {2601.14787},
 primaryClass = {astro-ph.GA},
       adsurl = {https://ui.adsabs.harvard.edu/abs/2026ApJ..1003..196P}
}

@ARTICLE{2026ApJ...997..326F,
       author = {{Feng}, Hai-Cheng and {Li}, Sha-Sha and {Sun}, Mouyuan and {Pinto}, Ciro and {Zhou}, Shuying and {Xu}, Yerong and {Bai}, J.~M. and {Dalla Bont{\`a}}, Elena and {Dong}, ZhongNan and {Kumari}, Neeraj and {Lin}, Jiaqi and {Liu}, H.~T. and {Lu}, Kai-Xing and {Ma}, Bin and {Mao}, Ji-Rong and {Nardini}, Emanuele and {Piconcelli}, Enrico and {Pintore}, Fabio and {Wang}, Jian-Guo and {Xiong}, Ding-Rong},
        title = "{Discovery of a Luminosity-dependent Continuum Lag in NGC 4151 from Photometric and Spectroscopic Continuum Reverberation Mapping}",
      journal = {\apj},
         year = 2026,
        month = feb,
       volume = {997},
       number = {2},
          eid = {326},
        pages = {326},
          doi = {10.3847/1538-4357/ae30db},
archivePrefix = {arXiv},
       eprint = {2512.18276},
 primaryClass = {astro-ph.GA},
       adsurl = {https://ui.adsabs.harvard.edu/abs/2026ApJ...997..326F}
}

@ARTICLE{2026MNRAS.546ag025K,
       author = {{Kynoch}, D. and {McHardy}, I.~M. and {Cackett}, E.~M. and {Gelbord}, J. and {Hern{\'a}ndez Santisteban}, J.~V. and {Horne}, K. and {Miller}, J.~A. and {Netzer}, H. and {Done}, C. and {Edelson}, R. and {Fausnaugh}, M.~M. and {Goad}, M.~R. and {Peterson}, B.~M. and {Vincentelli}, F.~M.},
        title = "{Intensive X-Ray/UVOIR continuum reverberation mapping of the Seyfert AGN MCG +08─11─11}",
      journal = {\mnras},
         year = 2026,
        month = mar,
       volume = {546},
       number = {3},
          eid = {stag025},
        pages = {stag025},
          doi = {10.1093/mnras/stag025},
archivePrefix = {arXiv},
       eprint = {2511.05342},
 primaryClass = {astro-ph.GA},
       adsurl = {https://ui.adsabs.harvard.edu/abs/2026MNRAS.546ag025K}
}

@ARTICLE{2024ApJ...974..271L,
       author = {{Lewin}, Collin and {Kara}, Erin and {Barth}, Aaron J. and {Cackett}, Edward M. and {De Rosa}, Gisella and {Homayouni}, Yasaman and {Horne}, Keith and {Kriss}, Gerard A. and {Landt}, Hermine and {Gelbord}, Jonathan and {Montano}, John and {Arav}, Nahum and {Bentz}, Misty C. and {Boizelle}, Benjamin D. and {Dalla Bont{\`a}}, Elena and {Brotherton}, Michael S. and {Dehghanian}, Maryam and {Ferland}, Gary J. and {Fian}, Carina and {Goad}, Michael R. and {Hern{\'a}ndez Santisteban}, Juan V. and {Ili{\'c}}, Dragana and {Kaastra}, Jelle and {Kaspi}, Shai and {Korista}, Kirk T. and {Kosec}, Peter and {Kova{\v{c}}evi{\'c}}, Andjelka and {Mehdipour}, Missagh and {Miller}, Jake A. and {Netzer}, Hagai and {Neustadt}, Jack M.~M. and {Panagiotou}, Christos and {Partington}, Ethan R. and {Popovi{\'c}}, Luka {\v{C}}. and {Sanmartim}, David and {Vestergaard}, Marianne and {Ward}, Martin J. and {Zaidouni}, Fatima},
        title = "{AGN STORM 2. VII. A Frequency-resolved Map of the Accretion Disk in Mrk 817: Simultaneous X-Ray Reverberation and UVOIR Disk Reprocessing Time Lags}",
      journal = {\apj},
         year = 2024,
        month = oct,
       volume = {974},
       number = {2},
          eid = {271},
        pages = {271},
          doi = {10.3847/1538-4357/ad6b08},
archivePrefix = {arXiv},
       eprint = {2409.09115},
 primaryClass = {astro-ph.HE},
       adsurl = {https://ui.adsabs.harvard.edu/abs/2024ApJ...974..271L}
}

@ARTICLE{2026MNRAS.546ag067D,
       author = {{Drewes}, Farin and {Vieliute}, Roberta and {Hern{\'a}ndez Santisteban}, Juan V. and {Horne}, Keith and {Barth}, Aaron J. and {Cackett}, Edward M. and {Romero Colmenero}, Encarni and {Goad}, Michael R. and {Kaspi}, Shai and {Landt}, Hermine and {Lira}, Paulina and {Netzer}, Hagai and {Vestergaard}, Marianne and {Winkler}, Hartmut},
        title = "{A phenomenological study of the accretion disc in the super-Eddington AGN I Zw 1}",
      journal = {\mnras},
         year = 2026,
        month = mar,
       volume = {546},
       number = {3},
          eid = {stag067},
        pages = {stag067},
          doi = {10.1093/mnras/stag067},
archivePrefix = {arXiv},
       eprint = {2601.05818},
 primaryClass = {astro-ph.GA},
       adsurl = {https://ui.adsabs.harvard.edu/abs/2026MNRAS.546ag067D}
}

@ARTICLE{Planck2020,
       author = {{Planck Collaboration} and {Aghanim}, N. and {Akrami}, Y. and {Ashdown}, M. and {Aumont}, J. and {Baccigalupi}, C. and {Ballardini}, M. and {Banday}, A.~J. and {Barreiro}, R.~B. and {Bartolo}, N. and {Basak}, S. and {Battye}, R. and {Benabed}, K. and {Bernard}, J. -P. and {Bersanelli}, M. and {Bielewicz}, P. and {Bock}, J.~J. and {Bond}, J.~R. and {Borrill}, J. and {Bouchet}, F.~R. and {Boulanger}, F. and {Bucher}, M. and {Burigana}, C. and {Butler}, R.~C. and {Calabrese}, E. and {Cardoso}, J. -F. and {Carron}, J. and {Challinor}, A. and {Chiang}, H.~C. and {Chluba}, J. and {Colombo}, L.~P.~L. and {Combet}, C. and {Contreras}, D. and {Crill}, B.~P. and {Cuttaia}, F. and {de Bernardis}, P. and {de Zotti}, G. and {Delabrouille}, J. and {Delouis}, J. -M. and {Di Valentino}, E. and {Diego}, J.~M. and {Dor{\'e}}, O. and {Douspis}, M. and {Ducout}, A. and {Dupac}, X. and {Dusini}, S. and {Efstathiou}, G. and {Elsner}, F. and {En{\ss}lin}, T.~A. and {Eriksen}, H.~K. and {Fantaye}, Y. and {Farhang}, M. and {Fergusson}, J. and {Fernandez-Cobos}, R. and {Finelli}, F. and {Forastieri}, F. and {Frailis}, M. and {Fraisse}, A.~A. and {Franceschi}, E. and {Frolov}, A. and {Galeotta}, S. and {Galli}, S. and {Ganga}, K. and {G{\'e}nova-Santos}, R.~T. and {Gerbino}, M. and {Ghosh}, T. and {Gonz{\'a}lez-Nuevo}, J. and {G{\'o}rski}, K.~M. and {Gratton}, S. and {Gruppuso}, A. and {Gudmundsson}, J.~E. and {Hamann}, J. and {Handley}, W. and {Hansen}, F.~K. and {Herranz}, D. and {Hildebrandt}, S.~R. and {Hivon}, E. and {Huang}, Z. and {Jaffe}, A.~H. and {Jones}, W.~C. and {Karakci}, A. and {Keih{\"a}nen}, E. and {Keskitalo}, R. and {Kiiveri}, K. and {Kim}, J. and {Kisner}, T.~S. and {Knox}, L. and {Krachmalnicoff}, N. and {Kunz}, M. and {Kurki-Suonio}, H. and {Lagache}, G. and {Lamarre}, J. -M. and {Lasenby}, A. and {Lattanzi}, M. and {Lawrence}, C.~R. and {Le Jeune}, M. and {Lemos}, P. and {Lesgourgues}, J. and {Levrier}, F. and {Lewis}, A. and {Liguori}, M. and {Lilje}, P.~B. and {Lilley}, M. and {Lindholm}, V. and {L{\'o}pez-Caniego}, M. and {Lubin}, P.~M. and {Ma}, Y. -Z. and {Mac{\'\i}as-P{\'e}rez}, J.~F. and {Maggio}, G. and {Maino}, D. and {Mandolesi}, N. and {Mangilli}, A. and {Marcos-Caballero}, A. and {Maris}, M. and {Martin}, P.~G. and {Martinelli}, M. and {Mart{\'\i}nez-Gonz{\'a}lez}, E. and {Matarrese}, S. and {Mauri}, N. and {McEwen}, J.~D. and {Meinhold}, P.~R. and {Melchiorri}, A. and {Mennella}, A. and {Migliaccio}, M. and {Millea}, M. and {Mitra}, S. and {Miville-Desch{\^e}nes}, M. -A. and {Molinari}, D. and {Montier}, L. and {Morgante}, G. and {Moss}, A. and {Natoli}, P. and {N{\o}rgaard-Nielsen}, H.~U. and {Pagano}, L. and {Paoletti}, D. and {Partridge}, B. and {Patanchon}, G. and {Peiris}, H.~V. and {Perrotta}, F. and {Pettorino}, V. and {Piacentini}, F. and {Polastri}, L. and {Polenta}, G. and {Puget}, J. -L. and {Rachen}, J.~P. and {Reinecke}, M. and {Remazeilles}, M. and {Renzi}, A. and {Rocha}, G. and {Rosset}, C. and {Roudier}, G. and {Rubi{\~n}o-Mart{\'\i}n}, J.~A. and {Ruiz-Granados}, B. and {Salvati}, L. and {Sandri}, M. and {Savelainen}, M. and {Scott}, D. and {Shellard}, E.~P.~S. and {Sirignano}, C. and {Sirri}, G. and {Spencer}, L.~D. and {Sunyaev}, R. and {Suur-Uski}, A. -S. and {Tauber}, J.~A. and {Tavagnacco}, D. and {Tenti}, M. and {Toffolatti}, L. and {Tomasi}, M. and {Trombetti}, T. and {Valenziano}, L. and {Valiviita}, J. and {Van Tent}, B. and {Vibert}, L. and {Vielva}, P. and {Villa}, F. and {Vittorio}, N. and {Wandelt}, B.~D. and {Wehus}, I.~K. and {White}, M. and {White}, S.~D.~M. and {Zacchei}, A. and {Zonca}, A.},
        title = "{Planck 2018 results. VI. Cosmological parameters}",
      journal = {\aap},
         year = 2020,
        month = sep,
       volume = {641},
          eid = {A6},
        pages = {A6},
          doi = {10.1051/0004-6361/201833910},
archivePrefix = {arXiv},
       eprint = {1807.06209},
 primaryClass = {astro-ph.CO},
       adsurl = {https://ui.adsabs.harvard.edu/abs/2020A&A...641A...6P}
}

@ARTICLE{2021ApJ...918...29M,
       author = {{Mart{\'\i}nez-Aldama}, Mary Loli and {Panda}, Swayamtrupta and {Czerny}, Bo{\.z}ena and {Marinello}, Murilo and {Marziani}, Paola and {Dultzin}, Deborah},
        title = "{The CaFe Project: Optical Fe II and Near-infrared Ca II Triplet Emission in Active Galaxies. II. The Driver(s) of the Ca II and Fe II and Its Potential Use as a Chemical Clock}",
      journal = {\apj},
         year = 2021,
        month = sep,
       volume = {918},
       number = {1},
          eid = {29},
        pages = {29},
          doi = {10.3847/1538-4357/ac03b6},
archivePrefix = {arXiv},
       eprint = {2101.06999},
 primaryClass = {astro-ph.GA},
       adsurl = {https://ui.adsabs.harvard.edu/abs/2021ApJ...918...29M}
}

@ARTICLE{panda2019,
       author = {{Panda}, Swayamtrupta and {Marziani}, Paola and {Czerny}, Bo{\.z}ena},
        title = "{The Quasar Main Sequence Explained by the Combination of Eddington Ratio, Metallicity, and Orientation}",
      journal = {\apj},
         year = 2019,
        month = sep,
       volume = {882},
       number = {2},
          eid = {79},
        pages = {79},
          doi = {10.3847/1538-4357/ab3292},
archivePrefix = {arXiv},
       eprint = {1905.01729},
 primaryClass = {astro-ph.HE},
       adsurl = {https://ui.adsabs.harvard.edu/abs/2019ApJ...882...79P}
}

@ARTICLE{kubota2018,
       author = {{Kubota}, Aya and {Done}, Chris},
        title = "{A physical model of the broad-band continuum of AGN and its implications for the UV/X relation and optical variability}",
      journal = {\mnras},
         year = 2018,
        month = oct,
       volume = {480},
       number = {1},
        pages = {1247-1262},
          doi = {10.1093/mnras/sty1890},
archivePrefix = {arXiv},
       eprint = {1804.00171},
 primaryClass = {astro-ph.HE},
       adsurl = {https://ui.adsabs.harvard.edu/abs/2018MNRAS.480.1247K}
}

@ARTICLE{2017ApJ...840...41E,
       author = {{Edelson}, R. and {Gelbord}, J. and {Cackett}, E. and {Connolly}, S. and {Done}, C. and {Fausnaugh}, M. and {Gardner}, E. and {Gehrels}, N. and {Goad}, M. and {Horne}, K. and {McHardy}, I. and {Peterson}, B.~M. and {Vaughan}, S. and {Vestergaard}, M. and {Breeveld}, A. and {Barth}, A.~J. and {Bentz}, M. and {Bottorff}, M. and {Brandt}, W.~N. and {Crawford}, S.~M. and {Dalla Bont{\`a}}, E. and {Emmanoulopoulos}, D. and {Evans}, P. and {Figuera Jaimes}, R. and {Filippenko}, A.~V. and {Ferland}, G. and {Grupe}, D. and {Joner}, M. and {Kennea}, J. and {Korista}, K.~T. and {Krimm}, H.~A. and {Kriss}, G. and {Leonard}, D.~C. and {Mathur}, S. and {Netzer}, H. and {Nousek}, J. and {Page}, K. and {Romero-Colmenero}, E. and {Siegel}, M. and {Starkey}, D.~A. and {Treu}, T. and {Vogler}, H.~A. and {Winkler}, H. and {Zheng}, W.},
        title = "{Swift Monitoring of NGC 4151: Evidence for a Second X-Ray/UV Reprocessing}",
      journal = {\apj},
         year = 2017,
        month = may,
       volume = {840},
       number = {1},
          eid = {41},
        pages = {41},
          doi = {10.3847/1538-4357/aa6890},
archivePrefix = {arXiv},
       eprint = {1703.06901},
 primaryClass = {astro-ph.HE},
       adsurl = {https://ui.adsabs.harvard.edu/abs/2017ApJ...840...41E}
}

@ARTICLE{2022ApJ...934L..37M,
       author = {{Montano}, John W. and {Guo}, Hengxiao and {Barth}, Aaron J. and {U}, Vivian and {Remigio}, Raymond and {Gonz{\'a}lez-Buitrago}, Diego H. and {Hern{\'a}ndez Santisteban}, Juan V.},
        title = "{Optical Continuum Reverberation in the Dwarf Seyfert Nucleus of NGC 4395}",
      journal = {\apjl},
         year = 2022,
        month = aug,
       volume = {934},
       number = {2},
          eid = {L37},
        pages = {L37},
          doi = {10.3847/2041-8213/ac7e54},
archivePrefix = {arXiv},
       eprint = {2205.13620},
 primaryClass = {astro-ph.GA},
       adsurl = {https://ui.adsabs.harvard.edu/abs/2022ApJ...934L..37M}
}

@ARTICLE{two_timescales2021,
       author = {{Vincentelli}, F.~M. and {McHardy}, I. and {Cackett}, E.~M. and {Barth}, A.~J. and {Horne}, K. and {Goad}, M. and {Korista}, K. and {Gelbord}, J. and {Brandt}, W. and {Edelson}, R. and {Miller}, J.~A. and {Pahari}, M. and {Peterson}, B.~M. and {Schmidt}, T. and {Baldi}, R.~D. and {Breedt}, E. and {Hern{\'a}ndez Santisteban}, J.~V. and {Romero-Colmenero}, E. and {Ward}, M. and {Williams}, D.~R.~A.},
        title = "{On the multiwavelength variability of Mrk 110: two components acting at different time-scales}",
      journal = {\mnras},
         year = 2021,
        month = jul,
       volume = {504},
       number = {3},
        pages = {4337-4353},
          doi = {10.1093/mnras/stab1033},
archivePrefix = {arXiv},
       eprint = {2104.04530},
 primaryClass = {astro-ph.HE},
       adsurl = {https://ui.adsabs.harvard.edu/abs/2021MNRAS.504.4337V}
}

@ARTICLE{1989ApJ...347..640R,
       author = {{Rees}, M.~J. and {Netzer}, Hagai and {Ferland}, G.~J.},
        title = "{Small Dense Broad-Line Regions in Active Nuclei}",
      journal = {\apj},
         year = 1989,
        month = dec,
       volume = {347},
        pages = {640},
          doi = {10.1086/168155},
       adsurl = {https://ui.adsabs.harvard.edu/abs/1989ApJ...347..640R}
}

@ARTICLE{2004ApJ...615..610B,
       author = {{Baldwin}, J.~A. and {Ferland}, G.~J. and {Korista}, K.~T. and {Hamann}, F. and {LaCluyz{\'e}}, A.},
        title = "{The Origin of Fe II Emission in Active Galactic Nuclei}",
      journal = {\apj},
         year = 2004,
        month = nov,
       volume = {615},
       number = {2},
        pages = {610-624},
          doi = {10.1086/424683},
archivePrefix = {arXiv},
       eprint = {astro-ph/0407404},
 primaryClass = {astro-ph},
       adsurl = {https://ui.adsabs.harvard.edu/abs/2004ApJ...615..610B}
}

@ARTICLE{2012ApJ...757...62N,
       author = {{Negrete}, C. Alenka and {Dultzin}, Deborah and {Marziani}, Paola and {Sulentic}, Jack W.},
        title = "{Broad-line Region Physical Conditions in Extreme Population A Quasars: A Method to Estimate Central Black Hole Mass at High Redshift}",
      journal = {\apj},
         year = 2012,
        month = sep,
       volume = {757},
       number = {1},
          eid = {62},
        pages = {62},
          doi = {10.1088/0004-637X/757/1/62},
archivePrefix = {arXiv},
       eprint = {1107.3188},
 primaryClass = {astro-ph.CO},
       adsurl = {https://ui.adsabs.harvard.edu/abs/2012ApJ...757...62N}
}

@ARTICLE{naddaf2021,
       author = {{Naddaf}, Mohammad-Hassan and {Czerny}, Bo{\.z}ena and {Szczerba}, Ryszard},
        title = "{The Picture of BLR in 2.5D FRADO: Dynamics and Geometry}",
      journal = {\apj},
         year = 2021,
        month = oct,
       volume = {920},
       number = {1},
          eid = {30},
        pages = {30},
          doi = {10.3847/1538-4357/ac139d},
archivePrefix = {arXiv},
       eprint = {2102.00336},
 primaryClass = {astro-ph.GA},
       adsurl = {https://ui.adsabs.harvard.edu/abs/2021ApJ...920...30N}
}

@ARTICLE{czhr2011,
       author = {{Czerny}, B. and {Hryniewicz}, K.},
        title = "{The origin of the broad line region in active galactic nuclei}",
      journal = {\aap},
         year = 2011,
        month = jan,
       volume = {525},
          eid = {L8},
        pages = {L8},
          doi = {10.1051/0004-6361/201016025},
archivePrefix = {arXiv},
       eprint = {1010.6201},
 primaryClass = {astro-ph.CO},
       adsurl = {https://ui.adsabs.harvard.edu/abs/2011A&A...525L...8C}
}

@ARTICLE{cackett2007,
       author = {{Cackett}, Edward M. and {Horne}, Keith and {Winkler}, Hartmut},
        title = "{Testing thermal reprocessing in active galactic nuclei accretion discs}",
      journal = {\mnras},
         year = 2007,
        month = sep,
       volume = {380},
       number = {2},
        pages = {669-682},
          doi = {10.1111/j.1365-2966.2007.12098.x},
archivePrefix = {arXiv},
       eprint = {0706.1464},
 primaryClass = {astro-ph},
       adsurl = {https://ui.adsabs.harvard.edu/abs/2007MNRAS.380..669C}
}

@ARTICLE{collier1999,
       author = {{Collier}, Stefan and {Horne}, Keith and {Wanders}, Ignaz and {Peterson}, Bradley M.},
        title = "{A new direct method for measuring the Hubble constant from reverberating accretion discs in active galaxies}",
      journal = {\mnras},
         year = 1999,
        month = jan,
       volume = {302},
       number = {1},
        pages = {L24-L28},
          doi = {10.1046/j.1365-8711.1999.02250.x},
archivePrefix = {arXiv},
       eprint = {astro-ph/9811278},
 primaryClass = {astro-ph},
       adsurl = {https://ui.adsabs.harvard.edu/abs/1999MNRAS.302L..24C}
}

@ARTICLE{panda2018,
       author = {{Panda}, Swayamtrupta and {Czerny}, Bo{\.z}ena and {Adhikari}, Tek P. and {Hryniewicz}, Krzysztof and {Wildy}, Conor and {Kuraszkiewicz}, Joanna and {{\'S}niegowska}, Marzena},
        title = "{Modeling of the Quasar Main Sequence in the Optical Plane}",
      journal = {\apj},
         year = 2018,
        month = oct,
       volume = {866},
       number = {2},
          eid = {115},
        pages = {115},
          doi = {10.3847/1538-4357/aae209},
archivePrefix = {arXiv},
       eprint = {1806.08571},
 primaryClass = {astro-ph.HE},
       adsurl = {https://ui.adsabs.harvard.edu/abs/2018ApJ...866..115P}
}

@ARTICLE{panda_cafe2021,
       author = {{Panda}, Swayamtrupta},
        title = "{The CaFe project: Optical Fe II and near-infrared Ca II triplet emission in active galaxies: simulated EWs and the co-dependence of cloud size and metal content}",
      journal = {\aap},
         year = 2021,
        month = jun,
       volume = {650},
          eid = {A154},
        pages = {A154},
          doi = {10.1051/0004-6361/202140393},
archivePrefix = {arXiv},
       eprint = {2004.13113},
 primaryClass = {astro-ph.GA},
       adsurl = {https://ui.adsabs.harvard.edu/abs/2021A&A...650A.154P}
}

@ARTICLE{peterson2004,
       author = {{Peterson}, B.~M. and {Ferrarese}, L. and {Gilbert}, K.~M. and {Kaspi}, S. and {Malkan}, M.~A. and {Maoz}, D. and {Merritt}, D. and {Netzer}, H. and {Onken}, C.~A. and {Pogge}, R.~W. and {Vestergaard}, M. and {Wandel}, A.},
        title = "{Central Masses and Broad-Line Region Sizes of Active Galactic Nuclei. II. A Homogeneous Analysis of a Large Reverberation-Mapping Database}",
      journal = {The Astrophysical Journal},
         year = 2004,
        month = oct,
       volume = {613},
       number = {2},
        pages = {682-699},
          doi = {10.1086/423269},
archivePrefix = {arXiv},
       eprint = {astro-ph/0407299},
 primaryClass = {astro-ph},
       adsurl = {https://ui.adsabs.harvard.edu/abs/2004ApJ...613..682P}
}

@ARTICLE{2012ApJ...758...67L,
       author = {{Lohfink}, Anne M. and {Reynolds}, Christopher S. and {Miller}, Jon M. and {Brenneman}, Laura W. and {Mushotzky}, Richard F. and {Nowak}, Michael A. and {Fabian}, Andrew C.},
        title = "{The Black Hole Spin and Soft X-Ray Excess of the Luminous Seyfert Galaxy Fairall 9}",
      journal = {\apj},
         year = 2012,
        month = oct,
       volume = {758},
       number = {1},
          eid = {67},
        pages = {67},
          doi = {10.1088/0004-637X/758/1/67},
archivePrefix = {arXiv},
       eprint = {1209.0468},
 primaryClass = {astro-ph.HE},
       adsurl = {https://ui.adsabs.harvard.edu/abs/2012ApJ...758...67L}
}

@ARTICLE{2015ApJ...801...38W,
       author = {{Woo}, Jong-Hak and {Yoon}, Yosep and {Park}, Songyoun and {Park}, Daeseong and {Kim}, Sang Chul},
        title = "{The Black Hole Mass-Stellar Velocity Dispersion Relation of Narrow-line Seyfert 1 Galaxies}",
      journal = {\apj},
         year = 2015,
        month = mar,
       volume = {801},
       number = {1},
          eid = {38},
        pages = {38},
        doi = {10.1088/0004-637X/801/1/38},
archivePrefix = {arXiv},
       eprint = {1412.7225},
 primaryClass = {astro-ph.GA},
       adsurl = {https://ui.adsabs.harvard.edu/abs/2015ApJ...801...38W}
}

@ARTICLE{hagen2023,
       author = {{Hagen}, Scott and {Done}, Chris},
        title = "{Modelling continuum reverberation in active galactic nuclei: a spectral-timing analysis of the ultraviolet variability through X-ray reverberation in Fairall 9}",
      journal = {\mnras},
         year = 2023,
        month = may,
       volume = {521},
       number = {1},
        pages = {251-268},
          doi = {10.1093/mnras/stad504},
archivePrefix = {arXiv},
       eprint = {2210.04924},
 primaryClass = {astro-ph.HE},
       adsurl = {https://ui.adsabs.harvard.edu/abs/2023MNRAS.521..251H}
}

@ARTICLE{clavel1989,
       author = {{Clavel}, J. and {Wamsteker}, W. and {Glass}, I.~S.},
        title = "{Hot Dust on the Outskirts of the Broad-Line Region in Fairall 9}",
      journal = {\apj},
         year = 1989,
        month = feb,
       volume = {337},
        pages = {236},
          doi = {10.1086/167100},
       adsurl = {https://ui.adsabs.harvard.edu/abs/1989ApJ...337..236C}
}

@ARTICLE{GuoH_etal_2022,
       author = {{Guo, H.} and {Barth}, Aaron J. and {Korista}, Kirk T. and {Goad}, Michael R. and {Cackett}, Edward M. and {Bentz}, Misty C. and {Brandt}, William N. and {Gonzalez-Buitrago}, D. and {Ferland}, Gary J. and {Gelbord}, Jonathan M. and {Ho}, Luis C. and {Horne}, Keith and {Joner}, Michael D. and {Kriss}, Gerard A. and {McHardy}, Ian and {Mehdipour}, Missagh and {Park}, Daeseong and {Remigio}, Raymond and {U}, Vivian and {Vestergaard}, Marianne},
        title = "{The Paschen Jump as a Diagnostic of the Diffuse Nebular Continuum Emission in Active Galactic Nuclei}",
      journal = {\apj},
         year = 2022,
        month = mar,
       volume = {927},
       number = {1},
          eid = {60},
        pages = {60},
          doi = {10.3847/1538-4357/ac4bc6},
archivePrefix = {arXiv},
       eprint = {2111.03090},
 primaryClass = {astro-ph.GA},
       adsurl = {https://ui.adsabs.harvard.edu/abs/2022ApJ...927...60G}
}

@ARTICLE{jaiswal2023,
       author = {{Jaiswal}, Vikram Kumar and {Prince}, Raj and {Panda}, Swayamtrupta and {Czerny}, Bo{\.z}ena},
        title = "{Modeling time delays from two reprocessors in active galactic nuclei}",
      journal = {\aap},
         year = 2023,
        month = feb,
       volume = {670},
          eid = {A147},
        pages = {A147},
          doi = {10.1051/0004-6361/202244352},
archivePrefix = {arXiv},
       eprint = {2206.12497},
 primaryClass = {astro-ph.GA},
       adsurl = {https://ui.adsabs.harvard.edu/abs/2023A&A...670A.147J}
}

@ARTICLE{patrick2011,
       author = {{Patrick}, A.~R. and {Reeves}, J.~N. and {Porquet}, D. and {Markowitz}, A.~G. and {Lobban}, A.~P. and {Terashima}, Y.},
        title = "{Iron line profiles in Suzaku spectra of bare Seyfert galaxies}",
      journal = {\mnras},
         year = 2011,
        month = mar,
       volume = {411},
       number = {4},
        pages = {2353-2370},
          doi = {10.1111/j.1365-2966.2010.17852.x},
archivePrefix = {arXiv},
       eprint = {1010.2080},
 primaryClass = {astro-ph.HE},
       adsurl = {https://ui.adsabs.harvard.edu/abs/2011MNRAS.411.2353P}
}

@ARTICLE{barvainis1987,
       author = {{Barvainis}, Richard},
        title = "{Hot Dust and the Near-Infrared Bump in the Continuum Spectra of Quasars and Active Galactic Nuclei}",
      journal = {\apj},
         year = 1987,
        month = sep,
       volume = {320},
        pages = {537},
          doi = {10.1086/165571},
       adsurl = {https://ui.adsabs.harvard.edu/abs/1987ApJ...320..537B}
}

@ARTICLE{wilkes1994,
       author = {{Wilkes}, Belinda J. and {Tananbaum}, Harvey and {Worrall}, D.~M. and {Avni}, Yoram and {Oey}, M.~S. and {Flanagan}, Joan},
        title = "{The Einstein Database of IPC X-Ray Observations of Optically Selected and Radio-selected Quasars. I.}",
      journal = {\apjs},
         year = 1994,
        month = may,
       volume = {92},
        pages = {53},
          doi = {10.1086/191959},
       adsurl = {https://ui.adsabs.harvard.edu/abs/1994ApJS...92...53W}
}

@ARTICLE{zheng1995,
       author = {{Zheng}, Wei and {Kriss}, Gerard A. and {Davidsen}, Arthur F. and {Kruk}, Jeffrey W.},
        title = "{Far-Ultraviolet Spectrum of Fairall 9 with the Hopkins Ultraviolet Telescope}",
      journal = {\apjl},
         year = 1995,
        month = nov,
       volume = {454},
        pages = {L11},
          doi = {10.1086/309769},
       adsurl = {https://ui.adsabs.harvard.edu/abs/1995ApJ...454L..11Z}
}

@ARTICLE{jaiswal2025,
       author = {{Jaiswal}, Vikram Kumar and {Mandal}, Amit Kumar and {Prince}, Raj and {Pandey}, Ashwani and {Naddaf}, Mohammad Hassan and {Czerny}, Bo{\.z}ena and {Panda}, Swayamtrupta and {Pozo Nu{\~n}ez}, Francisco},
        title = "{Application of the FRADO model of broad line region formation to Seyfert galaxy NGC 5548 and a first step toward determining the Hubble constant}",
      journal = {\aap},
         year = 2025,
        month = oct,
       volume = {702},
          eid = {A92},
        pages = {A92},
          doi = {10.1051/0004-6361/202452497},
archivePrefix = {arXiv},
       eprint = {2410.03597},
 primaryClass = {astro-ph.GA},
       adsurl = {https://ui.adsabs.harvard.edu/abs/2025A&A...702A..92J}
}

@ARTICLE{2010ApJ...712.1129M,
       author = {{Morgan}, Christopher W. and {Kochanek}, C.~S. and {Morgan}, Nicholas D. and {Falco}, Emilio E.},
        title = "{The Quasar Accretion Disk Size-Black Hole Mass Relation}",
      journal = {\apj},
         year = 2010,
        month = apr,
       volume = {712},
       number = {2},
        pages = {1129-1136},
          doi = {10.1088/0004-637X/712/2/1129},
archivePrefix = {arXiv},
       eprint = {1002.4160},
 primaryClass = {astro-ph.CO},
       adsurl = {https://ui.adsabs.harvard.edu/abs/2010ApJ...712.1129M}
}

@ARTICLE{2022ApJ...940...20G,
       author = {{Guo}, Hengxiao and {Barth}, Aaron J. and {Wang}, Shu},
        title = "{Active Galactic Nuclei Continuum Reverberation Mapping Based on Zwicky Transient Facility Light Curves}",
      journal = {\apj},
         year = 2022,
        month = nov,
       volume = {940},
       number = {1},
          eid = {20},
        pages = {20},
          doi = {10.3847/1538-4357/ac96ec},
archivePrefix = {arXiv},
       eprint = {2207.06432},
 primaryClass = {astro-ph.GA},
       adsurl = {https://ui.adsabs.harvard.edu/abs/2022ApJ...940...20G}
}

@ARTICLE{2023RMxAA..59..327C,
       author = {{Chatzikos}, M. and {Bianchi}, S. and {Camilloni}, F. and {Chakraborty}, P. and {Gunasekera}, C.~M. and {Guzm{\'a}n}, F. and {Milby}, J.~S. and {Sarkar}, A. and {Shaw}, G. and {van Hoof}, P.~A.~M. and {Ferland}, G.~J.},
        title = "{The 2023 Release of Cloudy}",
      journal = {\rmxaa},
         year = 2023,
        month = oct,
       volume = {59},
        pages = {327-343},
          doi = {10.22201/ia.01851101p.2023.59.02.12},
archivePrefix = {arXiv},
       eprint = {2308.06396},
 primaryClass = {astro-ph.GA},
       adsurl = {https://ui.adsabs.harvard.edu/abs/2023RMxAA..59..327C}
}

@ARTICLE{2014MNRAS.444.1469M,
       author = {{McHardy}, I.~M. and {Cameron}, D.~T. and {Dwelly}, T. and {Connolly}, S. and {Lira}, P. and {Emmanoulopoulos}, D. and {Gelbord}, J. and {Breedt}, E. and {Arevalo}, P. and {Uttley}, P.},
        title = "{Swift monitoring of NGC 5548: X-ray reprocessing and short-term UV/optical variability}",
      journal = {\mnras},
         year = 2014,
        month = oct,
       volume = {444},
       number = {2},
        pages = {1469-1474},
          doi = {10.1093/mnras/stu1636},
archivePrefix = {arXiv},
       eprint = {1407.6361},
 primaryClass = {astro-ph.HE},
       adsurl = {https://ui.adsabs.harvard.edu/abs/2014MNRAS.444.1469M}
}

@ARTICLE{pozo2024,
       author = {{Pozo Nu{\~n}ez}, F. and {Czerny}, B. and {Panda}, S. and {Kovacevic}, A. and {Brandt}, W. and {Horne}, K. and {LSST AGN Science Collaboration}},
        title = "{Reevaluating LSST's Capability for Time Delay Measurements in Quasar Accretion Disks}",
      journal = {Research Notes of the American Astronomical Society},
         year = 2024,
        month = feb,
       volume = {8},
       number = {2},
          eid = {47},
        pages = {47},
          doi = {10.3847/2515-5172/ad284a},
archivePrefix = {arXiv},
       eprint = {2402.10969},
 primaryClass = {astro-ph.CO},
       adsurl = {https://ui.adsabs.harvard.edu/abs/2024RNAAS...8...47P}
}

@ARTICLE{2026ApJ..1000..165L,
       author = {{Li}, Guodong and {Assef}, Roberto J. and {Brandt}, W.~N. and {Temple}, Matthew J. and {Bauer}, Franz E. and {Marculewicz}, Marcin and {Panda}, Swayamtrupta and {Peca}, Alessandro and {Ricci}, Claudio and {Richards}, Gordon T. and {Satheesh-Sheeba}, Sarath and {Tsai}, Chao-Wei and {Wu}, Jingwen and {Yoon}, Ilsang},
        title = "{Predicting Quasar Counts Detectable in the LSST Survey}",
      journal = {\apj},
         year = 2026,
        month = apr,
       volume = {1000},
       number = {2},
          eid = {165},
        pages = {165},
          doi = {10.3847/1538-4357/ae3164},
archivePrefix = {arXiv},
       eprint = {2512.08654},
 primaryClass = {astro-ph.GA},
       adsurl = {https://ui.adsabs.harvard.edu/abs/2026ApJ..1000..165L}
}

@ARTICLE{pozo2023,
       author = {{Pozo Nu{\~n}ez}, F. and {Bruckmann}, C. and {Deesamutara}, S. and {Czerny}, B. and {Panda}, S. and {Lobban}, A.~P. and {Pietrzy{\'n}ski}, G. and {Polsterer}, K.~L.},
        title = "{Modelling photometric reverberation mapping data for the next generation of big data surveys. Quasar accretion discs sizes with the LSST}",
      journal = {\mnras},
         year = 2023,
        month = jun,
       volume = {522},
       number = {2},
        pages = {2002-2018},
          doi = {10.1093/mnras/stad286},
archivePrefix = {arXiv},
       eprint = {2212.09161},
 primaryClass = {astro-ph.CO},
       adsurl = {https://ui.adsabs.harvard.edu/abs/2023MNRAS.522.2002P}
}

@ARTICLE{kammoun2023,
       author = {{Kammoun}, E.~S. and {Robin}, L. and {Papadakis}, I.~E. and {Dov{\v{c}}iak}, M. and {Panagiotou}, C.},
        title = "{Revisiting UV/optical continuum time lags in AGN}",
      journal = {\mnras},
         year = 2023,
        month = nov,
       volume = {526},
       number = {1},
        pages = {138-151},
          doi = {10.1093/mnras/stad2701},
archivePrefix = {arXiv},
       eprint = {2309.05392},
 primaryClass = {astro-ph.HE},
       adsurl = {https://ui.adsabs.harvard.edu/abs/2023MNRAS.526..138K}
}

@ARTICLE{eleonora_2022,
       author = {{Abdalla}, Elcio and {Abell{\'a}n}, Guillermo Franco and {Aboubrahim}, Amin and {Agnello}, Adriano and {Akarsu}, {\"O}zg{\"u}r and {Akrami}, Yashar and {Alestas}, George and {Aloni}, Daniel and {Amendola}, Luca and {Anchordoqui}, Luis A. and {Anderson}, Richard I. and {Arendse}, Nikki and {Asgari}, Marika and {Ballardini}, Mario and {Barger}, Vernon and {Basilakos}, Spyros and {Batista}, Ronaldo C. and {Battistelli}, Elia S. and {Battye}, Richard and {Benetti}, Micol and {Benisty}, David and {Berlin}, Asher and {de Bernardis}, Paolo and {Berti}, Emanuele and {Bidenko}, Bohdan and {Birrer}, Simon and {Blakeslee}, John P. and {Boddy}, Kimberly K. and {Bom}, Clecio R. and {Bonilla}, Alexander and {Borghi}, Nicola and {Bouchet}, Fran{\c{c}}ois R. and {Braglia}, Matteo and {Buchert}, Thomas and {Buckley-Geer}, Elizabeth and {Calabrese}, Erminia and {Caldwell}, Robert R. and {Camarena}, David and {Capozziello}, Salvatore and {Casertano}, Stefano and {Chen}, Geoff C. -F. and {Chluba}, Jens and {Chen}, Angela and {Chen}, Hsin-Yu and {Chudaykin}, Anton and {Cicoli}, Michele and {Copi}, Craig J. and {Courbin}, Fred and {Cyr-Racine}, Francis-Yan and {Czerny}, Bo{\.z}ena and {Dainotti}, Maria and {D'Amico}, Guido and {Davis}, Anne-Christine and {de Cruz P{\'e}rez}, Javier and {de Haro}, Jaume and {Delabrouille}, Jacques and {Denton}, Peter B. and {Dhawan}, Suhail and {Dienes}, Keith R. and {Di Valentino}, Eleonora and {Du}, Pu and {Eckert}, Dominique and {Escamilla-Rivera}, Celia and {Fert{\'e}}, Agn{\`e}s and {Finelli}, Fabio and {Fosalba}, Pablo and {Freedman}, Wendy L. and {Frusciante}, Noemi and {Gazta{\~n}aga}, Enrique and {Giar{\`e}}, William and {Giusarma}, Elena and {G{\'o}mez-Valent}, Adri{\`a} and {Handley}, Will and {Harrison}, Ian and {Hart}, Luke and {Hazra}, Dhiraj Kumar and {Heavens}, Alan and {Heinesen}, Asta and {Hildebrandt}, Hendrik and {Hill}, J. Colin and {Hogg}, Natalie B. and {Holz}, Daniel E. and {Hooper}, Deanna C. and {Hosseininejad}, Nikoo and {Huterer}, Dragan and {Ishak}, Mustapha and {Ivanov}, Mikhail M. and {Jaffe}, Andrew H. and {Jang}, In Sung and {Jedamzik}, Karsten and {Jimenez}, Raul and {Joseph}, Melissa and {Joudaki}, Shahab and {Kamionkowski}, Marc and {Karwal}, Tanvi and {Kazantzidis}, Lavrentios and {Keeley}, Ryan E. and {Klasen}, Michael and {Komatsu}, Eiichiro and {Koopmans}, L{\'e}on V.~E. and {Kumar}, Suresh and {Lamagna}, Luca and {Lazkoz}, Ruth and {Lee}, Chung-Chi and {Lesgourgues}, Julien and {Levi Said}, Jackson and {Lewis}, Tiffany R. and {L'Huillier}, Benjamin and {Lucca}, Matteo and {Maartens}, Roy and {Macri}, Lucas M. and {Marfatia}, Danny and {Marra}, Valerio and {Martins}, Carlos J.~A.~P. and {Masi}, Silvia and {Matarrese}, Sabino and {Mazumdar}, Arindam and {Melchiorri}, Alessandro and {Mena}, Olga and {Mersini-Houghton}, Laura and {Mertens}, James and {Milakovi{\'c}}, Dinko and {Minami}, Yuto and {Miranda}, Vivian and {Moreno-Pulido}, Cristian and {Moresco}, Michele and {Mota}, David F. and {Mottola}, Emil and {Mozzon}, Simone and {Muir}, Jessica and {Mukherjee}, Ankan and {Mukherjee}, Suvodip and {Naselsky}, Pavel and {Nath}, Pran and {Nesseris}, Savvas and {Niedermann}, Florian and {Notari}, Alessio and {Nunes}, Rafael C. and {{\'O} Colg{\'a}in}, Eoin and {Owens}, Kayla A. and {{\"O}z{\"u}lker}, Emre and {Pace}, Francesco and {Paliathanasis}, Andronikos and {Palmese}, Antonella and {Pan}, Supriya and {Paoletti}, Daniela and {Perez Bergliaffa}, Santiago E. and {Perivolaropoulos}, Leandros and {Pesce}, Dominic W. and {Pettorino}, Valeria and {Philcox}, Oliver H.~E. and {Pogosian}, Levon and {Poulin}, Vivian and {Poulot}, Gaspard and {Raveri}, Marco and {Reid}, Mark J. and {Renzi}, Fabrizio and {Riess}, Adam G. and {Sabla}, Vivian I. and {Salucci}, Paolo and {Salzano}, Vincenzo and {Saridakis}, Emmanuel N. and {Sathyaprakash}, Bangalore S. and {Schmaltz}, Martin and {Sch{\"o}neberg}, Nils and {Scolnic}, Dan and {Sen}, Anjan A. and {Sehgal}, Neelima and {Shafieloo}, Arman and {Sheikh-Jabbari}, M.~M. and {Silk}, Joseph and {Silvestri}, Alessandra and {Skara}, Foteini and {Sloth}, Martin S. and {Soares-Santos}, Marcelle and {Sol{\`a} Peracaula}, Joan and {Songsheng}, Yu-Yang and {Soriano}, Jorge F. and {Staicova}, Denitsa and {Starkman}, Glenn D. and {Szapudi}, Istv{\'a}n and {Teixeira}, Elsa M. and {Thomas}, Brooks and {Treu}, Tommaso and {Trott}, Emery and {van de Bruck}, Carsten and {Vazquez}, J. Alberto and {Verde}, Licia and {Visinelli}, Luca and {Wang}, Deng and {Wang}, Jian-Min and {Wang}, Shao-Jiang and {Watkins}, Richard and {Watson}, Scott and {Webb}, John K. and {Weiner}, Neal and {Weltman}, Amanda and {Witte}, Samuel J. and {Wojtak}, Rados{\l}aw and {Yadav}, Anil Kumar and {Yang}, Weiqiang and {Zhao}, Gong-Bo and {Zumalac{\'a}rregui}, Miguel},
        title = "{Cosmology intertwined: A review of the particle physics, astrophysics, and cosmology associated with the cosmological tensions and anomalies}",
      journal = {Journal of High Energy Astrophysics},
         year = 2022,
        month = jun,
       volume = {34},
        pages = {49-211},
          doi = {10.1016/j.jheap.2022.04.002},
archivePrefix = {arXiv},
       eprint = {2203.06142},
 primaryClass = {astro-ph.CO},
       adsurl = {https://ui.adsabs.harvard.edu/abs/2022JHEAp..34...49A}
}

@ARTICLE{netzer2024,
       author = {{Netzer}, Hagai and {Goad}, Michael R. and {Barth}, Aaron J. and {Cackett}, Edward M. and {Horne}, Keith and {Hu}, Chen and {Kara}, Erin and {Korista}, Kirk T. and {Kriss}, Gerard A. and {Lewin}, Collin and {Montano}, John and {Arav}, Nahum and {Behar}, Ehud and {Brotherton}, Michael S. and {Chelouche}, Doron and {De Rosa}, Gisella and {Dalla Bont{\`a}}, Elena and {Dehghanian}, Maryam and {Ferland}, Gary J. and {Fian}, Carina and {Homayouni}, Yasaman and {Ili{\'c}}, Dragana and {Kaspi}, Shai and {Kova{\v{c}}evi{\'c}}, Andjelka B. and {Landt}, Hermine and {{\v{C}}. Popovi{\'c}}, Luka and {Storchi-Bergmann}, Thaisa and {Wang}, Jian-Min and {Zaidouni}, Fatima},
        title = "{AGN STORM 2. X. The Origin of the Interband Continuum Delays in Mrk 817}",
      journal = {\apj},
         year = 2024,
        month = nov,
       volume = {976},
       number = {1},
          eid = {59},
        pages = {59},
          doi = {10.3847/1538-4357/ad8160},
archivePrefix = {arXiv},
       eprint = {2410.02652},
 primaryClass = {astro-ph.GA},
       adsurl = {https://ui.adsabs.harvard.edu/abs/2024ApJ...976...59N}
}

@ARTICLE{pandey2023,
       author = {{Pandey}, Ashwani and {Czerny}, Bo{\.z}ena and {Panda}, Swayamtrupta and {Prince}, Raj and {Jaiswal}, Vikram Kumar and {Martinez-Aldama}, Mary Loli and {Zaja{\v{c}}ek}, Michal and {{\'S}niegowska}, Marzena},
        title = "{Broad-line region in active galactic nuclei: Dusty or dustless?}",
      journal = {\aap},
         year = 2023,
        month = dec,
       volume = {680},
          eid = {A102},
        pages = {A102},
          doi = {10.1051/0004-6361/202347819},
archivePrefix = {arXiv},
       eprint = {2310.05089},
 primaryClass = {astro-ph.GA},
       adsurl = {https://ui.adsabs.harvard.edu/abs/2023A&A...680A.102P}
}

@ARTICLE{petrucci2020,
       author = {{Petrucci}, P. -O. and {Gronkiewicz}, D. and {Rozanska}, A. and {Belmont}, R. and {Bianchi}, S. and {Czerny}, B. and {Matt}, G. and {Malzac}, J. and {Middei}, R. and {De Rosa}, A. and {Ursini}, F. and {Cappi}, M.},
        title = "{Radiation spectra of warm and optically thick coronae in AGNs}",
      journal = {\aap},
         year = 2020,
        month = feb,
       volume = {634},
          eid = {A85},
        pages = {A85},
          doi = {10.1051/0004-6361/201937011},
archivePrefix = {arXiv},
       eprint = {2001.02026},
 primaryClass = {astro-ph.HE},
       adsurl = {https://ui.adsabs.harvard.edu/abs/2020A&A...634A..85P}
}

@ARTICLE{kammoun2021,
       author = {{Kammoun}, E.~S. and {Papadakis}, I.~E. and {Dov{\v{c}}iak}, M.},
        title = "{Modelling the UV/optical continuum time-lags in AGN}",
      journal = {\mnras},
         year = 2021,
        month = may,
       volume = {503},
       number = {3},
        pages = {4163-4171},
          doi = {10.1093/mnras/stab725},
archivePrefix = {arXiv},
       eprint = {2103.04892},
 primaryClass = {astro-ph.HE},
       adsurl = {https://ui.adsabs.harvard.edu/abs/2021MNRAS.503.4163K}
}

@ARTICLE{kinney1996,
       author = {{Kinney}, Anne L. and {Calzetti}, Daniela and {Bohlin}, Ralph C. and {McQuade}, Kerry and {Storchi-Bergmann}, Thaisa and {Schmitt}, Henrique R.},
        title = "{Template Ultraviolet to Near-Infrared Spectra of Star-forming Galaxies and Their Application to K-Corrections}",
      journal = {\apj},
         year = 1996,
        month = aug,
       volume = {467},
        pages = {38},
          doi = {10.1086/177583},
       adsurl = {https://ui.adsabs.harvard.edu/abs/1996ApJ...467...38K}
}

@ARTICLE{netzer2022,
       author = {{Netzer}, Hagai},
        title = "{Continuum reverberation mapping and a new lag-luminosity relationship for AGN}",
      journal = {\mnras},
         year = 2022,
        month = jan,
       volume = {509},
       number = {2},
        pages = {2637-2646},
          doi = {10.1093/mnras/stab3133},
archivePrefix = {arXiv},
       eprint = {2110.05512},
 primaryClass = {astro-ph.GA},
       adsurl = {https://ui.adsabs.harvard.edu/abs/2022MNRAS.509.2637N}
}

@ARTICLE{korista2001,
       author = {{Korista}, Kirk T. and {Goad}, Michael R.},
        title = "{The Variable Diffuse Continuum Emission of Broad-Line Clouds}",
      journal = {\apj},
         year = 2001,
        month = jun,
       volume = {553},
       number = {2},
        pages = {695-708},
          doi = {10.1086/320964},
archivePrefix = {arXiv},
       eprint = {astro-ph/0101117},
 primaryClass = {astro-ph},
       adsurl = {https://ui.adsabs.harvard.edu/abs/2001ApJ...553..695K}
}

@ARTICLE{2025MNRAS.544.4532O,
       author = {{Owen}, James E. and {Lin}, Douglas N.~C.},
        title = "{A dust condensation instability in AGN atmospheres: failed winds and the broad-line region}",
      journal = {\mnras},
         year = 2025,
        month = dec,
       volume = {544},
       number = {4},
        pages = {4532-4550},
          doi = {10.1093/mnras/staf1914},
archivePrefix = {arXiv},
       eprint = {2511.02920},
 primaryClass = {astro-ph.GA},
       adsurl = {https://ui.adsabs.harvard.edu/abs/2025MNRAS.544.4532O}
}

@ARTICLE{naddaf2022,
       author = {{Naddaf}, M.~H. and {Czerny}, B.},
        title = "{Radiation pressure on dust explaining the low ionized broad emission lines in active galactic nuclei. Dust as an important driver of line shape}",
      journal = {\aap},
         year = 2022,
        month = jul,
       volume = {663},
          eid = {A77},
        pages = {A77},
          doi = {10.1051/0004-6361/202142806},
archivePrefix = {arXiv},
       eprint = {2111.14963},
 primaryClass = {astro-ph.GA},
       adsurl = {https://ui.adsabs.harvard.edu/abs/2022A&A...663A..77N}
}

@ARTICLE{davidson1976,
       author = {{Davidson}, K.},
        title = "{Some remarks concerning Lyman-continuum emission in quasar spectra.}",
      journal = {\apj},
         year = 1976,
        month = aug,
       volume = {207},
        pages = {710-712},
          doi = {10.1086/154539},
       adsurl = {https://ui.adsabs.harvard.edu/abs/1976ApJ...207..710D}
}

@ARTICLE{hawley1978,
       author = {{Hawley}, S.~A. and {Phillips}, M.~M.},
        title = "{Spectrophotometry of Fairall 9 and Fairall 51: two extremes of the Seyfert 1 class.}",
      journal = {\apj},
         year = 1978,
        month = nov,
       volume = {225},
        pages = {780-783},
          doi = {10.1086/156542},
       adsurl = {https://ui.adsabs.harvard.edu/abs/1978ApJ...225..780H}
}

@ARTICLE{2001ApJ...553...47F,
       author = {{Freedman}, Wendy L. and {Madore}, Barry F. and {Gibson}, Brad K. and {Ferrarese}, Laura and {Kelson}, Daniel D. and {Sakai}, Shoko and {Mould}, Jeremy R. and {Kennicutt}, Jr., Robert C. and {Ford}, Holland C. and {Graham}, John A. and {Huchra}, John P. and {Hughes}, Shaun M.~G. and {Illingworth}, Garth D. and {Macri}, Lucas M. and {Stetson}, Peter B.},
        title = "{Final Results from the Hubble Space Telescope Key Project to Measure the Hubble Constant}",
      journal = {\apj},
         year = 2001,
        month = may,
       volume = {553},
       number = {1},
        pages = {47-72},
          doi = {10.1086/320638},
archivePrefix = {arXiv},
       eprint = {astro-ph/0012376},
 primaryClass = {astro-ph},
       adsurl = {https://ui.adsabs.harvard.edu/abs/2001ApJ...553...47F}
}

@ARTICLE{2019ApJ...875..133P,
       author = {{Panda}, Swayamtrupta and {Czerny}, Bo{\.z}ena and {Done}, Chris and {Kubota}, Aya},
        title = "{CLOUDY View of the Warm Corona}",
      journal = {\apj},
         year = 2019,
        month = apr,
       volume = {875},
       number = {2},
          eid = {133},
        pages = {133},
          doi = {10.3847/1538-4357/ab11cb},
archivePrefix = {arXiv},
       eprint = {1901.02962},
 primaryClass = {astro-ph.HE},
       adsurl = {https://ui.adsabs.harvard.edu/abs/2019ApJ...875..133P}
}

@ARTICLE{2025Univ...11...69M,
       author = {{Marziani}, Paola and {Garnica Luna}, Karla and {Floris}, Alberto and {del Olmo}, Ascensi{\'o}n and {Deconto-Machado}, Alice and {Buendia-Rios}, Tania M. and {Negrete}, C. Alenka and {Dultzin}, Deborah},
        title = "{Super-Eddington Accretion in Quasars}",
      journal = {Universe},
         year = 2025,
        month = feb,
       volume = {11},
       number = {2},
          eid = {69},
        pages = {69},
          doi = {10.3390/universe11020069},
archivePrefix = {arXiv},
       eprint = {2502.14713},
 primaryClass = {astro-ph.GA},
       adsurl = {https://ui.adsabs.harvard.edu/abs/2025Univ...11...69M}
}

@ARTICLE{jiang2021,
       author = {{Jiang}, Bo-Wei and {Marziani}, Paola and {Savi{\'c}}, {\DJ}or{\dj}e and {Shablovinskaya}, Elena and {Popovi{\'c}}, Luka {\v{C}}. and {Afanasiev}, Victor L. and {Czerny}, Bo{\.z}ena and {Wang}, Jian-Min and {del Olmo}, Ascensi{\'o}n and {D'Onofrio}, Mauro and {{\'S}niegowska}, Marzena and {Mazzei}, Paola and {Panda}, Swayamtrupta},
        title = "{Linear spectropolarimetric analysis of fairall 9 with VLT/FORS2}",
      journal = {\mnras},
         year = 2021,
        month = nov,
       volume = {508},
       number = {1},
        pages = {79-99},
          doi = {10.1093/mnras/stab2273},
archivePrefix = {arXiv},
       eprint = {2108.00983},
 primaryClass = {astro-ph.GA},
       adsurl = {https://ui.adsabs.harvard.edu/abs/2021MNRAS.508...79J}
}

@ARTICLE{fairall1977,
       author = {{Fairall}, A.~P.},
        title = "{150 southern compact and bright-nucleus galaxies.}",
      journal = {\mnras},
         year = 1977,
        month = aug,
       volume = {180},
        pages = {391-400},
          doi = {10.1093/mnras/180.3.391},
       adsurl = {https://ui.adsabs.harvard.edu/abs/1977MNRAS.180..391F}
}

@ARTICLE{noda2013,
       author = {{Noda}, Hirofumi and {Makishima}, Kazuo and {Nakazawa}, Kazuhiro and {Uchiyama}, Hideki and {Yamada}, Shin'ya and {Sakurai}, Soki},
        title = "{The Nature of Stable Soft X-Ray Emissions in Several Types of Active Galactic Nuclei Observed by Suzaku}",
      journal = {\pasj},
         year = 2013,
        month = feb,
       volume = {65},
          eid = {4},
        pages = {4},
          doi = {10.1093/pasj/65.1.4},
archivePrefix = {arXiv},
       eprint = {1208.3536},
 primaryClass = {astro-ph.CO},
       adsurl = {https://ui.adsabs.harvard.edu/abs/2013PASJ...65....4N}
}

@ARTICLE{koratkar1989,
       author = {{Koratkar}, Anuradha P. and {Gaskell}, C. Martin},
        title = "{Emission-Line Variability of Fairall 9: Determination of the Size of the Broad-Line Region and the Direction of Gas Motion}",
      journal = {\apj},
         year = 1989,
        month = oct,
       volume = {345},
        pages = {637},
          doi = {10.1086/167937},
       adsurl = {https://ui.adsabs.harvard.edu/abs/1989ApJ...345..637K}
}

@ARTICLE{rodriguez1997,
       author = {{Rodr{\'\i}guez-Pascual}, P.~M. and {Alloin}, D. and {Clavel}, J. and {Crenshaw}, D.~M. and {Horne}, K. and {Kriss}, G.~A. and {Krolik}, J.~H. and {Malkan}, M.~A. and {Netzer}, H. and {O'Brien}, P.~T. and {Peterson}, B.~M. and {Reichert}, G.~A. and {Wamsteker}, W. and {Alexander}, T. and {Barr}, P. and {Blandford}, R.~D. and {Bregman}, J.~N. and {Carone}, T.~E. and {Clements}, S. and {Courvoisier}, T. -J. and {De Robertis}, M.~M. and {Dietrich}, M. and {Dottori}, H. and {Edelson}, R.~A. and {Filippenko}, A.~V. and {Gaskell}, C.~M. and {Huchra}, J.~P. and {Hutchings}, J.~B. and {Kollatschny}, W. and {Koratkar}, A.~P. and {Korista}, K.~T. and {Laor}, A. and {MacAlpine}, G.~M. and {Martin}, P.~G. and {Maoz}, D. and {McCollum}, B. and {Morris}, S.~L. and {Perola}, G.~C. and {Pogge}, R.~W. and {Ptak}, R.~L. and {Recondo-Gonz{\'a}lez}, M.~C. and {Rodr{\'\i}guez-Espinoza}, J.~M. and {Rokaki}, E.~L. and {Santos-Lle{\'o}}, M. and {Sekiguchi}, K. and {Shull}, J.~M. and {Snijders}, M.~A.~J. and {Sparke}, L.~S. and {Stirpe}, G.~M. and {Stoner}, R.~E. and {Sun}, W. -H. and {Wagner}, S.~J. and {Wanders}, I. and {Wilkes}, J. and {Winge}, C. and {Zheng}, W.},
        title = "{Steps toward Determination of the Size and Structure of the Broad-Line Region in Active Galactic Nuclei. IX. Ultraviolet Observations of Fairall 9}",
      journal = {\apjs},
         year = 1997,
        month = may,
       volume = {110},
       number = {1},
        pages = {9-20},
          doi = {10.1086/312996},
       adsurl = {https://ui.adsabs.harvard.edu/abs/1997ApJS..110....9R}
}

@ARTICLE{chapman1985,
       author = {{Chapman}, G.~N.~F. and {Geller}, M.~J. and {Huchra}, J.~P.},
        title = "{The ultraviolet variability of Seyfert 1 galaxies.}",
      journal = {\apj},
         year = 1985,
        month = oct,
       volume = {297},
        pages = {151-165},
          doi = {10.1086/163512},
       adsurl = {https://ui.adsabs.harvard.edu/abs/1985ApJ...297..151C}
}

@ARTICLE{2024A&A...689A.321F,
       author = {{Floris}, A. and {Marziani}, P. and {Panda}, S. and {Sniegowska}, M. and {D'Onofrio}, M. and {Deconto-Machado}, A. and {del Olmo}, A. and {Czerny}, B.},
        title = "{Chemical abundances along the quasar main sequence}",
      journal = {\aap},
         year = 2024,
        month = sep,
       volume = {689},
          eid = {A321},
        pages = {A321},
          doi = {10.1051/0004-6361/202450458},
archivePrefix = {arXiv},
       eprint = {2405.04456},
 primaryClass = {astro-ph.GA},
       adsurl = {https://ui.adsabs.harvard.edu/abs/2024A&A...689A.321F}
}

\begin{appendix}
\label{ss:apendix}
\onecolumn

\section{BLR radius at different luminosity distances}
\label{dis:asump_blr}

\begin{figure}
    \centering
    \includegraphics[scale=0.4]{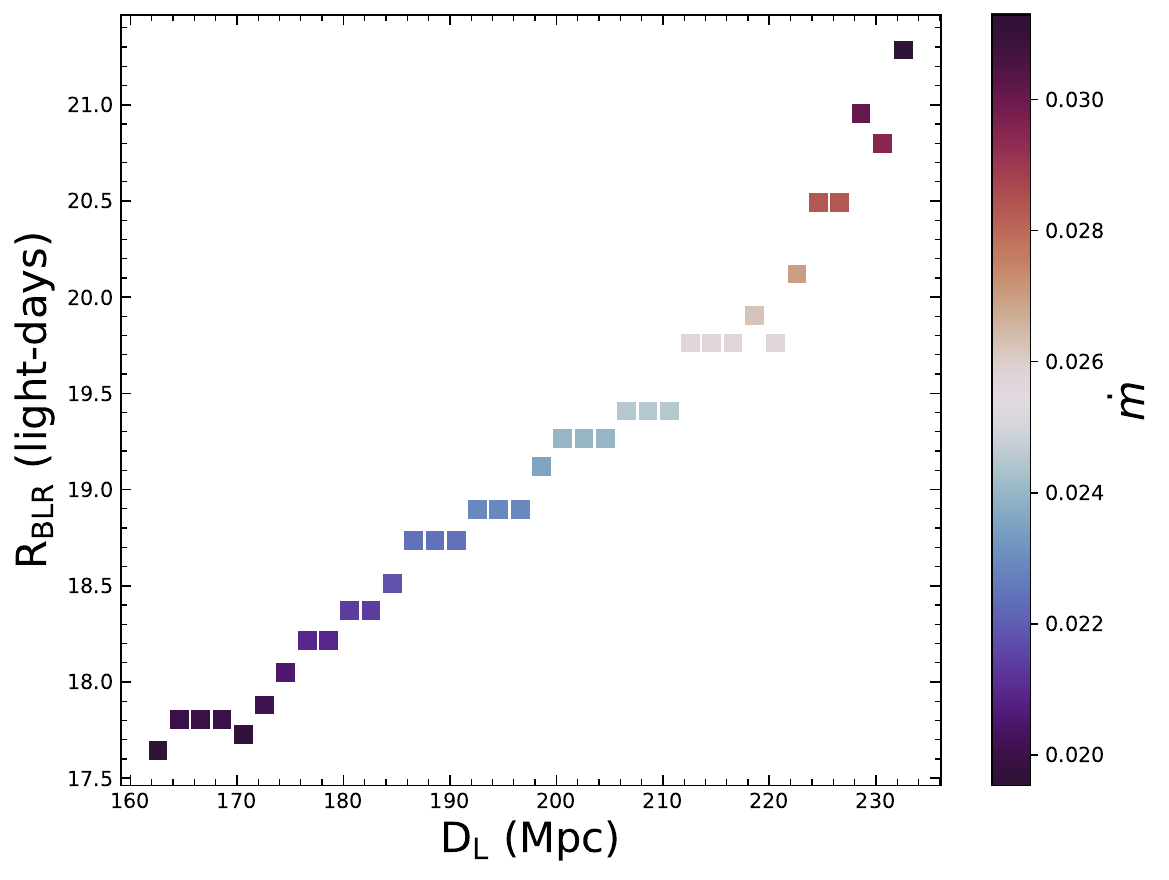}
    \includegraphics[scale=0.4]{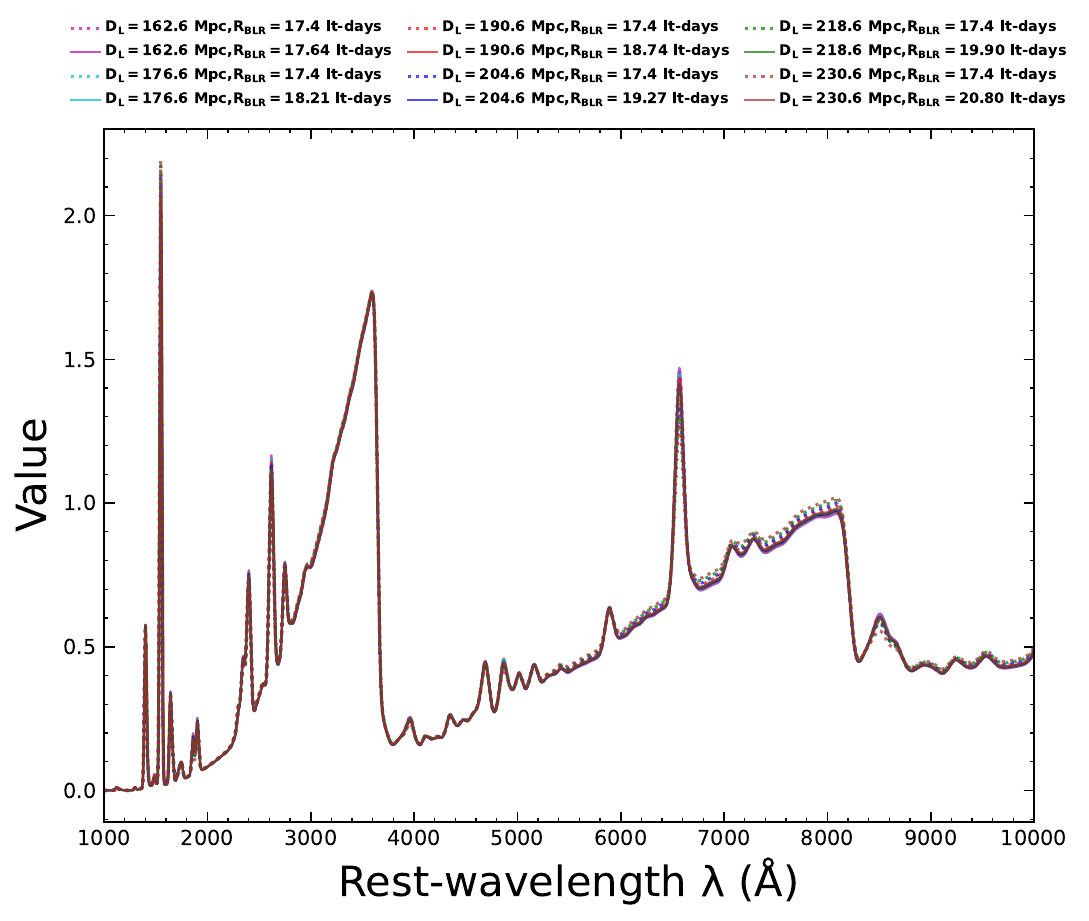}
    
    \caption{Impact of the BLR radius, $R_{\rm BLR}$ assumption on the model fitting. Left: Median values of $R_{\rm BLR}$ derived from $\psi_{\rm BLR}(t)$ at each luminosity distance, $D_{\rm L}$, plotted as a function of $D_{\rm L}$. Colors indicate different accretion rates, $\dot{m}$. Right: Comparison of {\tt Cloudy}-computed BLR emissivity profiles obtained using the fitted $R_{\rm BLR}$ values (solid lines) and the fixed fiducial $R_{\rm BLR}$ value (dotted lines). Different colors correspond to different luminosity distances spanning the full range of the model grid.
    } 
    \label{fig:blr_lg}
\end{figure}

In our current modeling framework, we treated $R_{\rm BLR}$ as a semi-free parameter. Specifically, $R_{\rm BLR}$ was fixed during the {\tt Cloudy} calculations, but was allowed to vary when constructing $\psi_{\rm BLR}(t)$ through its dependence on the accretion rate, $\dot{m}$. The left panel of Figure~\ref{fig:blr_lg} shows the median values of $R_{\rm BLR}$, derived from the constructed $\psi_{\rm BLR}(t)$ at each luminosity distance, $D_{\rm L}$, as a function of $D_{\rm L}$ and color-coded by $\dot{m}$. As expected, $R_{\rm BLR}$ increases with both $D_{\rm L}$ and $\dot{m}$, reflecting the larger BLR radius associated with higher intrinsic luminosities. Importantly, despite this systematic trend, the predicted variations in $R_{\rm BLR}$ remain within the observational uncertainties of the fiducial value inferred from the H$\beta$ BLR lag measurement across the full range of luminosity distances covered by our model grid, except at the highest accretion rate. This behavior indicates that linking $R_{\rm BLR}$ to $\dot{m}$ provides a physically consistent prescription for constructing $\psi_{\rm BLR}(t)$.

For the {\tt Cloudy} calculations, however, we adopted a fixed fiducial value of $R_{\rm BLR}=17.4$ light-days. To evaluate the impact of this assumption, the right panel of Figure~\ref{fig:blr_lg} compares BLR emissivity profiles computed using the fitted $R_{\rm BLR}$ values (solid lines) with those obtained using the fixed fiducial value (dotted lines) across a range of $D_{\rm L}$ spanning our model grid. The resulting emissivity profiles are nearly identical, demonstrating that adopting a fixed $R_{\rm BLR}$ in the {\tt Cloudy} calculations has a negligible effect on  our model-fitting results.

\section{BLR local density in {\tt Cloudy} computation}
\label{dis:cloud}

\begin{figure}[h]
    \centering
    \includegraphics[scale=0.8]{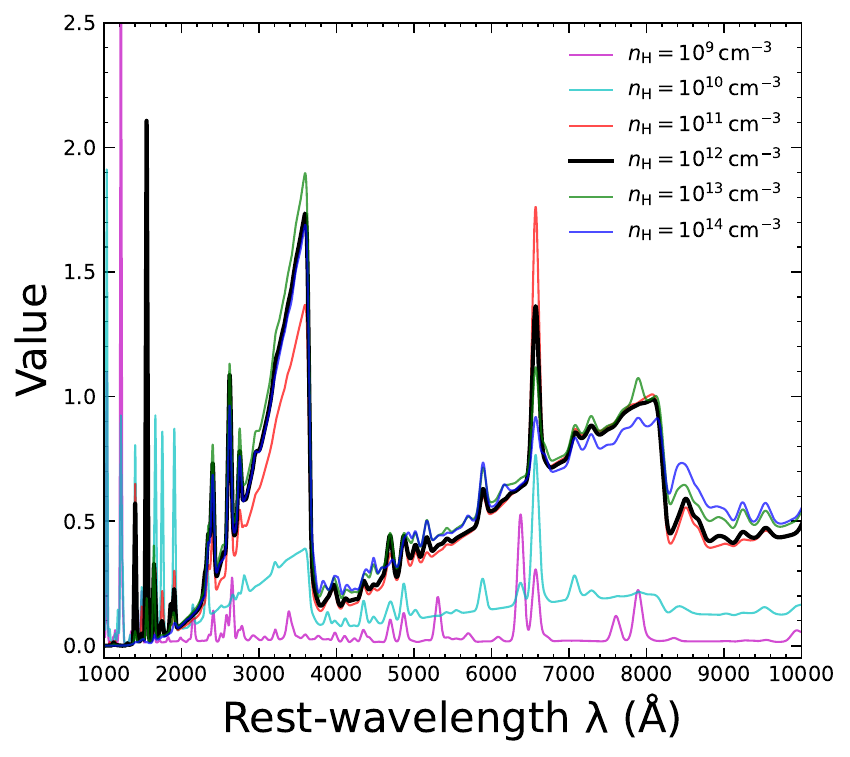}
    \caption{BLR emissivity profiles for different BLR gas densities. Emissivity profile of a BLR cloud centered at $\log r$ (cm) = 16.64, with an incident ionizing luminosity of $\log L$ (erg/s) = 44.97, a column density of $\log N_H$ (cm$^{-2}$) = 23.5, $f_{\rm BLR}=1$, and for a hydrogen density of $\log n_H$ (cm$^{-3}$) = 9, 10, 11, 12, 13 and 14 are shown by the solid magenta, cyan,  red, black, green, and blue lines, respectively.} 
    \label{fig:emv_den}
\end{figure}

The hydrogen gas density of the BLR is an important parameter that influences both the emission-line and continuum properties of AGN. Most of the observed broad-line emission is thought to originate from gas with densities in the range $n_{\rm H}\sim10^{9}-10^{11} \, \mathrm{cm^{-3}}$, while the BLR gas distribution is generally expected to be truncated below $n_{\rm H}\sim10^{9} \, \mathrm{cm^{-3}}$ to avoid the production of strong, broad [O III] emission lines \citep{1989ApJ...347..640R}. At the same time, substantially higher densities, ($n_{\rm H}\sim10^{11}-10^{13} \, \mathrm{cm^{-3}}$), are often required to explain the rich low-ionization emission spectrum observed in quasars, particularly the prominent Fe II features \citep{2004ApJ...615..610B, 2012ApJ...757...62N, 2019ApJ...875..133P, panda2019, panda_cafe2021}.

To investigate the impact of BLR density on our model predictions, Figure~\ref{fig:emv_den} presents the BLR emissivity profiles computed with {\tt Cloudy} for $n_{\rm H}=10^{9}, \, 10^{10}, \, 10^{11}, \, 10^{12}, \, 10^{13}$, and $10^{14} \, \mathrm{cm^{-3}}$. The figure demonstrates that the predicted emissivity profile depends sensitively on the assumed gas density. At the low-density end ($n_{\rm H}\sim10^{9} \, \mathrm{cm^{-3}}$), {\tt Cloudy} significantly underpredicts the AGN continuum contribution arising from free-free and free-bound recombination processes within the BLR, primarily because at lower density, the ionization parameter ($U$) is higher. Although a larger fraction of the gas is ionized, the recombination rate per unit volume is still low because there are fewer particles available to recombine. In contrast, at very high densities ($n_{\rm H}\sim10^{13}-10^{14} \, \mathrm{cm^{-3}}$), the emissivity profiles become nearly indistinguishable from one another, reflecting the fact that the gas becomes less ionized.

In our previous application to NGC~5548, \citet{jaiswal2025} found that a density of $n_{\rm H}\sim10^{11} \, \mathrm{cm^{-3}}$, representative of typical BLR conditions in AGNs, provided a satisfactory fit to both the lag-spectrum and the SED. Fairall~9, however, exhibits Fe II emission features that were not apparent in NGC~5548. Consistent with the expectation that stronger low-ionization emission favors denser BLR gas, we find that model with $n_{\rm H}=10^{12} \, \mathrm{cm^{-3}}$ yields a better fit to the combined lag-spectrum and SED data fitting, producing lower $\chi^{2}/{\rm dof}$ value than model with $n_{\rm H}=10^{11} \, \mathrm{cm^{-3}}$. Therefore, we adopt $n_{\rm H}=10^{12} \, \mathrm{cm^{-3}}$ as the fiducial BLR hydrogen gas density for Fairall~9 in our analysis presented in this work.

\end{appendix}

\end{document}